\documentclass[%
 reprint,
 amsmath,amssymb,
 aps,
]{revtex4-2}

\usepackage{graphicx}% Include figure files
\usepackage{subfigure}
\usepackage{dcolumn}% Align table columns on decimal point
\usepackage{bm}% bold math
\usepackage{braket}
\usepackage{physics}
\usepackage{comment}
\usepackage[normalem]{ulem}
\usepackage{xcolor}
\usepackage{appendix}
\begin{document}

\preprint{APS/123-QED}

\title{Quantum Point Contact with Local Two-body Loss}
% Force line breaks with \\

\author{Kensuke Kakimoto}
\affiliation{
Department of Electronic and Physical Systems,
Waseda University, Tokyo 169-8555, Japan
}
\author{Shun Uchino}%
\affiliation{
Department of Electronic and Physical Systems,
Waseda University, Tokyo 169-8555, Japan
}
\affiliation{
Department of Materials Science,
Waseda University, Tokyo 169-8555, Japan
}
%\affiliation{%


%\affiliation{
%\affiliation{
 %Third institution, the second for Charlie Author
%\affiliation{%


\date{\today}% It is always \today, today,

\begin{abstract}
Motivated by  recent advances in ultracold
atomic gas experiments,
we investigate a two-terminal mesoscopic system
in which two-body loss occurs locally at the center of
a one-dimensional chain.
By means of the self-consistent Born approximation in the Keldysh formalism,
we uncover mesoscopic current formulas that are
experimentally relevant and
applicable to the weak dissipation regime.
Although these formulas 
are analogous to those for systems with one-body loss,
it turns out that the channel transmittance and loss probability depend on the nonequilibrium occupation at the lossy site.
We demonstrate that this occupation dependence leads to a weaker suppression of 
currents in the presence of two-body loss  compared to one-body loss.
%An article usually includes an abstract, a concise summary of the work
%covered at length in the main body of the article. 
%\begin{description}
%use the optional argument of the \verb+\item+ command to give the category %of each item. 
%\end{description}
\end{abstract}

\maketitle

%\tableofcontents

%\textit{Introduction.}\textemdash
\section{Introduction}
Open quantum systems provide a natural arena for exploring the interplay between coherent dynamics and dissipative 
processes~\cite{breuer2002theory,wiseman2009quantum}. In contrast to isolated systems, where unitary evolution governs dynamics, coupling to an environment introduces decoherence and particle loss, fundamentally altering transport properties and steady-state 
behaviors~\cite{wiseman2009quantum,nazarov2009quantum}. Recent advances have revealed that such dissipation can induce nontrivial quantum phenomena, including nonequilibrium quantum criticality~\cite{Torre2010, Torre2012},
many-body quantum Zeno effect~\cite{Misra1977, Fisher2001, Facchi2002, Syassen2008, Facchi2008, Sun2023}, and unconventional renormalization-group flows in strongly correlated settings~\cite{ashida2017parity,Nakagawa2018}.

The development of highly controllable platforms, particularly ultracold atomic gases, has enabled systematic investigation of engineered dissipation in many-body systems~\cite{Daley2014, Muller2012, Sieber2016}. In these setups, particle losses realized via focused light fields and photo-association can be tuned with high precision, allowing experimental access to particle-loss processes~\cite{Barontini2013, Labouvie2015, Labouvie2016, Patil2015, Luschen2017,Mark2012, Syassen2008, Yan2013, Tomita2017, Tomita2019, Huang2025}. These capabilities have motivated a new class of transport experiments that probe the role of dissipation in mesoscopic conduction, with ultracold analogs of quantum point contacts (QPCs) serving as the core geometry~\cite{krinner2017two,RevModPhys.94.041001}.

Of particular interest is the recent realization of a two-terminal QPC setup exhibiting localized loss in a one-dimensional (1D) channel.
Previous theoretical and experimental efforts have primarily focused on one-body loss~\cite{PhysRevA.100.053605,PhysRevB.101.144301,
PhysRevLett.122.040402,PhysRevLett.123.193605,PhysRevLett.129.056802,Uchino2022,PhysRevLett.130.200404,PhysRevResearch.5.033095,PhysRevB.110.205419}, where dissipation effects can be effectively captured with introduction of an additional reservoir in which injection into the channel is prohibited.

In contrast, local two-body loss exhibits fundamentally different characteristics from
one-body loss, as it only occurs when two particles
simultaneously occupy a spatially restricted region.
This constraint renders two-body loss inherently a many-body effect, as explicitly captured by the nonequilibrium many-body formalism~\cite{Sieber2016,kamenev2023field,Stefanucci2024}.
This many-body nature suggests qualitatively distinct transport behavior
due to the nonlinear feedback between occupation and loss.
In particular, under nonequilibrium steady-state
conditions driven by chemical potential and temperature gradients, such dissipation may alter particle and energy currents in ways that elude straightforward extensions of
the Landauer-B\"{u}ttiker formalism~\cite{datta1997electronic,nazarov2009quantum}.
Despite its conceptual and experimental relevance,
a theoretical framework for quantum transport
incorporating two-body loss has remained largely unexplored.

In this work, we theoretically 
investigate two-terminal transport through a 1D 
chain
with localized two-body loss, modeling a dissipative QPC.
To this end,
we develop the Keldysh Green's function formalism combined with a
noise-field representation of Lindblad dynamics~\cite{Gardiner2004, Dolgirev2020, Jin2022}.
By applying the self-consistent Born approximation (SCBA)~\cite{Altland2023}
that is valid in a weak dissipative regime
and is also consistent with the quantum master equation approach,
we derive analytic expressions for mesoscopic particle
and energy currents.
We find that the effective dissipation strength becomes occupation-dependent,
resulting in weaker current suppression
compared to one-body loss.
Our prediction can be
tested in a two-terminal experiment with an ultracold Fermi gas,
where the superfluid reservoir case has recently been realized~\cite{Huang2025}.

The remainder of this paper is organized as follows.
In Sec. II, we introduces the two-terminal transport model with two-body loss. Section III presents the noise-field formalism consistent with the Lindblad master equation approach.  Sections IV and V analyze  single-site and multi-site cases, respectively. Finally, Sec. VI concludes the paper, and technical details of the theoretical analyses are provided in Appendices.

\begin{figure}
    \includegraphics[width=\linewidth]
    {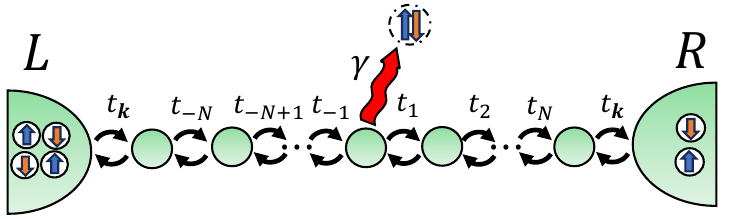}
    % Here is how to import EPS art
    \caption{
        Schematic of the mesoscopic transport setup:
        two normal reservoirs are connected via a
        1D chain representing a QPC.
        Particle, energy, and spin currents are induced by the difference in chemical potentials for each spin
        between the left and right reservoirs $\mu_{L/R, \uparrow/\downarrow}$.
        The local two-body loss, with rate $\gamma$, occurs at the central site when the spin-up and spin-down
        particles are simultaneously present.
    }
\label{system_model}
\end{figure}

%\textit{The model.}\textemdash
\section{The model}
We consider a system where two normal macroscopic reservoirs
are connected via a
QPC (Fig.~\ref{system_model}).
The corresponding Hamiltonian is given by
\begin{equation}
    H
    =   \sum_{s=L, R} \sum_{\sigma=\uparrow, \downarrow}
        H_{s\sigma}
        + H_T + \sum_{\alpha = 0, \pm} H_{1D}^{\alpha},
\end{equation}
\begin{equation}
    H_T
    =   -\sum_{\mathbf{k}, \sigma}
        t_{\mathbf{k}}
        \Big(
            \psi_{L\mathbf{k}\sigma}^{\dagger} d_{-N\sigma}
            +
            \psi_{R\mathbf{k}\sigma}^{\dagger} d_{N\sigma}
        \Big)
        +
        \text{H.c.},
\end{equation}
\begin{equation}
    H_{1D}^{\pm}
    =   -\sum_{\sigma} \sum_{i=1}^N
        t_{\pm i} \Big(
            d_{\pm i\sigma}^{\dagger} d_{\pm(i-1)\sigma}
            +
            d_{\pm(i-1)\sigma}^{\dagger} d_{\pm i\sigma}
        \Big).
\end{equation}
The Hamiltonian of the left (right) reservoir with
spin $\sigma$,
$
H_{L(R)\sigma}
=  \sum_{\mathbf{k}} (\epsilon_{\mathbf{k}}-\mu_{L(R)\sigma})
    \psi_{L(R)\mathbf{k}\sigma}^{\dagger} \psi_{L(R)\mathbf{k}\sigma}
$
is expressed with the kinetic energy 
$\epsilon_{\mathbf{k}}=\mathbf{k}^2/2m,$
the chemical potential $\mu_{L(R)\sigma}$, and
the field annihilation (creation) operator
$\psi_{L(R)\mathbf{k}\sigma}$ ($\psi^{\dagger}_{L(R)\mathbf{k}\sigma}$). 
We model a QPC with the 1D chain, which is composed of
the hopping term between different sites  $H_{1D}^{\pm}$ and 
the onsite energy term
$H_{1D}^0
=   \sum_{\sigma} \sum_{i=-N}^N
    \epsilon_i d_{i\sigma}^{\dagger} d_{i\sigma}$,
with
a potential of energy $\epsilon_i$ and annihilation (creation) operator
in the 1D chain $d_{i\sigma}$ $(d^{\dagger}_{i\sigma})$.
In addition, tunneling between reservoirs and 1D chain occurring at the edges of the 
1D chain is expressed in $H_T$ with the tunnel coupling strength $t_{\mathbf{k}}$.

We next consider two-body loss occurring at the central site
of the 1D chain and treat dissipation within the Markovian Lindblad master equation~\cite{breuer2002theory}. 
We note that such a Markovian treatment is relevant to ultracold atomic gases~\cite{Daley2014}
and has successfully been employed to interpret recent results on dissipative 
quantum transport~\cite{PhysRevLett.130.200404,Huang2025}.
Typically, the open quantum dynamics is analyzed by following 
the non-unitary  evolution of the density matrix,
which involves both
non-Hermitian and quantum jump terms in addition to the unitary dynamics. This treatment is in sharp contrast to the standard
approaches in many-body physics such as quantum field theory.
To enable a field-theoretic analysis of dissipation,
we adopt the Hamiltonian formalism incorporating noise 
fields~\cite{Gardiner2004, Pan2020, Dolgirev2020, Jin2022, Yang2024, Ferreira2024, Ishiyama2025}.
In this framework, two-body loss is captured by adding the term
\begin{equation}
    V_{\eta}
    =   d_{0\uparrow}^{\dagger} d_{0\downarrow}^{\dagger} \eta
        +
        \eta^{\dagger} d_{0\downarrow} d_{0\uparrow},
\label{coupling_term}
\end{equation}
which describes pair loss at the central site when
spin-up and spin-down
particles are simultaneously present.
Provided that noise fields $\eta$ and $\eta^{\dagger}$ satisfy the following statistical properties:
\begin{eqnarray}
    \langle \eta(\tau) \eta^{\dagger}(\tau') \rangle_{\eta}
    =   \gamma\delta(\tau-\tau'),\nonumber\\
    \langle \eta^{\dagger}(\tau) \eta(\tau') \rangle_{\eta}
    =   \langle \eta(\tau) \eta(\tau') \rangle_{\eta}
    =   0,
    \label{noise_average}
\end{eqnarray}
with the dissipation strength $\gamma$,
resulting dynamics
can reproduce those governed by the quantum master equation with Lindblad operator $d_{0\downarrow}d_{0\uparrow}$, which accounts for two-body loss (See also Appendix \ref{deriveGKSL}).
\begin{comment}
\begin{equation}
    \partial_{\tau} \rho
    =   -i\big[ H, \rho \big]
        +
        \gamma
        \left(
            d_{\downarrow} d_{\uparrow}
            \rho
            d_{\uparrow}^{\dagger} d_{\downarrow}^{\dagger}
            -
            \frac{
            \big\{
                d_{\uparrow}^{\dagger} d_{\downarrow}^{\dagger}
                d_{\downarrow} d_{\uparrow},
                \rho
            \big\}
            }{2}
        \right),
\end{equation}
\end{comment}

Based on the Hamiltonian comprised of two-terminal and
noise field terms, the particle and energy currents can be
expressed as
$
I
\equiv
\sum_{\sigma}[-\dot{N}_{L\sigma}+\dot{N}_{R\sigma}]/2
$
and
$
I_E
\equiv
    \sum_{\sigma}[
        -\dot{H}_{L\sigma}
        -
        \mu_{L\sigma} \dot{N}_{L\sigma}
        +
        \dot{H}_{R\sigma}
        +
        \mu_{R\sigma} \dot{N}_{R\sigma}
    ]/2
$, respectively.
These currents represent the net flow from the left to the right reservoir,
corresponding to the decrease in the left reservoir and the increase in the right reservoir.
As in closed quantum many-body systems,
these current expressions can be transformed into more convenient forms as follows:
\cite{Haug2008, Uchino2022}
\begin{eqnarray}
%\begin{split}
&    \dot{N}_{L(R)\sigma}
    =  \zeta\int \frac{d\omega}{2\pi}
        \sum_{\mathbf{k}}
        |t_{\mathbf{k}}|^2
        \operatorname{Re}
        \Big[
            G_{11(MM)\sigma}^R(\omega)
            g_{L(R)\sigma}^K(\mathbf{k}, \omega)
        \Big] \nonumber\\
 &       +
        \zeta\int \frac{d\omega}{2\pi}
        \sum_{\mathbf{k}}
        |t_{\mathbf{k}}|^2
        \operatorname{Re}
        \Big[
            G_{11(MM)\sigma}^K(\omega)
            g_{L(R)\sigma}^A(\mathbf{k}, \omega)
        \Big],
%\end{split}
\label{eq:particle-current}
\end{eqnarray}
%\begin{equation}
%\begin{split}
%    \dot{N}_{R\sigma}
%    &=  \zeta\int \frac{d\omega}{2\pi}
%        \sum_{\mathbf{k}}
%        |t_{\mathbf{k}}|^2
%        \operatorname{Re}
%        \Big[
%            G_{LL\sigma}^R(\omega)
%            g_{R\sigma}^K(\mathbf{k}, \omega)
%        \Big] \\
%        &+
%        \zeta\int \frac{d\omega}{2\pi}
%        \sum_{\mathbf{k}}
%        |t_{\mathbf{k}}|^2
%        \operatorname{Re}
%        \Big[
%            G_{LL\sigma}^K(\omega)
%            g_{R\sigma}^A(\mathbf{k}, \omega)
%        \Big],
%\end{split}
%\end{equation}
\begin{equation}
\begin{split}
    &\dot{H}_{L(R)\sigma}
    +
    \mu_{L(R)\sigma} \dot{N}_{L(R)\sigma} \\
    &=  \zeta\int \frac{d\omega}{2\pi}
        \sum_{\mathbf{k}} \epsilon_{\mathbf{k}}
        |t_{\mathbf{k}}|^2
        \operatorname{Re}
        \Big[
            G_{11(MM)\sigma}^R(\omega)
            g_{L(R)\sigma}^K(\mathbf{k}, \omega)
        \Big] \\
        &+
        \zeta\int \frac{d\omega}{2\pi}
        \sum_{\mathbf{k}} \epsilon_{\mathbf{k}}
        |t_{\mathbf{k}}|^2
        \operatorname{Re}
        \Big[
            G_{11(MM)\sigma}^K(\omega)
            g_{L(R)\sigma}^A(\mathbf{k}, \omega)
        \Big],
\end{split}
\label{eq:energy-current}
\end{equation}
%\begin{equation}
%\begin{split}
%    &\dot{H}_{R\sigma}
%    +
%    \mu_{R\sigma} \dot{N}_{R\sigma} \\
%    &=  \zeta\int \frac{d\omega}{2\pi}
%        \sum_{\mathbf{k}} \epsilon_{\mathbf{k}}
%        |t_{\mathbf{k}}|^2
%        \operatorname{Re}
%        \Big[
%            G_{LL\sigma}^R(\omega)
%            g_{R\sigma}^K(\mathbf{k}, \omega)
%        \Big] \\
%        &+
%        \zeta\int \frac{d\omega}{2\pi}
%        \sum_{\mathbf{k}} \epsilon_{\mathbf{k}}
%        |t_{\mathbf{k}}|^2
%        \operatorname{Re}
%        \Big[
%            G_{LL\sigma}^K(\omega)
%            g_{R\sigma}^A(\mathbf{k}, \omega)
%        \Big],
%\end{split}
%\end{equation}
where $\zeta=+1(-1)$ for bosons (fermions)
and $M=2N+1$ is the number of the site. Here,
$g_{L(R)\sigma}^{K/A}$ is the unperturbed
Keldysh/advanced Green's function of the
left (right) reservoir with spin $\sigma$ and is easily
evaluated in equilibrium in which fluctuation-dissipation theorem holds~\cite{kamenev2023field}.
In contrast, $G_{ij\sigma}^{K/R}$
is full Keldysh/retarded Green's function of the 1D chain and is 
extracted from
the corresponding contour-ordered Green's function
$G_{ij\sigma}^{C}(\tau, \tau')
= -i\langle
    T_C[d_{(i-N-1)\sigma}(\tau)d^{\dagger}_{(j-N-1)\sigma}(\tau')]
    \rangle
$ with contour-ordering $T_C$
discussed in
Appendix \ref{TwoBodyLossSingleSite} and \ref{TwoBodyLossMultiSite}.

%\textit{Noise field formalism.}\textemdash
\section{Noise field formalism}
The evaluation of the particle and energy currents~[Eqs.~\eqref{eq:particle-current} and \eqref{eq:energy-current}] reduces to computing the full Green's functions $G^{K/R}_{11(MM)\sigma}$.
Treating $V_{\eta}$ as a perturbation, 
we employ the Feynman diagram technique to systematically evaluate these functions.
Notably, the noise field formalism shares similarity with 
the treatment of disorder potentials~\cite{bruus2004many} in that 
both involve statistical averaging.
However, a complication in the present context arises from
the temporal correlations imposed by Eq.~\eqref{noise_average}.
The Keldysh formalism then provides a natural framework to 
incorporate these correlation effects in a systematic manner.

When the Lindblad operator involves two field operators, as in the case of
two-body loss and dephasing~\cite{breuer2002theory,Sieber2016},
the system-environment coupling effectively behaves as a two-body interaction
upon averaging over the noise field~\cite{Jin2022}.
However, this coupling
has a different causal structure from genuine interactions,
as it connects time arguments on the forward
and backward branches of the Keldysh contour~\cite{Sieber2016}.
This branch mixing leads to qualitative differences from
conventional two-body interactions
(See also Appendix \ref{DephasingModel} and \ref{TwoBodyLossSingleSite}).

In the dephasing model, where the Lindblad operator is $d^{\dagger}d$
and the associated noise operator is Hermitian,
 the SCBA
yields exact results~\cite{Dolgirev2020, Jin2022}.
This exactness arises from the symmetric contributions
of all components of the 
 contour-ordered Green's function
for the noise field $g_{\eta}^C$, 
%$g_{\eta}^T=g_{\eta}^<=g_{\eta}^>=g_{\eta}^{\tilde{T}}$.
which leads to the systematic cancellation of 
all higher-order Feynman diagrams beyond the SCBA
as shown in Appendix \ref{DephasingModel}.

We now turn to the case of two-body loss, where
the Lindblad operator is $d_{0\downarrow}d_{0\uparrow}$.
In contrast to the dephasing model, here 
the systematic cancellation of Feynman diagrams does not occur, 
preventing an 
exact evaluation via the Feynman diagram technique.
We find that this complication arises because
the noise averages specified in Eq.~\eqref{noise_average}
lead to asymmetric contributions from the time components of the contour-ordered
noise-field Green's functions, which is demonstrated in Appendix~\ref{TwoBodyLossSingleSite}.

%$2g_{\eta}^T=g_{\eta}^>=2g_{\eta}^{\tilde{T}}$
%and $g_{\eta}^<=0$.
%prevent the simplification of the problem based on causality,
%the property $g_{\eta}^<=0$ allows for a compact expression
%of the self-energy within the SCBA.

\begin{figure}
    \subfigure[]{
        \includegraphics[width=0.8\linewidth]{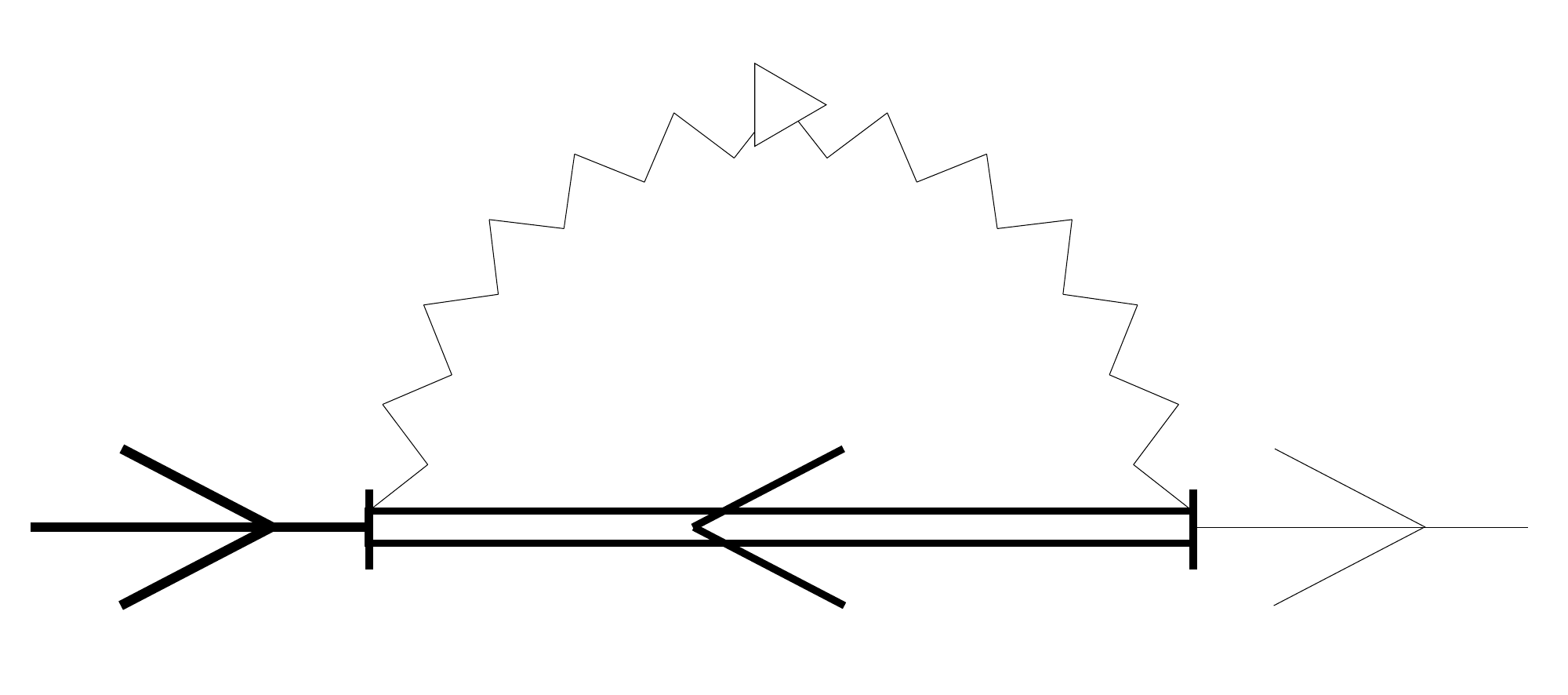}
    % Here is how to import EPS art
        \label{SCBA}
    }
    \subfigure[]{
        \includegraphics[width=0.45\linewidth]
        {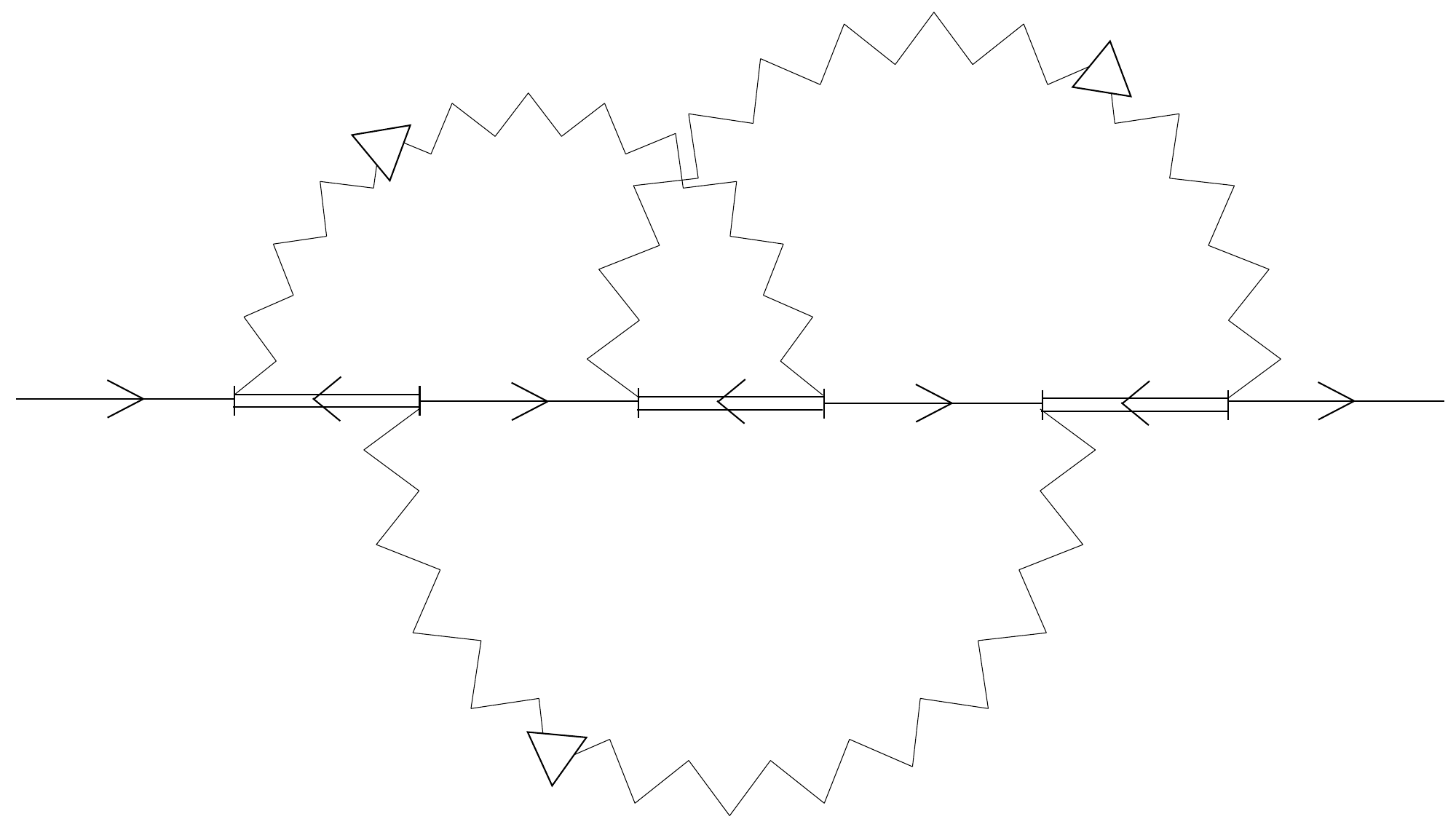}
        \label{diagram_cross_A}
    }
    \subfigure[]{
        \includegraphics[width=0.45\linewidth]
        {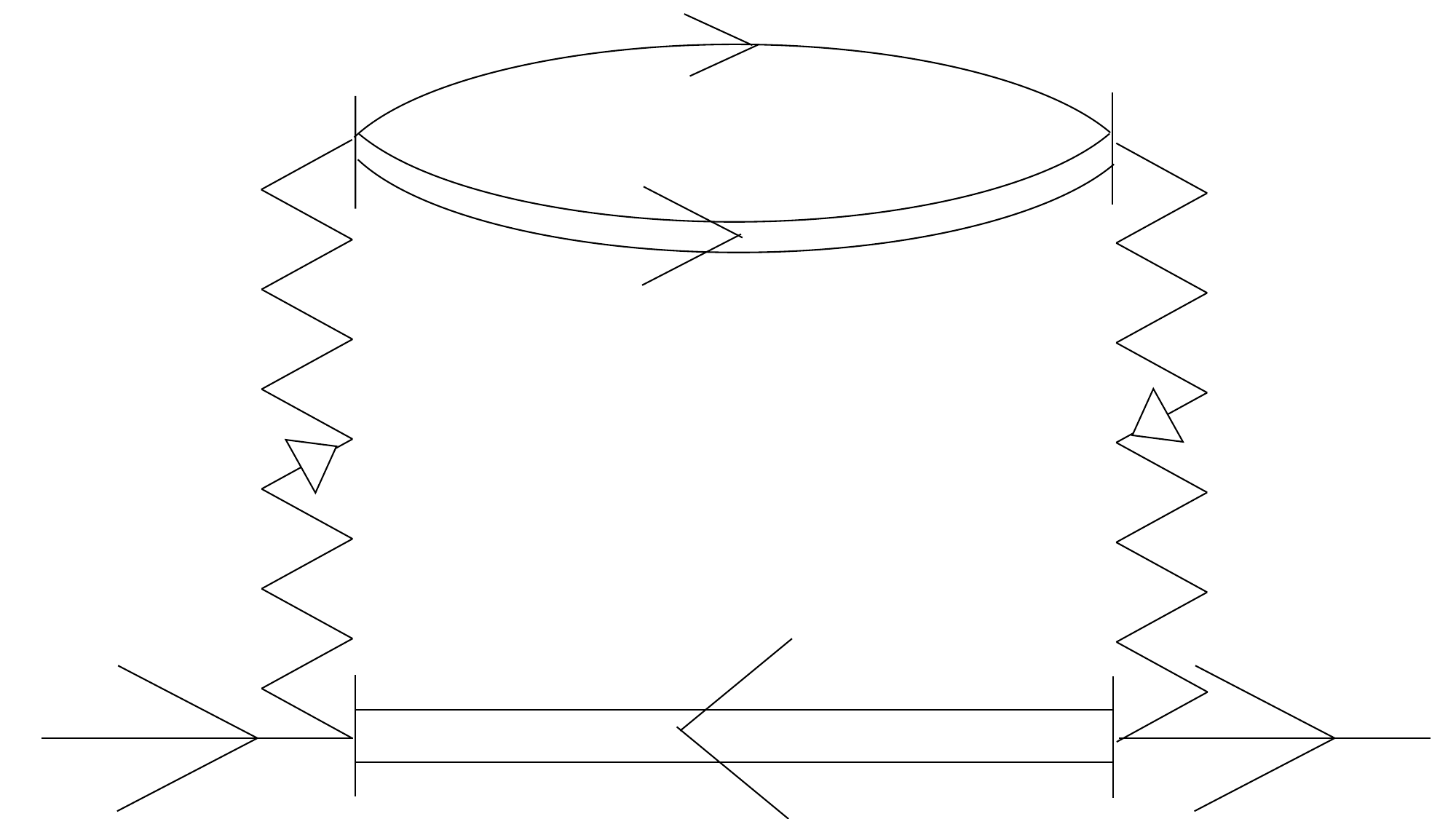}
        \label{diagram_cross_B}
    }
    \caption{
        (a) Feynman diagram corresponding to the SCBA.
        (b) Crossing diagram neglected in the SCBA.
        (c) Diagram to be discarded for consistency with the Lindblad master equation.
        The thin line represents the unperturbed
        Green's function of the particle with
        spin $\sigma$, $g_{\sigma}^C$,
        the thick line the full
        Green's function of the particle with
        spin $\sigma$, $G_{\sigma}^C$,
        the double line in Fig. \ref{SCBA}
        the full Green function
        of the particle with
        opposite spin $\bar{\sigma}$,
        $G_{\bar{\sigma}}^C$,
        the double line in
        Fig. \ref{diagram_cross_A} and \ref{diagram_cross_B}
        the unperturbed Green function
        of the particle with
        opposite spin $\bar{\sigma}$,
        $g_{\bar{\sigma}}^C$,
        and the wavy line the Green's function
        for the noise field,
        $g_{\eta}^C$.
    }
\end{figure}

To elucidate the essential differences in quantum transport
with and without dissipation, 
we focus on a regime in which dissipation is amenable to a perturbative treatment.
Such a treatment is justified when the dissipation strength is significantly smaller than the Fermi energy, tunneling rate between the reservoirs and the 1D chain, and the system
temperature.
We point out that this weakly dissipative regime is relevant to the recent experiment in which
local two-body loss is controlled via photo-association provided by an optical tweezer~\cite{Huang2025}.
When it comes to this regime, it is reasonable to neglect 
multiple scattering by the noise fields.
This leads to  
the SCBA, which accounts for processes where
 a spin-$\sigma$ particle interacts with an opposite-spin
 particle at via noise fields (See Fig. \ref{SCBA}),
while neglecting the crossing diagrams (such as Fig. \ref{diagram_cross_A}).
The SCBA does not also contain a process
in which spin-up and spin-down particles are transiently
created as pairs  by the noise field, propagate,
and subsequently annihilate, as illustrated  
in Fig.~\ref{diagram_cross_B}.
Physically, this diagram corresponds to an effective bath renormalization effect induced by the finite system.

We point out that for consistency with the Lindblad equation, such a diagram must rather be discarded.
To see this, we consider renormalization of the noise-field
Green's function $G_{\eta}^C$, which is also present
in Fig. \ref{diagram_cross_B},
\begin{equation}
    G_{\eta}^C
    =   g_{\eta}^C
        +
        g_{\eta}^C \circ
        i\zeta \Big\{ g_{\sigma}^C g_{\bar{\sigma}}^C \Big\} \circ
        g_{\eta}^C
        +
        \cdots,
\end{equation}
where $\circ$ denotes
$A \circ B = \int d\tau'' \, A(\tau, \tau'') B(\tau'', \tau')$.
This diagrammatic expansion can be obtained by considering the time evolution
of the heat bath
\begin{equation}
\begin{split}
    &\Braket{T_C\left[ \eta(\tau) \eta^{\dagger}(\tau') \right]} \\
    &=  \Braket{T_C\left[
            \exp\left( -i\int_C d\tau'' \, V_{\eta}(\tau'') \right)
            \eta(\tau) \eta^{\dagger}(\tau')
        \right]}_{0}.
\end{split}
\end{equation}
Here, $\Braket{\cdots}_0$ denotes the unperturbed expectation value and
the contribution of $\exp(-i\int_Cd\tau''V_{\eta}(\tau''))$
arises from the time evolution of the density matrix.
Such time evolution violates the Born approximation and the invariance of the environment, both of which are employed in deriving the Lindblad equation (see also Appendix \ref{deriveGKSL}).

We note that the Hartree-Fock theory bears resemblance to the SCBA,
which has successfully been applied to a system with two-body loss~\cite{Stefanucci2024}. In addition, 
the two-terminal quantum-dot system with two-body loss
has recently been analyzed with a variational approach,
yielding results consistent with the SCBA~\cite{Qu2025}.
It was shown in Ref.~\cite{Qu2025} that the mean field approach, i.e.,
the SCBA, remains valid up to around $\gamma/\Gamma \sim 1$, where
$\Gamma$ is the tunneling rate.
This implies that in the regime $\gamma/\Gamma \gg 1$,
effects beyond the SCBA are expected to emerge.
For example, the Kondo effect may arise from the contribution of
Fig. \ref{diagram_cross_A}, which was neglected in the SCBA.

%\textit{Single-site case}\textemdash
\section{Single-site result}
To clarify the essential role of two-body loss in two-terminal
transport, we first consider the simplest case of a single-site system, $M=1$.

To evaluate $G_{11\sigma}^{R/K}$
we harness the Dyson equation in the Keldysh formalism.
Under the SCBA, the contour-ordered Green's function
obeys
\begin{equation}
    G_{11\sigma}^C
    =   g_{11\sigma}^C
        +
        g_{11\sigma}^C
        \circ
        \Sigma_{11\sigma}^C
        \circ
        G_{11\sigma}^C,
\label{Dyson_eq}
\end{equation}
where the self-energy is given by
$
\Sigma_{11\sigma}^C(\tau, \tau')
=   i\zeta g_{\eta}^C(\tau, \tau')
    G_{11\bar{\sigma}}^C(\tau', \tau)
$.
By using the statistical properties of the noise fields,
 the retarded, advanced and Keldysh components
of the self-energy take the form:
$
\Sigma_{\sigma}^{R/A/K}(\tau, \tau')
=   i\zeta g_{\eta}^{R/A/K}(\tau, \tau')
    n_{0\bar{\sigma}}.
$
Here, $n_{0\bar{\sigma}}$ denotes the average occupation number of 
the opposite spin $\bar{\sigma}$ at the quantum dot,
satisfying the following self-consistent integral equation
(See Appendix \ref{TwoBodyLossSingleSite}):
\begin{equation}
    n_{0\sigma}
    =   \int \frac{d\omega}{2\pi}
        \frac{
            \Gamma_{\sigma}(\omega)
            [
                n_{L\sigma}(\omega)
                +
                n_{R\sigma}(\omega)
            ]
        }{
            [
                \omega
                -
                \epsilon_0
                -
                R_{\sigma}(\omega)
            ]^2
            +
            [
                \Gamma_{\sigma}(\omega)
                +
                \frac{\gamma}{2} n_{0\bar{\sigma}}
            ]^2
        },
\label{self-consistent_n}
\end{equation}
where
$n_{L(R)\sigma}(\omega)
=   [
        e^{(\omega-\Delta\mu_{L(R)\sigma})/T_{L(R)}}
        -
        \zeta
    ]^{-1}$
is the Fermi or Bose distribution function of the left (right) reservoir. The chemical potential difference
from the origin of the frequencies is defined as
$\Delta\mu_{L(R)\sigma}
=   \mu_{L(R)\sigma} - \sum_{s\sigma'}\mu_{s\sigma'}/4$
and  $T_{L(R)}$ denotes the reservoir temperature.
The quantities
$
    R_{\sigma}(\omega)
    =   2\sum_{\mathbf{k}}|t_{\mathbf{k}}|^2
        \operatorname{Re}
        [g_{s\sigma}^R(\mathbf{k}, \omega)]
$ and
$
    \Gamma_{\sigma}(\omega)
    =   -2\sum_{\mathbf{k}}|t_{\mathbf{k}}|^2
        \operatorname{Im}
        [g_{s\sigma}^R(\mathbf{k}, \omega)]
$ represent the energy shifts induced by the reservoir.
By using
$\operatorname{Im}[g_{s\sigma}^R]
\propto \delta(\omega+\mu-\epsilon_{\mathbf{k}})$
with $\mu=\sum_{s\sigma}\mu_{s\sigma}/4$,
the particle and energy currents are given by Appendix \ref{TwoBodyLossSingleSite}
\begin{equation}
%\begin{split}
    I
    = \sum_{\sigma}
        \int \frac{d\omega}{2\pi}
        \left[
            \mathcal{T}_ {\sigma}(\omega)
            +
            \frac{\mathcal{L}_ {\sigma}(\omega)}{2}
        \right] 
        \big[ n_{L\sigma}(\omega) - n_{R\sigma}(\omega) \big],
%\end{split}
\label{current_single_site}
\end{equation}
\begin{equation}
\begin{split}
    I_E
    =   &\sum_{\sigma}
        \int \frac{d\omega}{2\pi} (\omega+\mu)
        \left[
            \mathcal{T}_ {\sigma}(\omega)
            +
            \frac{\mathcal{L}_ {\sigma}(\omega)}{2}
        \right] \\
        &\quad\times
        \big[ n_{L\sigma}(\omega) - n_{R\sigma}(\omega) \big],
\end{split}
\label{energy_current_single_site}
\end{equation}
where $\mathcal{T}_{\sigma}(\omega)$ and $\mathcal{L}_{\sigma}(\omega)$
respectively represent the transmittance and loss probability defined as
\begin{equation}
    \mathcal{T}_{\sigma}(\omega)
    =   \frac{
            [\Gamma_{\sigma}(\omega)]^2
        }{
            [
                \omega
                -
                \epsilon_0
                -
                R_{\sigma}(\omega)
            ]^2
            +
            [
                \Gamma_{\sigma}(\omega)
                +
                \frac{\gamma}{2} n_{0\bar{\sigma}}
            ]^2
        },
\label{transmittance}
\end{equation}
\begin{equation}
    \mathcal{L}_{\sigma}(\omega)
    =   \frac{
            \Gamma_{\sigma}(\omega) \gamma n_{0\bar{\sigma}}
        }{
            [
                \omega
                -
                \epsilon_0
                -
                R_{\sigma}(\omega)
            ]^2
            +
            [
                \Gamma_{\sigma}(\omega)
                +
                \frac{\gamma}{2} n_{0\bar{\sigma}}
            ]^2
        }.
\label{loss_probability}
\end{equation}
%This result is consistent with one in Ref. \cite{Qu2025}.
%In contrast to assuming a tight-binding model as reservoirs,
%our model does not have a lower limit in the integral.
%The lower limit depending on the chemical potential
%difference between the left and right reservoirs does not
%exist, and "negative conductance" does not occur.

\begin{figure}
    \includegraphics[width=\linewidth]
    {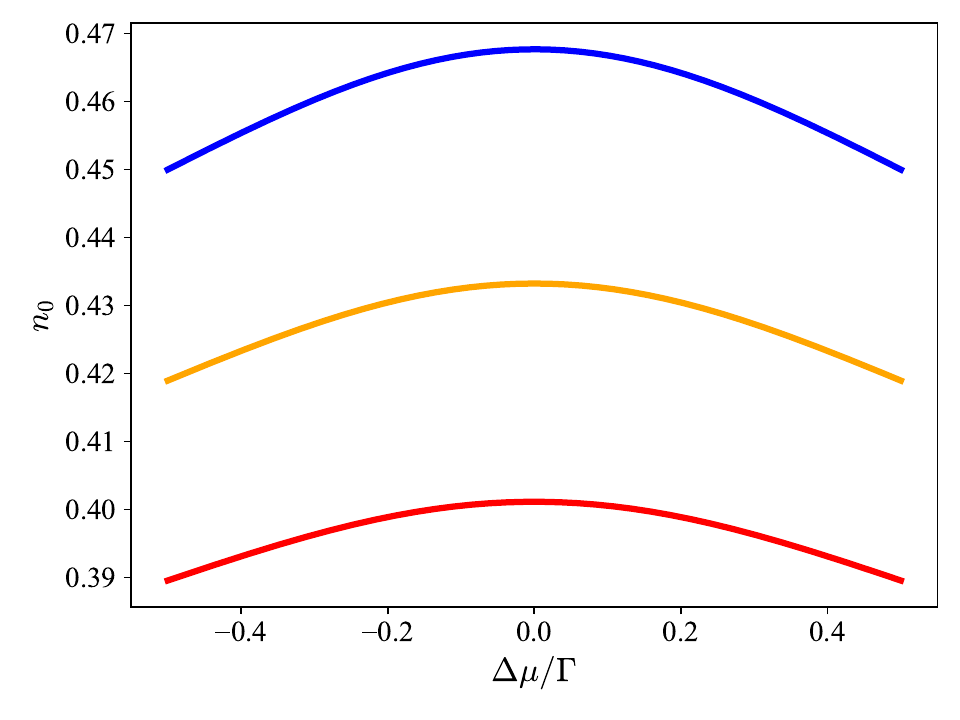}
    % Here is how to import EPS art
    \caption{
        The average number of particle at the central dot with
        two-body loss.
        The temperature is set to $T/\Gamma=0.1$.
        The blue, orange  and red curves correspond to 
        $\gamma/\Gamma=0.1, 0.5$ and $1.0$, respectively.
    }
\label{n_plot}
\end{figure}
We compare the above results with those for the one-body loss case~\cite{Uchino2022}.
A key similarity is that the formal expressions
for the particle and energy
currents remain identical for bosons and fermions
in both cases.
However,
a crucial difference lies in the structure of 
the effective dissipation: in the two-body loss case,
the effective loss strength is given by
$\gamma n_{0\bar{\sigma}}$~\cite{Qu2025}.
This occupation-dependent term  leads to differences
in the particle and energy currents
between one-body and two-body loss cases.

In the absence of two-body loss $\gamma=0$,
the reduction in particle number in the left reservoir
is fully compensated by  the increase in the right reservoir,
as captured by the currents in
Eqs.~(\ref{current_single_site}) and (\ref{energy_current_single_site}).
When two-body loss is present $\gamma \neq 0$,
an additional loss channel to the vacuum arises.
This corresponds to the particle-loss rate, which can also be
evaluated with the Keldysh formalism as follows
(See Appendix \ref{TwoBodyLossSingleSite} and \ref{TwoBodyLossMultiSite}):
\begin{equation}
    -\dot{N}
    =   -\sum_{\sigma}
        \left[
            \dot{N}_{L\sigma}+\dot{N}_{R\sigma}
        \right]
    =   2\gamma n_{0\uparrow} n_{0\downarrow}.
\label{particle_loss}
\end{equation}
This result is intuitively reasonable, since
 the particle loss rate increases with both the number of spin-up and spin-down particles
at the lossy site and the strength of the dissipation.
We note that Eq.~\eqref{particle_loss} remains valid regardless of the total number of sites.

For the sake of an analytic comparison between one-body and two-body loss effects, we now consider the case of 
two-component fermions at zero temperature.
Especially, we focus on particle transport near
the Fermi level, assuming
$R_{\sigma}(\omega) \to 0$ and
$\Gamma_{\sigma}(\omega) \to \Gamma$ with
$\epsilon = 0$, $\mu_{L(R)\sigma} = \mu_{L(R)}$ and
$\Delta\mu = \mu_L - \mu_R\ll\frac{\mu_L+\mu_R}{2}$~\footnote{Notice that $\epsilon=0$ represents that the channel transmittance at the Fermi level is 1 in the absence of dissipation.}.
Under these assumptions,
spin-up and spin-down particles are equally populated
$n_{0\uparrow(\downarrow)}=n_0
=(\sqrt{1+\gamma/\Gamma}-1)/(\gamma/\Gamma)$,
and the particle current follows an Ohmic relation $I = G\Delta\mu$ with conductance $G$.
The resulting  expressions for the conductance
are
\begin{align}
    G_1
    =   \frac{2}{2\pi}\cdot
        \frac{1}{1 + \gamma/2\Gamma},
    \quad
    G_2
    =   \frac{2}{2\pi}\cdot
        \frac{2}{1 + \sqrt{1 + \gamma/\Gamma}},
\label{conductance}
\end{align}
where $G_1$ and $G_2$ correspond to
the conductance in the one-body and two-body loss cases,
respectively.
Here, $1/2\pi$ is the conductance quantum in natural units and
the prefactor $2$ accounts for spin 
degeneracy~\footnote{Here,  the conductance in neutral systems is given.
In the case of electric systems, the conductance quantum becomes
$e^2/2\pi$ in natural units.}.
We note that the above expression of $G_1$ is obtained, provided that 
dissipation is spin-independent, i.e.,
$\gamma_{\uparrow(\downarrow)}=\gamma$.
These expressions demonstrate that $G_2>G_1$, i.e., 
the particle current is more strongly suppressed by one-body loss than two-body loss.

We emphasize that these essential results persist under a finite bias at 
nonzero temperature $T_{L(R)}=T>0$,
whose numerical results are shown in
Fig.~\ref{n_plot}.
As in the zero-temperature case,  $n_0$ decreases with 
increasing $\gamma$.
Notably, at finite temperature,
 $n_0$ also decreases with increasing
$|\Delta \mu|$, a feature absent at zero temperature.

\begin{figure}
    \includegraphics[width=\linewidth]
    {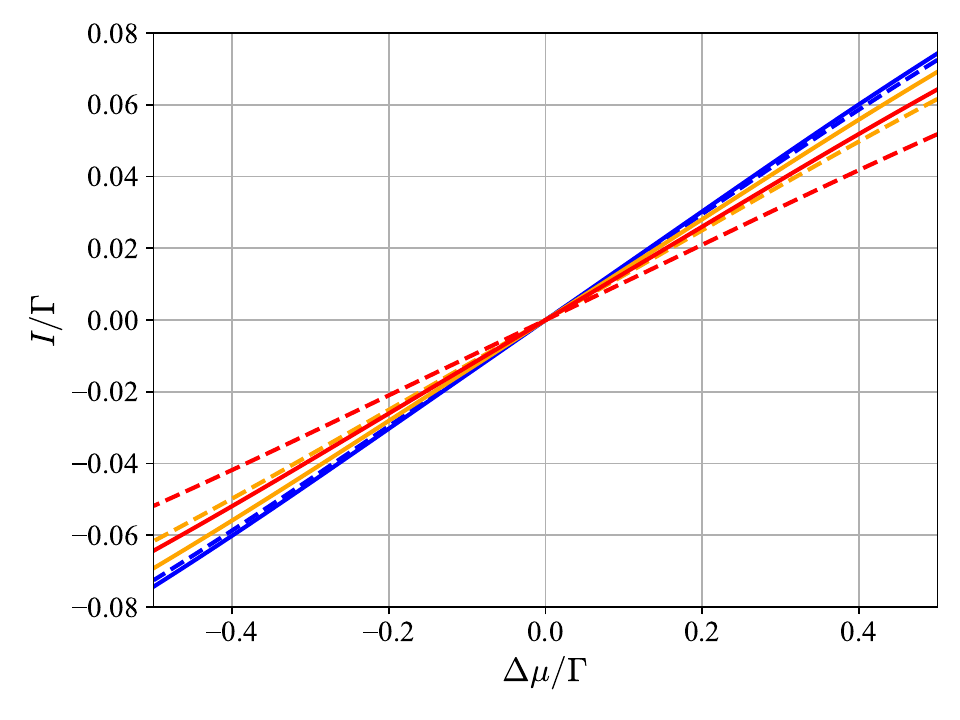}
    % Here is how to import EPS art
    \caption{
        Comparison between one- and two-body loss
        effects on the particle current.
        The temperature is set to $T/\Gamma=0.1$.
        The blue, orange  and red curves correspond to
        $\gamma/\Gamma=0.1, 0.5$ and $1.0$, respectively.
        The solid ones represent results with two-body loss,
        while the dashed ones indicate those with one-body loss.
    }
\label{I_plot}
\end{figure}

Figure~\ref{I_plot} compares particle currents under one-body and two-body losses.
For $\gamma/\Gamma\in [0.1,1]$, 
the occupation 
$n_0 < 1$ (Fig. \ref{n_plot}), and the denominator of
$\mathcal{T}_{\sigma}(\omega) + \mathcal{L}_{\sigma}(\omega)/2$
decreases more rapidly than  the numerator.
Consequently, the particle current in the presence of
two-body loss exceeds that in one-body loss.
This indicates that
the inequality $G_2>G_1$ remains  
robust even under thermal fluctuations.

To enable a more direct comparison between theory and experiment,
we also examine $I$–$\dot{N}$ characteristics.
This is due to the fact that
in cold-atom experiments, the dissipation strength is usually obtained by
exponential fitting of the time dependence of the particle number, yet the effective two-body dissipation strength cannot be directly extracted 
from this method in experiments.

\begin{figure}
    \includegraphics[width=\linewidth]
    {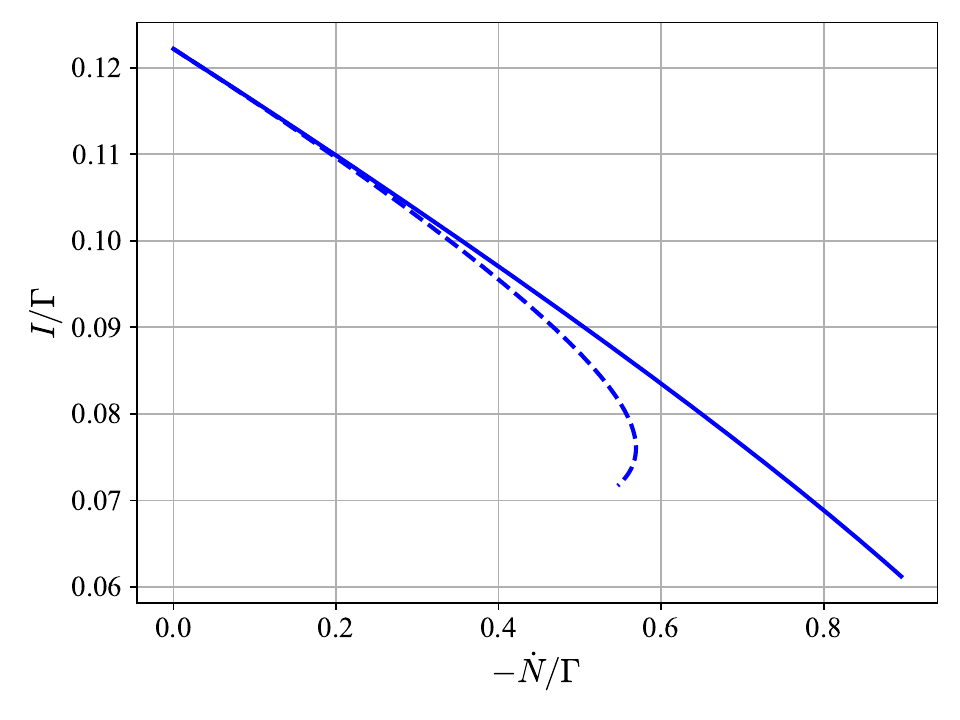}
    % Here is how to import EPS art
    \caption{
        Comparison of particle current and particle-loss rate
        in the cases of one-body and two-body loss.
        The temperature and chemical potential deference are set to
        $T/\Gamma=0.1$ and $\Delta\mu/\Gamma=0.4$, respectively.
        We vary $\gamma/\Gamma$ from $0$ to $10$.
        The solid line represents results with two-body loss,
        while the dashed one indicates those with one-body loss.
    }
\label{I_vs_dN_plot}
\end{figure}

Figure~\ref{I_vs_dN_plot} is the comparison
of $I$-$\dot{N}$ characteristics between one-body
and two-body losses.
Here, we assume the spin-dependent one-body
loss such that $\gamma_{\uparrow}=0$ and $\gamma_{\downarrow}=\gamma$, which is 
 consistent with the experimental condition in Ref.~\cite{Huang2025}.
This comparison demonstrates that the particle current with two-body loss is always
larger than that with one-body loss under the same particle-loss rate.
It is remarkable that the similar trend is also observed in the experiment with superfluid reservoirs~\cite{Huang2025}.

We can also discuss effects of an interaction in the channel.
For fermions, the Hamiltonian of the system
is reduced to the Anderson model~\cite{bruus2004many}, where
$
    H_{1D}^0
    \to
    \sum_{\sigma}
    \epsilon_0 d_{0\sigma}^{\dagger} d_{0\sigma}
    +
    U d_{0\uparrow}^{\dagger} d_{0\uparrow}
    d_{0\downarrow}^{\dagger} d_{0\downarrow}.
$
In the two-terminal experiments with normal reservoirs in 
ultracold atomic gases,
the interaction is weak 
such that the renormalization effect associated with the
Kondo effect is negligible~\cite{krinner2017two}.
In such a regime, 
the following mean-field approximation
is allowed~\cite{yamada1975perturbation}:
$
    H_{1D}^0 \approx
    \sum_{\sigma}
    ( \epsilon_0 + U n_{0\bar{\sigma}})
    d_{0\sigma}^{\dagger} d_{0\sigma}.
$
The resultant expressions of the particle and energy currents
correspond to ones by considering the energy-shift
$\epsilon_0 \to \epsilon_0 + U n_{0\bar{\sigma}}$.

%\textit{Multi-site case.}\textemdash
\section{Multi-site result}
We now turn to the multi-site case.
Under the symmetry of the system
$\epsilon_{-i}=\epsilon_i$ and
$t_{-i}=t_i$ for $i=1, 2, \cdots, N$,
Green's function obeys the following relation:
$G_{ij\sigma}^R(\omega)
=   G_{ji\sigma}^R(\omega)
=   G_{(M-j+1)(M-i+1)\sigma}^R(\omega)$.
By using above, we obtain (See Appendix \ref{TwoBodyLossMultiSite})
\begin{equation}
%\begin{split}
    I
    =  \sum_{\sigma} \int \frac{d\omega}{2\pi}
        \Big[
            \mathcal{T}_ {\sigma}^M(\omega)
            +
            \frac{
                \mathcal{L}_ {\sigma}^M(\omega)
            }{2}
        \Big] 
        [n_{L\sigma}(\omega) - n_{R\sigma}(\omega)],
%\end{split}
\end{equation}

\begin{equation}
\begin{split}
    I_E
    &=  \sum_{\sigma} \int \frac{d\omega}{2\pi}
        (\omega + \mu)
        \Big[
            \mathcal{T}_ {\sigma}^M(\omega)
            +
            \frac{
                \mathcal{L}_ {\sigma}^M(\omega)
            }{2}
        \Big] \\
        &\quad\times
        [n_{L\sigma}(\omega) - n_{R\sigma}(\omega)],
\end{split}
\end{equation}
where
\begin{align}
    &\mathcal{T}_ {\sigma}^M(\omega)
    =   \Big[ \Gamma_{\sigma}(\omega) \Big]^2
        \Big| G_{1M\sigma}^R(\omega) \Big|^2, \\
    &\mathcal{L}_ {\sigma}^M(\omega)
    =   \Big[
            \Gamma_{\sigma}(\omega)
            \gamma n_{0\bar{\sigma}}
        \Big]
        \Big| G_{1\frac{M+1}{2}\sigma}^R(\omega) \Big|^2.
\end{align}
These expressions are analogous to the single-site case ($M=1$).

%\textit{Conclusion.}\textemdash
\section{Conclusion}
By developing a Keldysh field-theoretical framework that systematically incorporates two-body loss and reveals a class of Feynman diagrams consistent with the Lindblad formalism,
we have investigated quantum transport through the 1D 
chain subject to localized two-body loss.
Our formulation allows for the systematic derivation of observables while preserving dissipative dynamics.
In particular, we obtain compact and universal current formulas
that are independent of system size and structurally resemble those for one-body loss.
For fermions,
in contrast to the one-body loss case, the dissipation strength in our model
acquires an occupation-dependent character, leading to a weaker suppression
of the particle current in the weakly dissipative regime. 
This mechanism may naturally account for a recent observation in a superfluid junction~\cite{Huang2025}. For direct quantitative comparison, 
similar experiments using normal (non-superfluid) reservoirs would be highly desirable. Such setups are feasible in ultracold atomic systems by tuning
 interactions via Feshbach resonances. 
Another promising experimental platform is
an optical lattice system, where arbitrary system geometries
can be engineered with a quantum gas microscope~\cite{gross2017} and two-body loss 
has been implemented with photoassociation~\cite{Tomita2017}.
Finally, it would be theoretically compelling to extend the present 
field-theory framework to the strongly dissipative regime,
where a dissipation-induced Kondo effect has been predicted for the single-cite case~\cite{stefanini2024,Qu2025,sonner2025}.
Extending it to multi-site systems and elucidating crossover from weak to strong dissipation within a field-theoretic analysis
remain an important direction for future work.

\section*{acknowledgment}
We are also grateful to T. Esslinger, P. Fabritius, T. Giamarchi
M.-Z. Huang, J. Mohan, M. Talebi, A.-M. Visuri, and
S. Wili for discussions.
This work is supported by JST PRESTO~(JPMJPR235) and
JSPS KAKENHI~(JP25K07191).

\appendix
\onecolumngrid

\section{Keldysh formalism}
\begin{figure}
    \includegraphics[width=\linewidth]
    {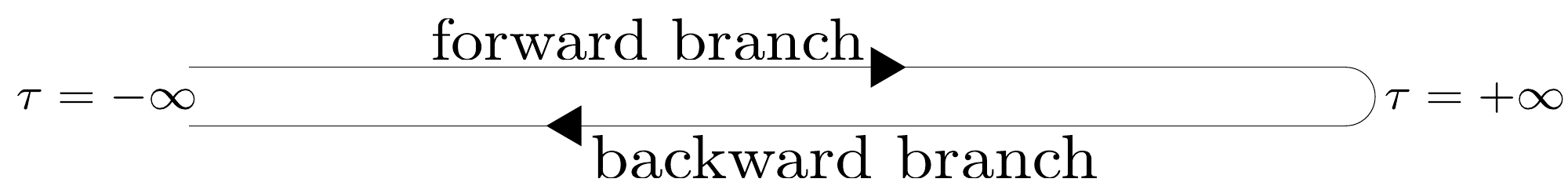}
    \caption{
       Keldysh contour.
    }
    \label{Keldysh_contour}
\end{figure}

In order to perform field-theoretic calculations, we now divide  $\hat{H}$ into
an unperturbed part $\hat{H}_0$ 
and
a perturbed part $\hat{V}$ and
construct the interaction picture
of a system operator $\hat{A}$
\begin{equation}
    \hat{A}_I(\tau)
    =   e^{i(\hat{H}_0+\hat{H}_B)\tau}
        \hat{A}_S
        e^{-i(\hat{H}_0+\hat{H}_B)\tau},
\end{equation}
where the subscripts $S$ and $I$ denote the
Schr{\"o}dinger and interaction pictures.
In the Keldysh formalism, we consider a contour-ordered Green's function
of system operators $\hat{A}$ and $\hat{B}$, which is defined on the contour
illustrated in Fig.~\ref{Keldysh_contour} and is given by
\begin{equation}
\begin{split}
    &G^C(\tau, \tau') \\
    &=  -i\Braket{T_C\left[
            \hat{A}_H(\tau) \hat{B}_H(\tau')
        \right]} \\
    &=  \sum_{n=0}^{\infty}
        \frac{(-i)^{n+1}}{n!}
        \int_C d\tau_1 \cdots \int_C d\tau_n
        \Braket{T_C\left[
            \left(
                \hat{V}_I(\tau_1)
                +
                \hat{V}_{\eta, I}(\tau_1)
            \right)
            \cdots
            \left(
                \hat{V}_I(\tau_n)
                +
                \hat{V}_{\eta, I}(\tau_n)
            \right)
            \hat{A}_I(\tau) \hat{B}_I(\tau')
        \right]}_0.
\end{split}
\end{equation}
Here, the subscript $H$ denotes the
Heisenberg picture and
$\Braket{\cdots}_0$ is the unperturbed expectation value.
In addition, $T_C$ denotes contour-ordered product.
When the $\hat{H}_0$ is bilinear in field operators, which 
holds in our system,
Wick's theorem applies, and 
$G^C$ can be computed
by using a Dyson series expansion associated within
the Feynman diagram technique.

In practical calculations of physical quantities,
it is convenient to work with the 
retarded, advanced and Keldysh components of Green's functions.
These components are related by the following expressions:
\begin{align}
    \left( \begin{matrix}
        G^K(\tau, \tau') & G^R(\tau, \tau') \\
        G^A(\tau, \tau') & 0
    \end{matrix} \right)
    &=
    \left( \begin{matrix}
        -i
        \Braket{\left[
            \hat{A}_H(\tau), \hat{B}_H(\tau')
        \right]_{\zeta}} &
        -i\theta(\tau-\tau')
        \Braket{
            \left[ \hat{A}_H(\tau), \hat{B}_H(\tau') \right]_{-\zeta}
        } \\
        i\theta(\tau'-\tau)\Braket{
            \left[ \hat{A}_H(\tau), \hat{B}_H(\tau') \right]_{-\zeta}
        } &
        0
    \end{matrix} \right), \\
    \left( \begin{matrix}
        G^T(\tau, \tau') & G^<(\tau, \tau') \\
        G^>(\tau, \tau') & G^{\tilde{T}}(\tau, \tau')
    \end{matrix} \right)
    &=
    \left( \begin{matrix}
        G^C(\tau^+, {\tau'}^+) & G^C(\tau^+, {\tau'}^-) \\
        G^C(\tau^-, {\tau'}^+) & G^C(\tau^-, {\tau'}^-)
    \end{matrix} \right) \\
    &=
    \left( \begin{matrix}
        -i\Braket{T\left[
            \hat{A}_H(\tau) \hat{B}_H(\tau')
        \right]} &
        -i\Braket{
            \hat{A}_H(\tau) \hat{B}_H(\tau')
        } \\
        -i\zeta\Braket{
            \hat{B}_H(\tau') \hat{A}_H(\tau)
        } &
        -i\Braket{\tilde{T}\left[
            \hat{A}_H(\tau) \hat{B}_H(\tau')
        \right]}
    \end{matrix} \right) \\
    &=
    \frac{1}{2}
    \left( \begin{matrix}
        G^K(\tau, \tau') + G^R(\tau, \tau') + G^A(\tau, \tau') &
        G^K(\tau, \tau') - G^R(\tau, \tau') + G^A(\tau, \tau') \\
        G^K(\tau, \tau') + G^R(\tau, \tau') - G^A(\tau, \tau') &
        G^K(\tau, \tau') - G^R(\tau, \tau') - G^A(\tau, \tau')
    \end{matrix} \right).
    \label{time-RAK}
\end{align}
Here, $\zeta=+1(-1)$ for bosons (fermions),
$[A, B]_{\zeta}=AB + \zeta BA$.
The time argument $\tau^{+(-)}$
denotes times on the forward (backward) branch of the
Keldysh contour (see Fig.~\ref{Keldysh_contour}).
The symbol $T(\tilde{T})$ denotes (anti-)time-ordered product,
while
$G^{<(>)}$ refers to  lesser (greater) Green's function.

\section{Hamiltonian formalism with noise fields}
In this section,
we review the Hamiltonian formalism with noise fields analogous to "white noise"
in stochastic processes. 
In this formalism,
we start with the Hamiltonian of a system coupled to an environment
$(\hbar=1)$
\begin{equation}
    \hat{H}_{\text{tot}}
    =   \hat{H}
        +
        \hat{H}_{B}
        +
        \hat{V}_{\eta},
\end{equation}
Here,  $\hat{H}$ is the system Hamiltonian and
$\hat{H}_{B}$ is the Hamiltonian for a heat bath defined by
\begin{equation}
    \hat{H}_{B}
    =   \int_{-\infty}^{\infty} d\omega \,
        \omega \hat{\eta}^{\dagger}(\omega) \hat{\eta}(\omega),
\label{Hamiltonian_Bath}
\end{equation}
with the commutation relations
\begin{equation}
    \left[
        \hat{\eta}(\omega), \hat{\eta}^{\dagger}(\omega')
    \right]
    =   \delta(\omega-\omega'),
    \quad
    \left[
        \hat{\eta}(\omega), \hat{\eta}(\omega')
    \right]
    =
    \left[
        \hat{\eta}^{\dagger}(\omega), \hat{\eta}^{\dagger}(\omega')
    \right]
    =   0.
\end{equation}
The energy spectrum of the heat bath is assumed to be continuous.
The coupling between the system and the heat bath is given by
\begin{equation}
    \hat{V}_{\eta}
    =   \int_{-\infty}^{\infty} d\omega
        \left[
            \gamma(\omega) \hat{L}^{\dagger} \hat{\eta}(\omega)
            +
            \gamma^*(\omega) \hat{\eta}^{\dagger}(\omega) \hat{L}
        \right],
\end{equation}
where $\hat{L}$ is the Lindblad operator and
$\gamma(\omega)$ is a coupling
between the system and bath, which in general depends on frequencies.

We now consider the interaction picture under
the unperturbed Hamiltonian $\hat{H}+\hat{H}_{B}$.
In this interaction picture,
the Lindblad operator is
$\hat{L}(\tau)
=   e^{i(\hat{H}+\hat{H}_B)\tau}
    \hat{L}
    e^{-i(\hat{H}+\hat{H}_B)\tau}$,
and the noise field is obtained by using
the Baker-Campbell-Hausdorff formula
\begin{equation}
    \hat{\eta}(\omega, \tau)
    =   e^{i(\hat{H}+\hat{H}_B)\tau}
        \hat{\eta}(\omega)
        e^{-i(\hat{H}+\hat{H}_B)\tau}
    =   e^{i\hat{H}_B\tau}
        \hat{\eta}(\omega)
        e^{-i\hat{H}_B\tau}
    =   \hat{\eta}(\omega) e^{-i\omega\tau}.
\end{equation}
Then, the coupling term is expressed as
\begin{equation}
    \hat{V}_{\eta}(\tau)
    =   \hat{L}^{\dagger}(\tau) \hat{\eta}(\tau)
        +
        \hat{\eta}^{\dagger}(\tau) \hat{L}(\tau),
\label{couple_term}
\end{equation}
where
\begin{equation}
    \hat{\eta}(\tau)
    =   \int_{-\infty}^{\infty} d\omega \,
        \gamma(\omega)
        \hat{\eta}(\omega) e^{-i\omega\tau}.
\end{equation}
The operators $\hat{\eta}(t)$
obeys the following commutation relations
\begin{equation}
    \left[
        \hat{\eta}(\tau),
        \hat{\eta}^{\dagger}(\tau')
    \right]
    =   \int_{-\infty}^{\infty} d\omega
        |\gamma(\omega)|^2
        e^{-i\omega(\tau-\tau')},
\label{comuutator_eta}
\end{equation}
and
\begin{equation}
    \big[
        \hat{\eta}(\tau),
        \hat{\eta}(\tau')
    \big]
    =
    \left[
        \hat{\eta}^{\dagger}(\tau),
        \hat{\eta}^{\dagger}(\tau')
    \right]
    =   0.
\end{equation}
Equation (\ref{comuutator_eta}) indicates that
the commutator depends on the past history $\tau'(\neq \tau)$,
resulting in non-Markovian dynamics of the system.

We next take the Markovian limit that neglects the memories of the heat bath. This corresponds to taking the frequency-independent coupling,
\begin{equation}
    \gamma(\omega)
    =   \sqrt{\frac{\gamma}{2\pi}}.
    \label{eq:markov}
\end{equation}
In this limit, the noise fields $\hat{\eta}(\omega)$ and
the Lindblad operator $\hat{L}$ are coupled
with equal weight over the entire energy spectra, and it follows that
\begin{equation}
    \left[
        \hat{\eta}(\tau), \hat{\eta}^{\dagger}(\tau')
    \right]
    =   \gamma\delta(\tau-\tau'),
    \quad
    \big[
        \hat{\eta}(\tau), \hat{\eta}(\tau')
    \big]
    =
    \left[
        \hat{\eta}^{\dagger}(\tau), \hat{\eta}^{\dagger}(\tau')
    \right]
    =   0,
\end{equation}
where
\begin{equation}
    \hat{\eta}(\tau)
    =   \sqrt{\gamma}
        \int_{-\infty}^{\infty} \frac{d\omega}{\sqrt{2\pi}}
        \hat{\eta}(\omega) e^{-i\omega \tau}.
\end{equation}

%Based on the above consideration, 
%we consider the interpretation of the noise fields.
%The new operator $\hat{\eta}(\tau)$ is
%a superposition of $\hat{\eta}(\omega)$,
%which is an annihilation operator
%of Bose fields in the heat bath with energy $\omega$.
%We can interpret the Bose fields as
%particles outside the system or photons
%giving rise to dissipation.
%Although both interpretations have reasonable justifications,
%the former interpretation is more straightforward
%for the calculations.
%To obtain results consistent with other theories,
In addition to the commutation relation, 
the noise field formalism considers 
 statistical averages on noise fields.
 In the case of two-body loss concerned in the main text,
 we consider the following averages:
\begin{equation}
    \Braket{\hat{\eta}(\tau)\hat{\eta}^{\dagger}(\tau')}_{\eta}
    =   \gamma\delta(\tau-\tau'),
    \quad
    \Braket{\hat{\eta}^{\dagger}(\tau)\hat{\eta}(\tau')}_{\eta}
    =
    \Braket{\hat{\eta}(\tau)\hat{\eta}(\tau')}_{\eta}
    =
    \Braket{\hat{\eta}^{\dagger}(\tau)\hat{\eta}^{\dagger}(\tau')}_{\eta}
    =
    \Braket{\hat{\eta}(\tau)}_{\eta}
    =
    \Braket{\hat{\eta}^{\dagger}(\tau)}_{\eta}
    =   0.
    \label{eq:noise-avarage}
\end{equation}
The above averages of the noise fields correspond to the
assumption that the heat bath is in the vacuum state
$\Braket{\cdots}_{\eta} = \Braket{\text{vac}|\cdots|\text{vac}}$
with $\hat{\eta}(\tau)\ket{\text{vac}}=0$.
Physically, this indicates that
the population of particles outside the system
is absent and therefore no particle injection from outside.

\section{Derivation of Lindblad master equation}
\label{deriveGKSL}
In this section, we derive the Lindblad master equation
using the noise fields.
The density operator of the total system obeys the von Neumann equation
in the Schr{\"o}dinger picture
\begin{equation}
    \frac{\partial}{\partial \tau} \hat{\rho}_{\text{tot}}(\tau)
    =   -i\left[
            \hat{H}_{\text{tot}},
            \hat{\rho}_{\text{tot}}(\tau)
        \right].
\end{equation}
By dividing the Hamiltonian into an unperturbed part
$\hat{H}+\hat{H}_B$ and a perturbed part $\hat{V}_{\eta}$,
we construct the interaction picture
\begin{equation}
    \frac{\partial}{\partial \tau} \hat{\rho}_{\text{tot}, I}(\tau)
    =   -i\left[
            \hat{V}_{\eta}(\tau),
            \hat{\rho}_{\text{tot}, I}(\tau)
        \right],
\label{EOM}
\end{equation}
where
$\hat{\rho}_{\text{tot}, I}(\tau)
=   e^{i(\hat{H}+\hat{H}_{B})\tau}
    \hat{\rho}_{\text{tot}}(\tau)
    e^{-i(\hat{H}+\hat{H}_{B})\tau}$
and $\hat{V}_{\eta}(\tau)$ is defined as
Eq. (\ref{couple_term}).
Here, the subscript $I$ denotes the interaction picture.
By formally solving Eq. (\ref{EOM})
\begin{equation}
    \hat{\rho}_{\text{tot}, I}(\tau)
    =   \hat{\rho}_{\text{tot}, I}(\tau_0)
        -
        i \int_{\tau_0}^{\tau} d\tau'
        \left[
            \hat{V}_{\eta}(\tau'),
            \hat{\rho}_{\text{tot}, I}(\tau')
        \right],
\end{equation}
Eq. (\ref{EOM}) can be rewritten
\begin{equation}
    \frac{\partial}{\partial \tau} \hat{\rho}_{\text{tot}, I}(\tau)
    =   -i\left[
            \hat{V}_{\eta}(\tau),
            \hat{\rho}_{\text{tot}, I}(\tau_0)
        \right]
        -
        \int_{\tau_0}^{\tau} d\tau'
        \left[
            \hat{V}_{\eta}(\tau),
            \left[
                \hat{V}_{\eta}(\tau'),
                \hat{\rho}_{\text{tot}, I}(\tau')
            \right]
        \right],
\label{EOM_2}
\end{equation}
where $\tau_0$ is the initial time.

The density operator
of the system is defined as
\begin{equation}
    \hat{\rho}_I(\tau)
    =
    \operatorname{Tr}_B
    \left\{
        \hat{\rho}_{\text{tot}, I}(\tau)
    \right\}.
\end{equation}
The partial trace over the environment
is associated with the noise average as follows:
\begin{equation}
    \Braket{\cdots}_{\eta}
    =
    \operatorname{Tr}_B
    \left\{
        \hat{\rho}_B(\tau_0)
        (\cdots)
    \right\}.
\end{equation}
Here, $\hat{\rho}_B$ is the density operator
of the heat bath.
To obtain an equation of motion for $\hat{\rho}_I(\tau)$,
we consider the following conditions~\cite{breuer2002theory}:
\begin{enumerate}
    \item Initial state:
    $\hat{\rho}_{\text{tot}}(\tau_0)
    = \hat{\rho}(\tau_0) \otimes \hat{\rho}_B(\tau_0)$
    \item Born approximation:
    $\hat{\rho}_{\text{tot}}(\tau)
    \approx \hat{\rho}(\tau) \otimes \hat{\rho}_B(\tau)$
    \item Invariance of the environment:
    $\hat{\rho}_B(\tau)
    \approx \hat{\rho}_B(\tau_0)$
\end{enumerate}
By taking the noise average of (\ref{EOM_2}), we obtain
\begin{equation}
    \frac{\partial}{\partial \tau} \hat{\rho}_I(\tau)
    =   -\int_{\tau_0}^{\tau} d\tau' \,
        \operatorname{Tr}_B
        \left\{
        \left[
            \hat{V}_{\eta}(\tau),
            \left[
                \hat{V}_{\eta}(\tau'),
                \hat{\rho}_I(\tau') \otimes
                \hat{\rho}_B(\tau_0)
            \right]
        \right]
        \right\},
        \label{eq:system-density-matrix}
\end{equation}
where we use
$\operatorname{Tr}_B\left\{\hat{\rho}_B(\tau_0) \right\}=1$
and
\begin{equation}
\begin{split}
    \operatorname{Tr}_B
    \left\{
    \left[
        \hat{V}_{\eta}(\tau), \hat{\rho}_{\text{tot}, I}(\tau_0)
    \right]
    \right\}
    &=
    \operatorname{Tr}_B
    \left\{\hat{\eta}(\tau) \hat{\rho}_B(\tau_0) \right\}
    \hat{L}^{\dagger}(\tau) \hat{\rho}(\tau_0)
    -
    \operatorname{Tr}_B
    \left\{ \hat{\rho}_B(\tau_0) \hat{\eta}(\tau) \right\}
    \hat{\rho}(\tau_0) \hat{L}^{\dagger}(\tau) \\
    &\quad+
    \operatorname{Tr}_B
    \left\{ \hat{\eta}^{\dagger}(\tau) \hat{\rho}_B(\tau_0) \right\}
    \hat{L}(\tau) \hat{\rho}(\tau_0)
    -
    \operatorname{Tr}
    \left\{ \hat{\rho}_B(\tau_0) \hat{\eta}^{\dagger}(\tau) \right\}
    \hat{\rho}(\tau_0) \hat{L}(\tau) \\
    &=
    \Braket{\hat{\eta}(\tau)}
    \hat{L}^{\dagger}(\tau) \hat{\rho}(\tau_0)
    -
    \Braket{\hat{\eta}(\tau)}
    \hat{\rho}(\tau_0) \hat{L}^{\dagger}(\tau) \\
    &\quad+
    \Braket{\hat{\eta}^{\dagger}(\tau)}
    \hat{L}(\tau) \hat{\rho}(\tau_0)
    -
    \Braket{\hat{\eta}^{\dagger}(\tau)}
    \hat{\rho}(\tau_0) \hat{L}(\tau) \\
    &=  0.
    \label{eq:first}
\end{split}
\end{equation}
The above equation follows from the cyclic property of the trace
$\operatorname{Tr}_B\{AB\}=\operatorname{Tr}_B\{BA\}$.
The right hand side of Eq.~\eqref{eq:system-density-matrix} 
is rewritten as
\begin{equation}
\begin{split}
    &\operatorname{Tr}_B
    \left\{
    \left[
        \hat{V}_{\eta}(\tau),
        \left[
            \hat{V}_{\eta}(\tau'),
            \hat{\rho}_I(\tau') \otimes
            \hat{\rho}_B(\tau_0)
        \right]
    \right]
    \right\} \\
    &=
    \Braket{
    \hat{\eta}(\tau) \hat{\eta}^{\dagger}(\tau')
    }_{\eta}
    \hat{L}^{\dagger}(\tau) \hat{L}(\tau') \hat{\rho}_I(\tau')
    +
    \Braket{
    \hat{\eta}(\tau') \hat{\eta}^{\dagger}(\tau)
    }_{\eta}
    \hat{\rho}_I(\tau') \hat{L}^{\dagger}(\tau') \hat{L}(\tau) \\
    &\quad-
    \Braket{
    \hat{\eta}(\tau') \hat{\eta}^{\dagger}(\tau)
    }_{\eta}
    \hat{L}(\tau) \hat{\rho}_I(\tau') \hat{L}^{\dagger}(\tau')
    -
    \Braket{
    \hat{\eta}(\tau) \hat{\eta}^{\dagger}(\tau')
    }_{\eta}
    \hat{L}(\tau') \hat{\rho}_I(\tau') \hat{L}^{\dagger}(\tau) \\
    &=
    \gamma\delta(\tau-\tau')
    \left[
        \left\{
            \hat{L}^{\dagger}(\tau) \hat{L}(\tau),
            \hat{\rho}_I(\tau)
        \right\}
        -
        2\hat{L}(\tau) \hat{\rho}_I(\tau) \hat{L}^{\dagger}(\tau)
    \right].
    \label{eq:second}
\end{split}
\end{equation}
Finally, by using the following identity
\begin{alignat}{2}
    \int_{\tau_0}^{\tau} d\tau' \, \delta(\tau-\tau')
    &=  \int_{-\infty}^{\tau} d\tau' \, \delta(\tau-\tau')
        && \quad (\delta(\tau)=0 \, \text{at} \, \tau \neq 0) \notag \\
    &=  \int_0^{\infty} ds \, \delta(s)
        && \quad (s=\tau-\tau') \notag \\
    &=  \frac{1}{2} \int_{-\infty}^{\infty} ds \,
        \delta(s)
        && \quad (\text{even function:} \,
        \delta(s)=\delta(-s)) \notag \\
    &=  \frac{1}{2},
\end{alignat}
we obtain the Lindblad master equation in
the interaction picture
\begin{equation}
    \frac{\partial}{\partial \tau} \hat{\rho}_I(\tau)
    =
    \gamma
    \left[
        \hat{L}(\tau) \hat{\rho}_I(\tau) \hat{L}^{\dagger}(\tau)
        -
        \frac{\left\{
            \hat{L}^{\dagger}(\tau) \hat{L}(\tau),
            \hat{\rho}_I(\tau)
        \right\}}{2}
    \right].
\end{equation}
Since the time derivative of the density matrix in
the interaction picture is associated with that in 
in the Schr{\"o}dinger picture as follows:
\begin{equation}
\begin{split}
    \frac{\partial}{\partial \tau} \hat{\rho}_I(\tau)
    &=  \frac{\partial}{\partial \tau}
        \left(
            e^{i(\hat{H}+\hat{H}_B)\tau}
            \hat{\rho}(\tau)
            e^{-i(\hat{H}+\hat{H}_B)\tau}
        \right) \\
    &=  e^{i(\hat{H}+\hat{H}_B)\tau}
        \left(
            \frac{\partial}{\partial \tau} \hat{\rho}(\tau)
            +
            i\left[
                \hat{H} + \hat{H}_B, \hat{\rho}(\tau)
            \right]
        \right)
        e^{-i(\hat{H}+\hat{H}_B)\tau} \\
    &=  e^{i(\hat{H}+\hat{H}_B)\tau}
        \left(
            \frac{\partial}{\partial \tau} \hat{\rho}(\tau)
            +
            i\left[
                \hat{H}, \hat{\rho}(\tau)
            \right]
        \right)
        e^{-i(\hat{H}+\hat{H}_B)\tau}.
\end{split}
\end{equation}
we obtain the well-known Lindblad master equation
\begin{equation}
    \frac{\partial}{\partial \tau} \hat{\rho}(\tau)
    =
    -i\left[
        \hat{H}, \hat{\rho}(\tau)
    \right]
    +
    \gamma
    \left[
        \hat{L} \hat{\rho}(\tau) \hat{L}^{\dagger}
        -
        \frac{\left\{
            \hat{L}^{\dagger} \hat{L},
            \hat{\rho}(\tau)
        \right\}}{2}
    \right].
\label{Lindblad_eq}
\end{equation}

%To summarize, by using the noise-field formalism discussed 
%in the previous section and assuming some conditions that are standard for derivation
%of the quantum master equation, 
%the Lindblad-type master equation has indeed been obtained.

It is instructive to clarify the assumptions underlying the above derivation.
First, the initial separability between the system and the bath
is a standard assumption in both quantum many-body and open quantum systems,
allowing for a perturbative treatment.
Second, the Born approximation presumes that this separability holds throughout the evolution,
 which is reasonable when the system-bath coupling is weak.
Third, the time-invariance of the environment is naturally satisfied 
for a large bath in thermal equilibrium or in vacuum.
In conventional derivations of 
the Lindblad master equation~\cite{breuer2002theory},
the Markov and rotating-wave approximations are also imposed.
In our case, however, these approximations are not explicitly required, as
the Markov approximation is incorporated through the assumption
of frequency-independent coupling in Eq.~\eqref{eq:markov},
and the rotating-wave approximation is implemented 
via Eq.~\eqref{couple_term}.

It should also be emphasized that the Born approximation is essential
for deriving  Eqs.~\eqref{eq:first} and \eqref{eq:second}.
Without this condition,  Eq.~\eqref{eq:noise-avarage} cannot
be applied directly,
and the temporal correlations deviates from the delta function due to bath renormalization effects induced by
the finite quantum system. This point is particularly suggestive
when performing field-theoretic calculations based on Feynman diagram techniques,
since perturbative expansions naturally generate diagrams 
corresponding to such bath renormalizations.
To maintain consistency with the Lindblad master equation,
it is necessary to discard these diagrams, a consideration
that becomes especially crucial
when analyzing two-body loss processes. 

To be concrete, we consider the renormalized Green’s function of the noise field.
The Green's function is formally expressed as
\begin{equation}
\begin{split}
    G_{\eta}^C(\tau, \tau')
    &=  -i\operatorname{Tr}\left[
        \hat{\rho}_{\text{tot}}(\tau_0)
        T_C\left\{
            \exp\left( -i \int_C d\tau'' \, \hat{V}_{\eta}(\tau'') \right)
            \hat{\eta}(\tau) \hat{\eta}^{\dagger}(\tau')
        \right\} \right] \\
    &=  -i\operatorname{Tr}\left[
        T_C\left\{
            \hat{\rho}_{\text{tot}}(\tau_0)
            \exp\left( -i \int_C d\tau'' \, \hat{V}_{\eta}(\tau'') \right)
            \hat{\eta}(\tau) \hat{\eta}^{\dagger}(\tau')
        \right\}
        \right] \\
    &=  -i\operatorname{Tr}\left[
        T_C\left\{
            \hat{S}(\infty, \tau_0) \hat{\rho}_{\text{tot}}(\tau_0) \hat{S}^{\dagger}(\infty, \tau_0)
            \hat{\eta}(\tau) \hat{\eta}^{\dagger}(\tau')
        \right\}
        \right] \\
    &=  -i\operatorname{Tr}\left[
        T_C\left\{
            \hat{\rho}_{\text{tot}}(\infty)
            \hat{\eta}(\tau) \hat{\eta}^{\dagger}(\tau')
        \right\}
        \right].
\end{split}
\end{equation}
Here, $\hat{S}$ is the time evolution operator defined as follows:
\begin{equation}
    \hat{S}(\tau, \tau')
    =   \sum_{n=0}^{\infty} \frac{(-i)^n}{n!} \int_{\tau'}^{\tau} d\tau_1 \cdots \int_{\tau'}^{\tau} d\tau_n \,
        T\left[ \hat{V}_{\eta}(\tau_1) \cdots \hat{V}_{\eta}(\tau_n) \right],
\end{equation}
where $T$ is the time-ordered operator.
Under the Born approximation and the invariance of the environment, $G_{\eta}^C$ must be $g_{\eta}^C$
\begin{equation}
    G_{\eta}^C(\tau, \tau')
    =   -i\operatorname{Tr}\left[
        T_C\left\{
            \hat{\rho}(\infty) \otimes \hat{\rho}_B(\tau_0)
            \hat{\eta}(\tau) \hat{\eta}^{\dagger}(\tau')
        \right\}
        \right]
    =   -i\operatorname{Tr}_B\left[
        T_C\left\{
            \hat{\rho}_B(\tau_0)
            \hat{\eta}(\tau) \hat{\eta}^{\dagger}(\tau')
        \right\}
        \right]
    =   g_{\eta}^C(\tau, \tau'),
    \label{eq:renormalization-eta}
\end{equation}
where we use the partial trace over the system and the identity
$\operatorname{Tr}_S[\hat{\rho}(\tau)]=1$.
Equation~\eqref{eq:renormalization-eta}
asserts that the bath renormalization is absent.

When the Lindbald operator $\hat{L}$ involves
two field operators, as in the case of two-particle loss,
terms containing $\hat{L}$ in Eq.~\eqref{Lindblad_eq} effectively
behave like interaction terms.
The third term on the right-hand side of Eq. (\ref{Lindblad_eq})
$-\gamma\{ \hat{L}^{\dagger} \hat{L}, \hat{\rho}(t) \}/2$,
corresponds to
the so-called non-Hermitian evolution term and plays the role of 
an effective interaction with a complex amplitude in the context of two-body loss.
The second term,
$\gamma\hat{L}\hat{\rho}(t)\hat{L}^{\dagger}$, is 
referred to as the quantum jump term and
is often omitted in  non-Hermitian approaches to many-body physics.
However, retaining this jump term is essential to ensure the completely-positive 
trace-preserving nature of the evolution of the density matrix.
The noise-field formalism  offers an alternative representation
that consistently incorporates
the full non-unitary dynamics described by Eq.~\eqref{Lindblad_eq}.

\section{Frequency tuning}
In multi-terminal transport system,
the origin of frequencies is defined by the average of
reservoir's chemical potentials
\begin{equation}
    \mu
    =   \frac{\sum_i \mu_i}{\sum_i 1},
\end{equation}
where $i$ denotes the reservoir $i$.
When calculating reservoir's Green's functions,
the distribution function of the reservoir $i$ incorporates
the energy shift depending on $\mu_i - \mu$.
In this section, we review the method to compute this energy shift.

\begin{figure}
    \includegraphics[width=\linewidth]
    {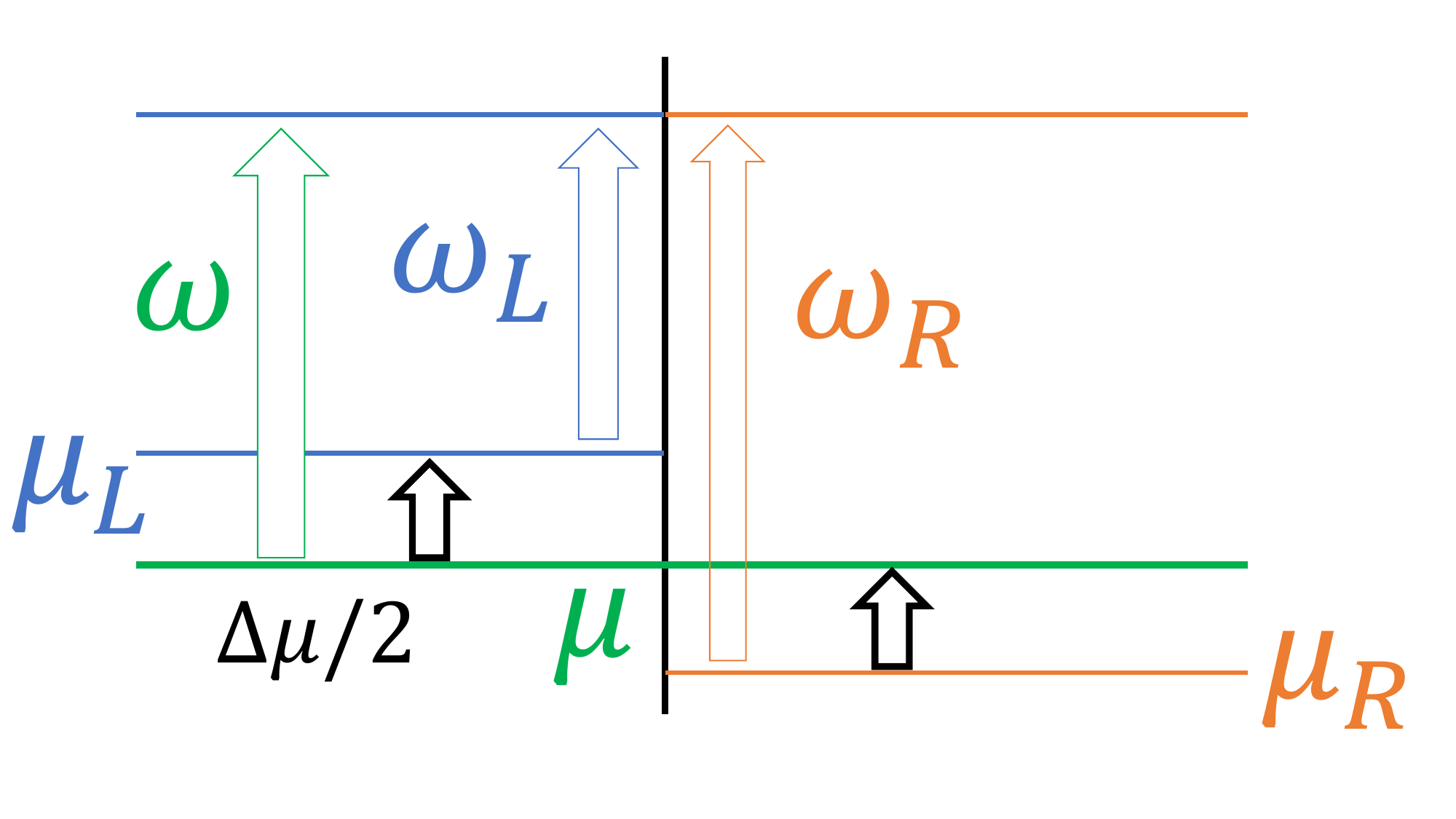}
    \caption{
       Concept of frequency tuning.
    }
    \label{frequency_tuning}
\end{figure}

To elucidate the essential features, we focus on a two-terminal system.
The typical expression of left (right)
reservoir's Keldysh Green's function
\begin{equation}
    g_{L(R)}^K(\mathbf{k}, \omega_{L(R)})
    =   -2\pi i\delta(\omega_{L(R)} - (\epsilon_{\mathbf{k}}-\mu_{L(R)}))
        n_{L(R)}(\omega_{L(R)}),
\end{equation}
with $\epsilon_{\mathbf{k}}=\mathbf{k}^2/2m$
and
$n_{L(R)}(\omega_{L(R)})
=   \frac{1}{e^{\omega_{L(R)}/T_{L(R)}} - \zeta}$,
implies that frequency $\omega_{L(R)}$ is measured from $\mu_{L(R)}$.
When connecting left and right reservoirs with a quantum point contact,
we measure the frequency $\omega$ from averaged chemical potential $\mu$, and
the energy shifts are considered as Fig.~\ref{frequency_tuning}, leading to
\begin{equation}
    \omega_{L/R}
    =   \omega \mp \frac{\Delta\mu}{2},
\end{equation}
with $\Delta\mu=\mu_L - \mu_R$.
Thus, we obtain
\begin{equation}
    g_{L/R}^K(\mathbf{k}, \omega)
    =   -2\pi i\delta( (\omega \mp \Delta\mu/2)  - \epsilon_{\mathbf{k}}-\mu_{L/R}))
        n_{L/R}(\omega),
\label{green_eq}
\end{equation}
with
$n_{L/R}(\omega)
=   \frac{1}{e^{(\omega \mp \Delta\mu/2)/T_{L/R}} - \zeta}$.

To calculate the above results systematically,
we start with the following canonical Hamiltonian
\begin{align}
    \hat{H}
    =   \sum_{s=L, R} \sum_{\mathbf{k}} (\epsilon_{\mathbf{k}} - \mu)
        \hat{\psi}_ {s\mathbf{k}}^{\dagger} \hat{\psi}_ {s\mathbf{k}},
\end{align}
where the energy is measured from $\mu$.
The system dynamics is described by $\hat{H}$ while
the equilibrium expectation value of Green's function (\ref{green_eq})
is obtained by grand-canonical Hamiltonian
\begin{align}
    \hat{K}
    =   \sum_{s=L, R} \sum_{\mathbf{k}} (\epsilon_{\mathbf{k}} - \mu_s)
        \hat{\psi}_ {s\mathbf{k}}^{\dagger} \hat{\psi}_ {s\mathbf{k}}.
\end{align}
Considering the difference between $\hat{H}$ and $\hat{K}$
\begin{equation}
    \hat{H}_{\Delta}
    =   \hat{K} - \hat{H}
    =   -\frac{\Delta\mu}{2} \left( \hat{N}_L - \hat{N}_R \right),
\end{equation}
the annihilation operator of the interaction picture can be rewritten as
\begin{align}
        \hat{\psi}_ {L/R\mathbf{k}, I}(\tau)
    &=  e^{i\hat{H}\tau}
        \hat{\psi}_ {L/R\mathbf{k}, S}
        e^{-i\hat{H}\tau} \\
    &=  e^{i\hat{K}\tau}
        e^{-i\hat{H}_{\Delta}\tau}
        \hat{\psi}_ {L/R\mathbf{k}, S}
        e^{i\hat{H}_{\Delta}\tau}
        e^{-i\hat{K}\tau} \\
    &=  e^{\mp i\Delta\mu \tau/2} e^{i\hat{K}\tau}
        \hat{\psi}_ {L/R\mathbf{k}, S}
        e^{-i\hat{K}\tau} \\
    &\equiv
        e^{\mp i\Delta\mu\tau/2} \hat{\psi}_ {L/R\mathbf{k}, K}(\tau),
\end{align}
where the subscripts $S, I$ and $K$ denote
the Schr{\"o}dinger picture,
the interaction picture constructed by $\hat{H}$,
and the interaction picture constructed by $\hat{K}$,
respectively.
Thus, we obtain
\begin{align}
    \Big[
        g_{L/R}^C(\mathbf{k}, \tau, \tau')
    \Big]_ I
    &\equiv
        -i\Braket{T_C\left[
            \hat{\psi}_{L/R\mathbf{k}, I}(\tau)
            \hat{\psi}_{L/R\mathbf{k}, I}^{\dagger}(\tau')
        \right]}_0 \\
    &=  -i\Braket{T_C\left[
            \hat{\psi}_{L/R\mathbf{k}, K}(\tau)
            \hat{\psi}_{L/R\mathbf{k}, K}^{\dagger}(\tau')
        \right]}_0
        e^{\mp i\Delta\mu(\tau-\tau')/2} \\
    &\equiv
        \Big[
            g_{L/R}^C(\mathbf{k}, \tau, \tau')
        \Big]_ K
        e^{\mp i\Delta\mu(\tau-\tau')/2},
\end{align}
where $[g]_{I(K)}$ is the Green's function constructed by
$\hat{\psi}_{I(K)}$.
When calculating the following integral,
the expression of grand-canonical Hamiltonian is given by
\begin{align}
    &\int_{-\infty}^{\infty} d\tau_1 \, f(\tau, \tau_1)
    \left[
        g_{L/R}^K(\mathbf{k}, \tau_1, \tau')
    \right]_ I \notag \\
    &=  \int_{-\infty}^{\infty} d\tau_1
        \int_{-\infty}^{\infty} \frac{d\omega}{2\pi}
        \int_{-\infty}^{\infty} \frac{d\omega_{L/R}}{2\pi}
        f(\omega)
        \left[
            g_{L/R}^K(\mathbf{k}, \omega_{L/R})
        \right]_ K
        e^{-i\omega(\tau-\tau_1)} e^{-i(\omega_{L/R} \pm \Delta\mu/2)(\tau_1-\tau')} \notag \\
    &=  \int_{-\infty}^{\infty} \frac{d\omega}{2\pi}
        f(\omega)
        \left[
            g_{L/R}^K(\mathbf{k}, \omega \mp \Delta\mu/2)
        \right]_ K
        e^{-i\omega(\tau-\tau')},
\end{align}
where $f$ is a function which arises in the calculation, and
$\Big[
    g_{L/R}^K(\mathbf{k}, \omega \mp \Delta\mu/2)
\Big]_ K$ corresponds Eq. (\ref{green_eq}).
The energy shifts of the retarded and advanced components
of reservoir's Green's function can be obtained
in the same manner.

\section{Analysis of dephasing model}
\label{DephasingModel}
To see how the noise field formalism works, 
we here analyze the dephasing model, which has also been
examined in Refs.~\cite{Dolgirev2020,Jin2022}.
For the sake of simplicity, we omit indices of site and spin.
Then, the perturbed Hamiltonian is given by
\begin{equation}
    \hat{V}_{\eta}
    =   \hat{d}^{\dagger} \hat{d} \hat{\eta}.
\label{couple_dephase}
\end{equation}
Here, $\hat{d}(\hat{d}^{\dagger})$ is an
annihilation (creation) operator of the system and
$\hat{\eta}$ is a real noise field,
obeying the following condition:
\begin{equation}
    \Braket{\hat{\eta}(\tau)\hat{\eta}(\tau')}_{\eta}
    =   \gamma\delta(\tau-\tau'),
    \quad
    \Braket{\hat{\eta}(\tau)}_{\eta} = 0.
\end{equation}
The contour-ordered Green's function satisfies
\begin{equation}
\begin{split}
    G^C(\tau, \tau')
    &=  \sum_{n=0}^{\infty}
        \frac{(-i)^{n+1}}{n!}
        \int_C d\tau_1 \cdots \int_C d\tau_n
        \Braket{T_C\left[
            \hat{V}_{\eta}(\tau_1)
            \cdots
            \hat{V}_{\eta}(\tau_n)
            \hat{d}(\tau) \hat{d}^{\dagger}(\tau')
        \right]}_0, \\
    &=  g^C(\tau, \tau')
        +
        \int_C d\tau_1 \int d\tau_2 \,
        g^C(\tau, \tau_1)
        \Sigma^C(\tau_1, \tau_2)
        G^C(\tau_2, \tau'),
\end{split}
\label{Dyson_eq_dephase}
\end{equation}
where $g^C$ is an unperturbed Green's function
and $\Sigma^C$ is a self-energy.

\begin{figure}
    \subfigure[]{
        \includegraphics[width=0.45\linewidth]
        {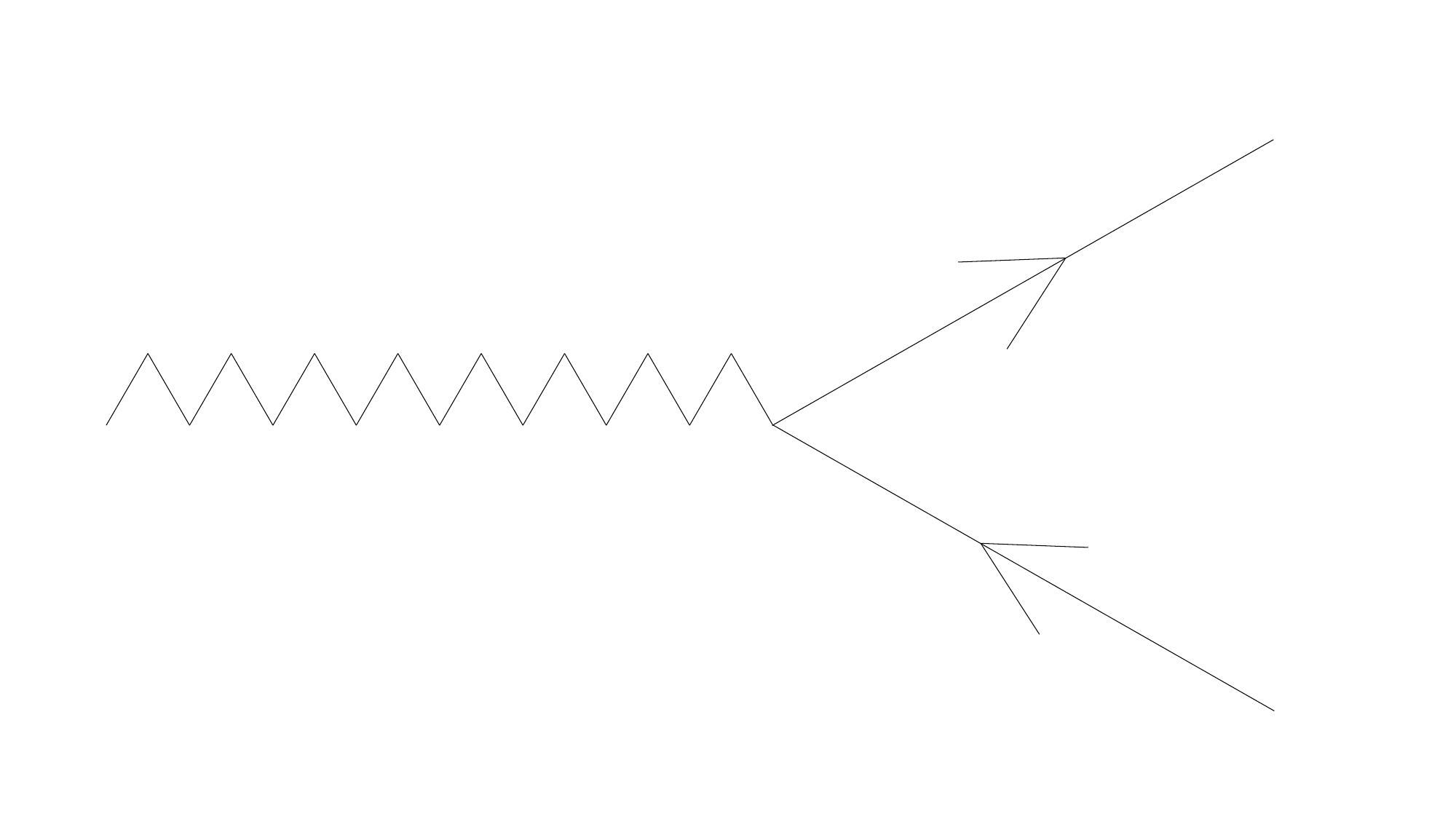}
        \label{diagram_couple_dephase}
    }
    \subfigure[]{
        \includegraphics[width=0.45\linewidth]
        {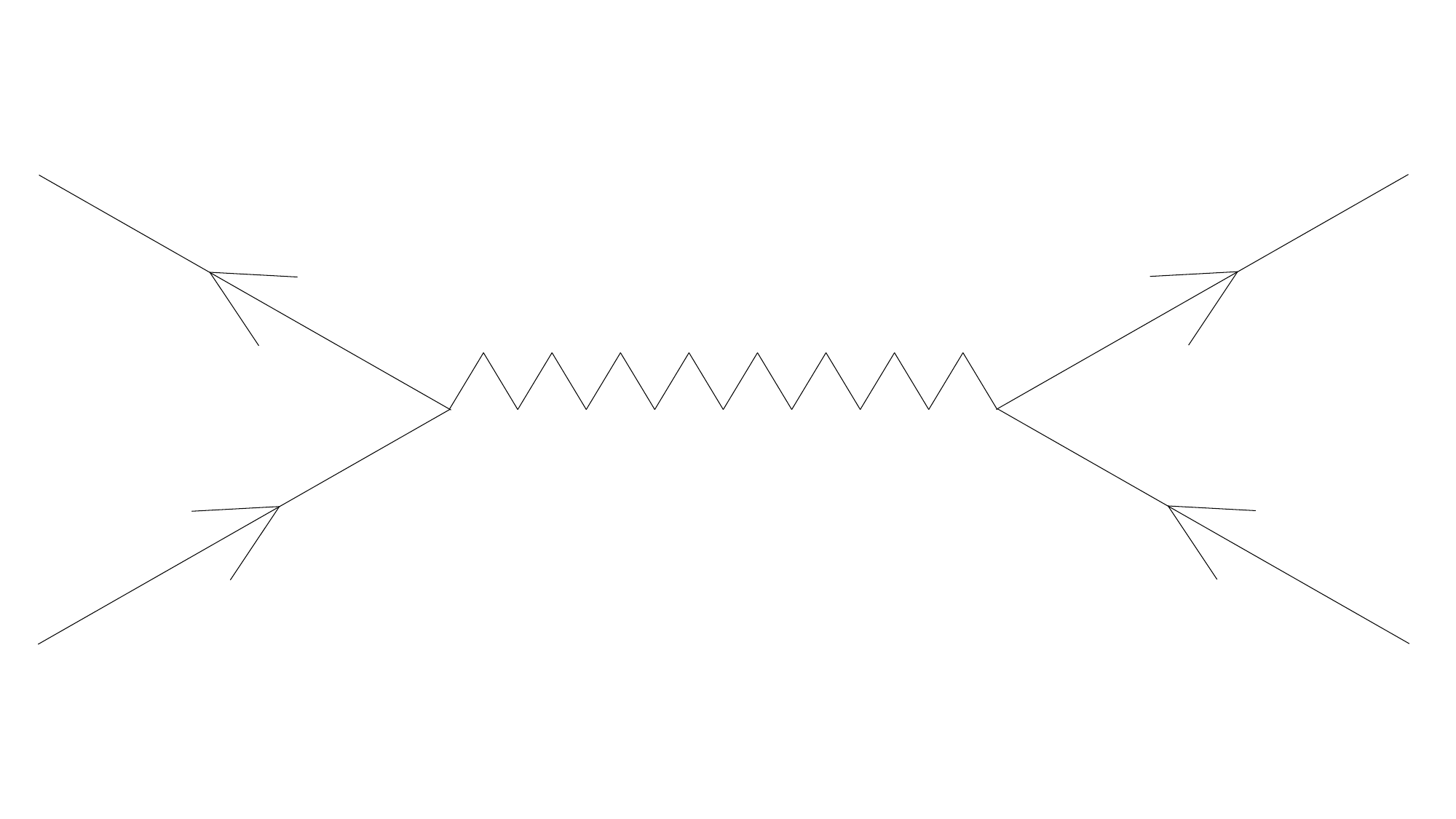}
        \label{diagram_interaction_dephase}
    }
    \caption{
        (a) Feynman diagram corresponding to the
        coupling term~(\ref{couple_dephase}).
        The solid lines represent
         $\hat{d}$ and $\hat{d}^{\dagger}$,
         and wavy line represents $\hat{\eta}$.
        (b) The term generated from
        the noise average of pair of~\eqref{couple_dephase}.
    }
\end{figure}

In terms of Feynman diagrams, the coupling term (\ref{couple_dephase}) is
depicted as Fig. S\ref{diagram_couple_dephase}.
Considering the noise average of pair of  (\ref{couple_dephase}),
we obtain the Feynman diagram depicted in
Fig.~S\ref{diagram_interaction_dephase}, which
is similar to the diagram of the two-body interaction.
A crucial difference between genuine
interactions and Fig.~S\ref{diagram_interaction_dephase} is that
the latter connects time arguments on
the forward and backward
branches of the Keldysh contour while
the former only connects time argument on
the same branch.

We now examine typical diagrams of the dephasing model.
At first, we verify that the contour-ordered Green's function
for the tadpole diagram
\begin{equation}
\begin{split}
    G_{\text{tad}}^C(\tau, \tau')
    &=  \int_C d\tau_1 \int_C d\tau_2 \,
        A^C(\tau, \tau_1)
        g_{\eta}^C(\tau_1, \tau_2) C^C(\tau_2, \tau_2)
        B^C(\tau_1, \tau'),
\end{split}
\end{equation}
vanishes as follows:
\begin{equation}
\begin{split}
    G_{\text{tad}}^T(\tau, \tau')
    &=  \frac{1}{2} \int d\tau_1 \int d\tau_2 \,
        A^T(\tau, \tau_1) B^T(\tau_1, \tau')
        \Big[
            g_{\eta}^T(\tau_1, \tau_2)
            -
            g_{\eta}^<(\tau_1, \tau_2)
        \Big]
        C^K(\tau_2, \tau_2) \\
        &\quad-
        \frac{1}{2} \int d\tau_1 \int d\tau_2 \,
        A^<(\tau, \tau_1) B^>(\tau_1, \tau')
        \Big[
            g_{\eta}^>(\tau_1, \tau_2)
            -
            g_{\eta}^{\tilde{T}}(\tau_1, \tau_2)
        \Big]
        C^K(\tau_2, \tau_2) \\
    &=  0,
\end{split}
\end{equation}
\begin{equation}
\begin{split}
    G_{\text{tad}}^<(\tau, \tau')
    &=  \frac{1}{2} \int d\tau_1 \int d\tau_2 \,
        A^T(\tau, \tau_1) B^<(\tau_1, \tau')
        \Big[
            g_{\eta}^T(\tau_1, \tau_2)
            -
            g_{\eta}^<(\tau_1, \tau_2)
        \Big]
        C^K(\tau_2, \tau_2) \\
        &\quad-
        \frac{1}{2} \int d\tau_1 \int d\tau_2 \,
        A^<(\tau, \tau_1) B^{\tilde{T}}(\tau_1, \tau')
        \Big[
            g_{\eta}^>(\tau_1, \tau_2)
            -
            g_{\eta}^{\tilde{T}}(\tau_1, \tau_2)
        \Big]
        C^K(\tau_2, \tau_2) \\
    &=  0,
\end{split}
\end{equation}
\begin{equation}
\begin{split}
    G_{\text{tad}}^>(\tau, \tau')
    &=  \frac{1}{2} \int d\tau_1 \int d\tau_2 \,
        A^>(\tau, \tau_1) B^T(\tau_1, \tau')
        \Big[
            g_{\eta}^T(\tau_1, \tau_2)
            -
            g_{\eta}^<(\tau_1, \tau_2)
        \Big]
        C^K(\tau_2, \tau_2) \\
        &\quad-
        \frac{1}{2} \int d\tau_1 \int d\tau_2 \,
        A^{\tilde{T}}(\tau, \tau_1) B^>(\tau_1, \tau')
        \Big[
            g_{\eta}^>(\tau_1, \tau_2)
            -
            g_{\eta}^{\tilde{T}}(\tau_1, \tau_2)
        \Big]
        C^K(\tau_2, \tau_2) \\
    &=  0,
\end{split}
\end{equation}
\begin{equation}
\begin{split}
    G_{\text{tad}}^{\tilde{T}}(\tau, \tau')
    &=  \frac{1}{2} \int d\tau_1 \int d\tau_2 \,
        A^>(\tau, \tau_1) B^<(\tau_1, \tau')
        \Big[
            g_{\eta}^T(\tau_1, \tau_2)
            -
            g_{\eta}^<(\tau_1, \tau_2)
        \Big]
        C^K(\tau_2, \tau_2) \\
        &\quad-
        \frac{1}{2} \int d\tau_1 \int d\tau_2 \,
        A^{\tilde{T}}(\tau, \tau_1) B^{\tilde{T}}(\tau_1, \tau')
        \Big[
            g_{\eta}^>(\tau_1, \tau_2)
            -
            g_{\eta}^{\tilde{T}}(\tau_1, \tau_2)
        \Big]
        C^K(\tau_2, \tau_2) \\
    &=  0,
\end{split}
\end{equation}
where the equal-time Green's function can be written as 
\begin{equation}
    C^T(\tau_2, \tau_2)
    =   C^{\tilde{T}}(\tau_2, \tau_2)
    =   \frac{1}{2} C^K(\tau_2, \tau_2),
\end{equation}
and
\begin{equation}
\begin{split}
    \left( \begin{matrix}
        g_{\eta}^T(\tau, \tau') & g_{\eta}^<(\tau, \tau') \\
        g_{\eta}^>(\tau, \tau') & g_{\eta}^{\tilde{T}}(\tau, \tau')
    \end{matrix} \right)
    &=   
    \left( \begin{matrix}
        -i\Braket{T\left[
            \hat{\eta}(\tau) \hat{\eta}(\tau')
        \right]}_{\eta} &
        -i\Braket{
            \hat{\eta}(\tau) \hat{\eta}(\tau')
        }_{\eta} \\
        -i\Braket{
            \hat{\eta}(\tau') \hat{\eta}(\tau)
        }_{\eta} &
        -i\Braket{\tilde{T}\left[
            \hat{\eta}(\tau) \hat{\eta}(\tau')
        \right]}_{\eta}
    \end{matrix} \right)
    =
    -i\gamma\delta(\tau - \tau')
    \left( \begin{matrix}
        1 & 1 \\
        1 & 1
    \end{matrix} \right),
\end{split}
\label{Green_noise}
\end{equation}
with $\theta(0)=1/2$.

\begin{figure}
    \subfigure[]{
        \includegraphics[width=0.3\linewidth]
        {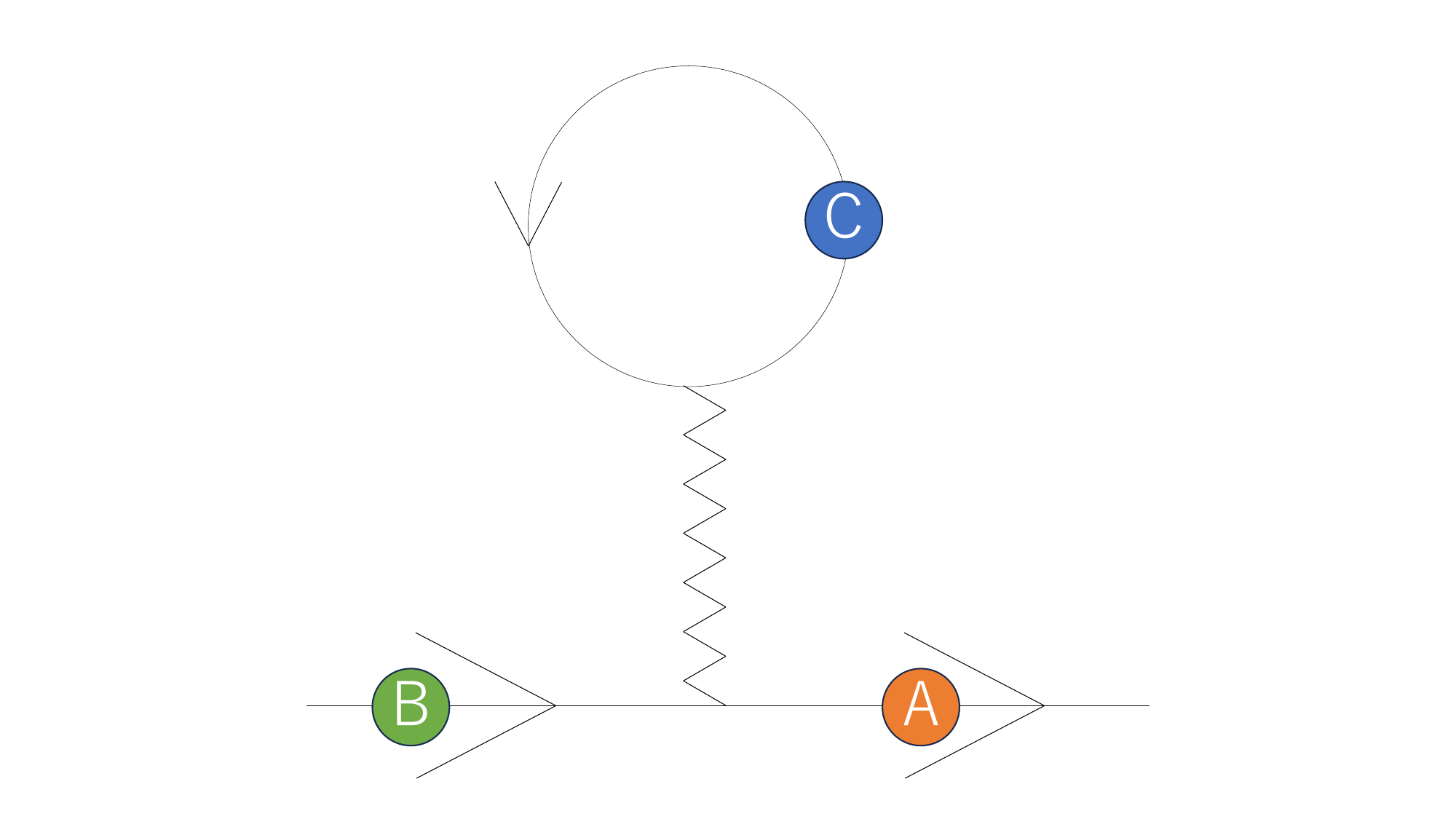}
        \label{tadpole}
    }
    \subfigure[]{
        \includegraphics[width=0.3\linewidth]
        {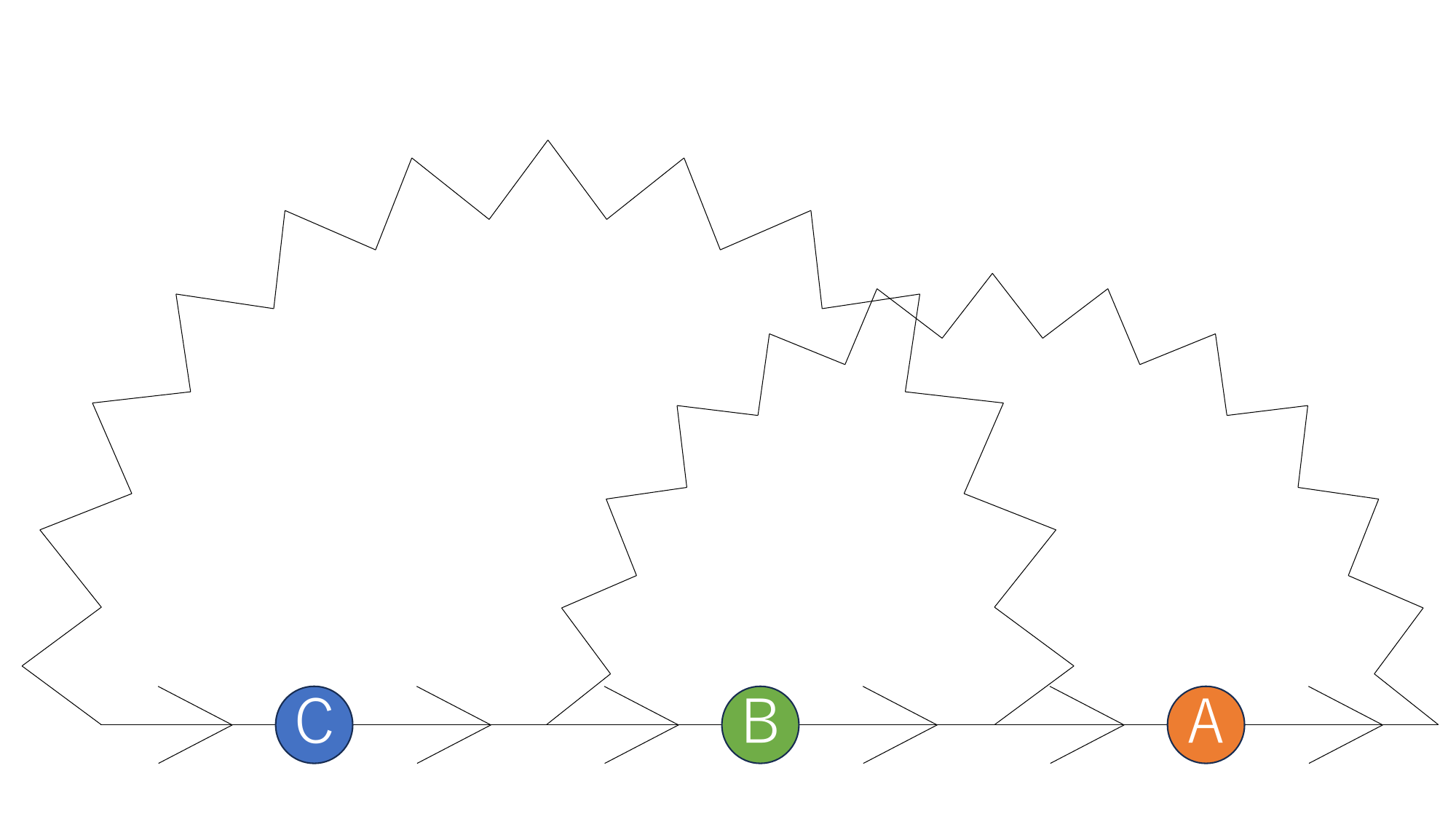}
        \label{diagram_cross_A_dephase}
    }
    \subfigure[]{
        \includegraphics[width=0.3\linewidth]
        {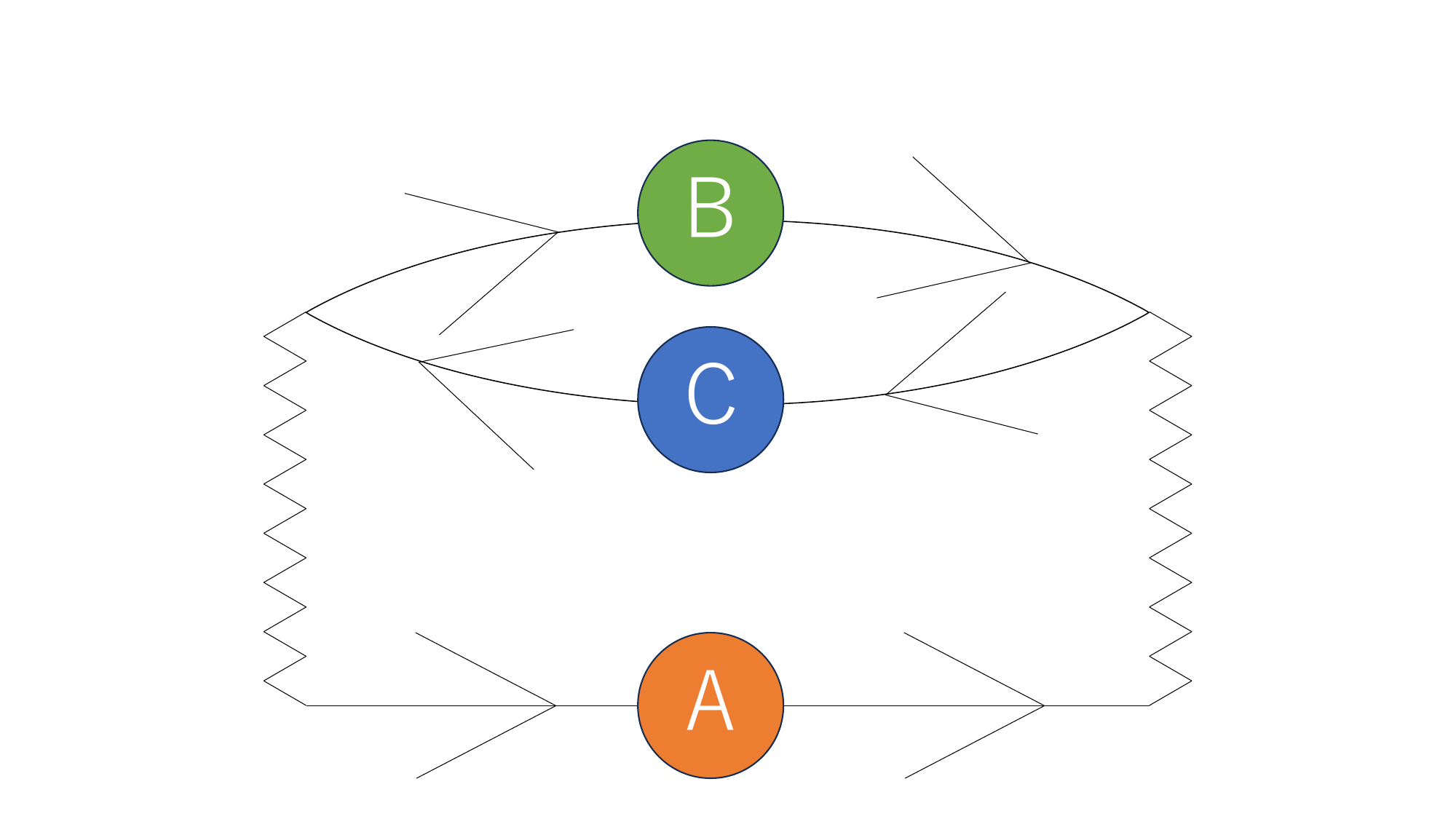}
        \label{diagram_cross_B_dephase}
    }
    \caption{
        (a) Tadpole diagram.
        (b) Self-energy of the crossing diagram.
        (c) Diagram associated with bath renormalization by
        $\hat{d}$ and $\hat{d}^{\dagger}$.
    }
\end{figure}

For the crossing diagram shown in Fig. S\ref{diagram_cross_A_dephase},
the self-energy is given by
\begin{equation}
\begin{split}
    \Sigma_{\ref{diagram_cross_A_dephase}}^T(\tau, \tau')
    &=  \int_C d\tau_1 \int_C d\tau_2 \,
        A^C(\tau^+, \tau_1)
        B^C(\tau_1, \tau_2)
        C^C(\tau_2, {\tau'}^+)
        g_{\eta}^C(\tau^+, \tau_2)
        g_{\eta}^C(\tau_1, {\tau'}^+) \\
    &=  \int_{-\infty}^{\infty} d\tau_1
        \int_{-\infty}^{\infty} d\tau_2 \,
        A^T(\tau, \tau_1)
        B^T(\tau_1, \tau_2)
        C^T(\tau_2, \tau')
        g_{\eta}^T(\tau, \tau_2)
        g_{\eta}^T(\tau_1, \tau') \\
        &\quad-
        \int_{-\infty}^{\infty} d\tau_1
        \int_{-\infty}^{\infty} d\tau_2 \,
        A^T(\tau, \tau_1)
        B^<(\tau_1, \tau_2)
        C^>(\tau_2, \tau')
        g_{\eta}^<(\tau, \tau_2)
        g_{\eta}^T(\tau_1, \tau') \\
        &\quad-
        \int_{-\infty}^{\infty} d\tau_1
        \int_{-\infty}^{\infty} d\tau_2 \,
        A^<(\tau, \tau_1)
        B^>(\tau_1, \tau_2)
        C^T(\tau_2, \tau')
        g_{\eta}^T(\tau, \tau_2)
        g_{\eta}^>(\tau_1, \tau') \\
        &\quad+
        \int_{-\infty}^{\infty} d\tau_1
        \int_{-\infty}^{\infty} d\tau_2 \,
        A^<(\tau, \tau_1)
        B^{\tilde{T}}(\tau_1, \tau_2)
        C^>(\tau_2, \tau')
        g_{\eta}^<(\tau, \tau_2)
        g_{\eta}^>(\tau_1, \tau') \\
    &=  -\gamma^2
        A^T(\tau, \tau')
        \Big[
            B^T(\tau', \tau)
            C^T(\tau, \tau')
            -
            B^<(\tau', \tau)
            C^>(\tau, \tau')
        \Big] \\
        &\quad-
        \gamma^2
        A^<(\tau, \tau')
        \Big[
            B^{\tilde{T}}(\tau', \tau)
            C^>(\tau, \tau')
            -
            B^>(\tau', \tau)
            C^T(\tau, \tau')
        \Big],
\end{split}
\end{equation}

\begin{equation}
\begin{split}
    \Sigma_{\ref{diagram_cross_A_dephase}}^<(\tau, \tau')
    &=  \int_C d\tau_1 \int_C d\tau_2 \,
        A^C(\tau^+, \tau_1)
        B^C(\tau_1, \tau_2)
        C^C(\tau_2, {\tau'}^-)
        g_{\eta}^C(\tau^+, \tau_2)
        g_{\eta}^C(\tau_1, {\tau'}^-) \\
    &=  \int_{-\infty}^{\infty} d\tau_1
        \int_{-\infty}^{\infty} d\tau_2 \,
        A^T(\tau, \tau_1)
        B^T(\tau_1, \tau_2)
        C^<(\tau_2, \tau')
        g_{\eta}^T(\tau, \tau_2)
        g_{\eta}^<(\tau_1, \tau') \\
        &\quad-
        \int_{-\infty}^{\infty} d\tau_1
        \int_{-\infty}^{\infty} d\tau_2 \,
        A^T(\tau, \tau_1)
        B^<(\tau_1, \tau_2)
        C^{\tilde{T}}(\tau_2, \tau')
        g_{\eta}^<(\tau, \tau_2)
        g_{\eta}^<(\tau_1, \tau') \\
        &\quad-
        \int_{-\infty}^{\infty} d\tau_1
        \int_{-\infty}^{\infty} d\tau_2 \,
        A^<(\tau, \tau_1)
        B^>(\tau_1, \tau_2)
        C^<(\tau_2, \tau')
        g_{\eta}^T(\tau, \tau_2)
        g_{\eta}^{\tilde{T}}(\tau_1, \tau') \\
        &\quad+
        \int_{-\infty}^{\infty} d\tau_1
        \int_{-\infty}^{\infty} d\tau_2 \,
        A^<(\tau, \tau_1)
        B^{\tilde{T}}(\tau_1, \tau_2)
        C^{\tilde{T}}(\tau_2, \tau')
        g_{\eta}^<(\tau, \tau_2)
        g_{\eta}^{\tilde{T}}(\tau_1, \tau') \\
    &=  -\gamma^2
        A^T(\tau, \tau')
        \Big[
            B^T(\tau', \tau)
            C^<(\tau, \tau')
            -
            B^<(\tau', \tau)
            C^{\tilde{T}}(\tau, \tau')
        \Big] \\
        &\quad-
        \gamma^2
        A^<(\tau, \tau')
        \Big[
            B^{\tilde{T}}(\tau', \tau)
            C^{\tilde{T}}(\tau, \tau')
            -
            B^>(\tau', \tau)
            C^<(\tau, \tau')
        \Big],
\end{split}
\end{equation}

\begin{equation}
\begin{split}
    \Sigma_{\ref{diagram_cross_A_dephase}}^>(\tau, \tau')
    &=  \int_C d\tau_1 \int_C d\tau_2 \,
        A^C(\tau^-, \tau_1)
        B^C(\tau_1, \tau_2)
        C^C(\tau_2, {\tau'}^+)
        g_{\eta}^C(\tau^-, \tau_2)
        g_{\eta}^C(\tau_1, {\tau'}^+) \\
    &=  \int_{-\infty}^{\infty} d\tau_1
        \int_{-\infty}^{\infty} d\tau_2 \,
        A^>(\tau, \tau_1)
        B^T(\tau_1, \tau_2)
        C^T(\tau_2, \tau')
        g_{\eta}^>(\tau, \tau_2)
        g_{\eta}^T(\tau_1, \tau') \\
        &\quad-
        \int_{-\infty}^{\infty} d\tau_1
        \int_{-\infty}^{\infty} d\tau_2 \,
        A^>(\tau, \tau_1)
        B^<(\tau_1, \tau_2)
        C^>(\tau_2, \tau')
        g_{\eta}^{\tilde{T}}(\tau, \tau_2)
        g_{\eta}^T(\tau_1, \tau') \\
        &\quad-
        \int_{-\infty}^{\infty} d\tau_1
        \int_{-\infty}^{\infty} d\tau_2 \,
        A^{\tilde{T}}(\tau, \tau_1)
        B^>(\tau_1, \tau_2)
        C^T(\tau_2, \tau')
        g_{\eta}^>(\tau, \tau_2)
        g_{\eta}^>(\tau_1, \tau') \\
        &\quad+
        \int_{-\infty}^{\infty} d\tau_1
        \int_{-\infty}^{\infty} d\tau_2 \,
        A^{\tilde{T}}(\tau, \tau_1)
        B^{\tilde{T}}(\tau_1, \tau_2)
        C^>(\tau_2, \tau')
        g_{\eta}^{\tilde{T}}(\tau, \tau_2)
        g_{\eta}^>(\tau_1, \tau') \\
    &=  -\gamma^2
        A^>(\tau, \tau')
        \Big[
            B^T(\tau', \tau)
            C^T(\tau, \tau')
            -
            B^<(\tau', \tau)
            C^>(\tau, \tau')
        \Big] \\
        &\quad-
        \gamma^2
        A^{\tilde{T}}(\tau, \tau')
        \Big[
            B^{\tilde{T}}(\tau', \tau)
            C^>(\tau, \tau')
            -
            B^>(\tau', \tau)
            C^T(\tau, \tau')
        \Big],
\end{split}
\end{equation}

\begin{equation}
\begin{split}
    \Sigma_{\ref{diagram_cross_A_dephase}}^{\tilde{T}}(\tau, \tau')
    &=  \int_C d\tau_1 \int_C d\tau_2 \,
        A^C(\tau^-, \tau_1)
        B^C(\tau_1, \tau_2)
        C^C(\tau_2, {\tau'}^-)
        g_{\eta}^C(\tau^-, \tau_2)
        g_{\eta}^C(\tau_1, {\tau'}^-) \\
    &=  \int_{-\infty}^{\infty} d\tau_1
        \int_{-\infty}^{\infty} d\tau_2 \,
        A^>(\tau, \tau_1)
        B^T(\tau_1, \tau_2)
        C^<(\tau_2, \tau')
        g_{\eta}^>(\tau, \tau_2)
        g_{\eta}^<(\tau_1, \tau') \\
        &\quad-
        \int_{-\infty}^{\infty} d\tau_1
        \int_{-\infty}^{\infty} d\tau_2 \,
        A^>(\tau, \tau_1)
        B^<(\tau_1, \tau_2)
        C^{\tilde{T}}(\tau_2, \tau')
        g_{\eta}^{\tilde{T}}(\tau, \tau_2)
        g_{\eta}^<(\tau_1, \tau') \\
        &\quad-
        \int_{-\infty}^{\infty} d\tau_1
        \int_{-\infty}^{\infty} d\tau_2 \,
        A^{\tilde{T}}(\tau, \tau_1)
        B^>(\tau_1, \tau_2)
        C^<(\tau_2, \tau')
        g_{\eta}^>(\tau, \tau_2)
        g_{\eta}^{\tilde{T}}(\tau_1, \tau') \\
        &\quad+
        \int_{-\infty}^{\infty} d\tau_1
        \int_{-\infty}^{\infty} d\tau_2 \,
        A^{\tilde{T}}(\tau, \tau_1)
        B^{\tilde{T}}(\tau_1, \tau_2)
        C^{\tilde{T}}(\tau_2, \tau')
        g_{\eta}^{\tilde{T}}(\tau, \tau_2)
        g_{\eta}^{\tilde{T}}(\tau_1, \tau') \\
    &=  -\gamma^2
        A^>(\tau, \tau')
        \Big[
            B^T(\tau', \tau)
            C^<(\tau, \tau')
            -
            B^<(\tau', \tau)
            C^{\tilde{T}}(\tau, \tau')
        \Big] \\
        &\quad-
        \gamma^2
        A^{\tilde{T}}(\tau, \tau')
        \Big[
            B^{\tilde{T}}(\tau', \tau)
            C^{\tilde{T}}(\tau, \tau')
            -
            B^>(\tau', \tau)
            C^<(\tau, \tau')
        \Big].
\end{split}
\end{equation}
Using the relations of each component (\ref{time-RAK}),
we obtain
\begin{align}
    \Sigma_{\ref{diagram_cross_A_dephase}}^K(\tau, \tau')
    &=
    \frac{1}{2}
    \Big(
        \Sigma^T(\tau, \tau')
        +
        \Sigma^<(\tau, \tau')
        +
        \Sigma^>(\tau, \tau')
        +
        \Sigma^{\tilde{T}}(\tau, \tau')
    \Big) \notag \\
    &=  -\gamma^2
        \Big(
            A^R(\tau, \tau') B^R(\tau', \tau) C^K(\tau, \tau')
            +
            A^R(\tau, \tau') B^K(\tau', \tau) C^A(\tau, \tau')
            +
            A^K(\tau, \tau') B^A(\tau', \tau) C^A(\tau, \tau')
        \Big) \notag \\
    &=  0, \\
    \Sigma_{\ref{diagram_cross_A_dephase}}^R(\tau, \tau')
    &=
    \frac{1}{2}
    \Big(
        \Sigma^T(\tau, \tau')
        -
        \Sigma^<(\tau, \tau')
        +
        \Sigma^>(\tau, \tau')
        -
        \Sigma^{\tilde{T}}(\tau, \tau')
    \Big) \notag \\
    &=  -\gamma^2
        A^R(\tau, \tau') B^R(\tau', \tau) C^R(\tau, \tau') \notag \\
    &=  0, \\
    \Sigma_{\ref{diagram_cross_A_dephase}}^A(\tau, \tau')
    &=
    \frac{1}{2}
    \Big(
        \Sigma^T(\tau, \tau')
        +
        \Sigma^<(\tau, \tau')
        -
        \Sigma^>(\tau, \tau')
        -
        \Sigma^{\tilde{T}}(\tau, \tau')
    \Big) \notag \\
    &=  -\gamma^2
        A^A(\tau, \tau') B^A(\tau', \tau) C^A(\tau, \tau') \notag \\
    &=  0,
\end{align}
with the causality
$A^R(\tau, \tau') B^R(\tau', \tau)
=   B^A(\tau', \tau) C^A(\tau, \tau')
=   A^R(\tau, \tau') C^A(\tau, \tau')
=   0$
in the integral of Eq. (\ref{Dyson_eq_dephase}).
Thus, the crossing diagram is shown to vanish.

We next consider the diagram shown in
Fig. S\ref{diagram_cross_B_dephase},
representing the effective renormalization of the noise field properties
by the fields of the system $\hat{d}$ and $\hat{d}^{\dagger}$.
As in the case of the crossing diagram,
the self-energy is given by
\begin{equation}
\begin{split}
    &\Sigma_{\ref{diagram_cross_B_dephase}}^T(\tau, \tau') \\
    &=  \int_C d\tau_1 \int_C d\tau_2 \,
        A^C(\tau^+, {\tau'}^+)
        B^C(\tau_1, \tau_2)
        C^C(\tau_2, \tau_1)
        g_{\eta}^C(\tau^+, \tau_1)
        g_{\eta}^C(\tau_2, {\tau'}^+) \\
    &=  \int_{-\infty}^{\infty} d\tau_1
        \int_{-\infty}^{\infty} d\tau_2 \,
        A^T(\tau, \tau')
        B^T(\tau_1, \tau_2)
        C^T(\tau_2, \tau_1)
        g_{\eta}^T(\tau, \tau_1)
        g_{\eta}^T(\tau_2, \tau') \\
        &\quad-
        \int_{-\infty}^{\infty} d\tau_1
        \int_{-\infty}^{\infty} d\tau_2 \,
        A^T(\tau, \tau')
        B^<(\tau_1, \tau_2)
        C^>(\tau_2, \tau_1)
        g_{\eta}^T(\tau, \tau_1)
        g_{\eta}^>(\tau_2, \tau') \\
        &\quad-
        \int_{-\infty}^{\infty} d\tau_1
        \int_{-\infty}^{\infty} d\tau_2 \,
        A^T(\tau, \tau')
        B^>(\tau_1, \tau_2)
        C^<(\tau_2, \tau_1)
        g_{\eta}^<(\tau, \tau_1)
        g_{\eta}^T(\tau_2, \tau') \\
        &\quad+
        \int_{-\infty}^{\infty} d\tau_1
        \int_{-\infty}^{\infty} d\tau_2 \,
        A^T(\tau, \tau')
        B^{\tilde{T}}(\tau_1, \tau_2)
        C^{\tilde{T}}(\tau_2, \tau_1)
        g_{\eta}^<(\tau, \tau_1)
        g_{\eta}^>(\tau_2, \tau') \\
    &=  -\gamma^2
        A^T(\tau, \tau')
        \Big[
            B^T(\tau, \tau')
            C^T(\tau', \tau)
            -
            B^<(\tau, \tau')
            C^>(\tau', \tau)
            -
            B^>(\tau, \tau')
            C^<(\tau', \tau)
            +
            B^{\tilde{T}}(\tau, \tau')
            C^{\tilde{T}}(\tau', \tau)
        \Big],
\end{split}
\end{equation}

\begin{equation}
\begin{split}
    &\Sigma_{\ref{diagram_cross_B_dephase}}^<(\tau, \tau') \\
    &=  \int_C d\tau_1 \int_C d\tau_2 \,
        A^C(\tau^+, {\tau'}^-)
        B^C(\tau_1, \tau_2)
        C^C(\tau_2, \tau_1)
        g_{\eta}^C(\tau^+, \tau_1)
        g_{\eta}^C(\tau_2, {\tau'}^-) \\
    &=  \int_{-\infty}^{\infty} d\tau_1
        \int_{-\infty}^{\infty} d\tau_2 \,
        A^<(\tau, \tau')
        B^T(\tau_1, \tau_2)
        C^T(\tau_2, \tau_1)
        g_{\eta}^T(\tau, \tau_1)
        g_{\eta}^<(\tau_2, \tau') \\
        &\quad-
        \int_{-\infty}^{\infty} d\tau_1
        \int_{-\infty}^{\infty} d\tau_2 \,
        A^<(\tau, \tau')
        B^<(\tau_1, \tau_2)
        C^>(\tau_2, \tau_1)
        g_{\eta}^T(\tau, \tau_1)
        g_{\eta}^{\tilde{T}}(\tau_2, \tau') \\
        &\quad-
        \int_{-\infty}^{\infty} d\tau_1
        \int_{-\infty}^{\infty} d\tau_2 \,
        A^<(\tau, \tau')
        B^>(\tau_1, \tau_2)
        C^<(\tau_2, \tau_1)
        g_{\eta}^<(\tau, \tau_1)
        g_{\eta}^<(\tau_2, \tau') \\
        &\quad+
        \int_{-\infty}^{\infty} d\tau_1
        \int_{-\infty}^{\infty} d\tau_2 \,
        A^<(\tau, \tau')
        B^{\tilde{T}}(\tau_1, \tau_2)
        C^{\tilde{T}}(\tau_2, \tau_1)
        g_{\eta}^<(\tau, \tau_1)
        g_{\eta}^{\tilde{T}}(\tau_2, \tau') \\
    &=  -\gamma^2
        A^<(\tau, \tau')
        \Big[
            B^T(\tau, \tau')
            C^T(\tau', \tau)
            -
            B^<(\tau, \tau')
            C^>(\tau', \tau)
            -
            B^>(\tau, \tau')
            C^<(\tau', \tau)
            +
            B^{\tilde{T}}(\tau, \tau')
            C^{\tilde{T}}(\tau', \tau)
        \Big],
\end{split}
\end{equation}

\begin{equation}
\begin{split}
    &\Sigma_{\ref{diagram_cross_B_dephase}}^>(\tau, \tau') \\
    &=  \int_C d\tau_1 \int_C d\tau_2 \,
        A^C(\tau^-, {\tau'}^+)
        B^C(\tau_1, \tau_2)
        C^C(\tau_2, \tau_1)
        g_{\eta}^C(\tau^-, \tau_1)
        g_{\eta}^C(\tau_2, {\tau'}^+) \\
    &=  \int_{-\infty}^{\infty} d\tau_1
        \int_{-\infty}^{\infty} d\tau_2 \,
        A^>(\tau, \tau')
        B^T(\tau_1, \tau_2)
        C^T(\tau_2, \tau_1)
        g_{\eta}^>(\tau, \tau_1)
        g_{\eta}^T(\tau_2, \tau') \\
        &\quad-
        \int_{-\infty}^{\infty} d\tau_1
        \int_{-\infty}^{\infty} d\tau_2 \,
        A^>(\tau, \tau')
        B^<(\tau_1, \tau_2)
        C^>(\tau_2, \tau_1)
        g_{\eta}^>(\tau, \tau_1)
        g_{\eta}^>(\tau_2, \tau') \\
        &\quad-
        \int_{-\infty}^{\infty} d\tau_1
        \int_{-\infty}^{\infty} d\tau_2 \,
        A^>(\tau, \tau')
        B^>(\tau_1, \tau_2)
        C^<(\tau_2, \tau_1)
        g_{\eta}^{\tilde{T}}(\tau, \tau_1)
        g_{\eta}^T(\tau_2, \tau') \\
        &\quad+
        \int_{-\infty}^{\infty} d\tau_1
        \int_{-\infty}^{\infty} d\tau_2 \,
        A^>(\tau, \tau')
        B^{\tilde{T}}(\tau_1, \tau_2)
        C^{\tilde{T}}(\tau_2, \tau_1)
        g_{\eta}^{\tilde{T}}(\tau, \tau_1)
        g_{\eta}^>(\tau_2, \tau') \\
    &=  -\gamma^2
        A^>(\tau, \tau')
        \Big[
            B^T(\tau, \tau')
            C^T(\tau', \tau)
            -
            B^<(\tau, \tau')
            C^>(\tau', \tau)
            -
            B^>(\tau, \tau')
            C^<(\tau', \tau)
            +
            B^{\tilde{T}}(\tau, \tau')
            C^{\tilde{T}}(\tau', \tau)
        \Big],
\end{split}
\end{equation}

\begin{equation}
\begin{split}
    &\Sigma_{\ref{diagram_cross_B_dephase}}^{\tilde{T}}(\tau, \tau') \\
    &=  \int_C d\tau_1 \int_C d\tau_2 \,
        A^C(\tau^-, {\tau'}^-)
        B^C(\tau_1, \tau_2)
        C^C(\tau_2, \tau_1)
        g_{\eta}^C(\tau^-, \tau_1)
        g_{\eta}^C(\tau_2, {\tau'}^-) \\
    &=  \int_{-\infty}^{\infty} d\tau_1
        \int_{-\infty}^{\infty} d\tau_2 \,
        A^{\tilde{T}}(\tau, \tau')
        B^T(\tau_1, \tau_2)
        C^T(\tau_2, \tau_1)
        g_{\eta}^>(\tau, \tau_1)
        g_{\eta}^<(\tau_2, \tau') \\
        &\quad-
        \int_{-\infty}^{\infty} d\tau_1
        \int_{-\infty}^{\infty} d\tau_2 \,
        A^{\tilde{T}}(\tau, \tau')
        B^<(\tau_1, \tau_2)
        C^>(\tau_2, \tau_1)
        g_{\eta}^>(\tau, \tau_1)
        g_{\eta}^{\tilde{T}}(\tau_2, \tau') \\
        &\quad-
        \int_{-\infty}^{\infty} d\tau_1
        \int_{-\infty}^{\infty} d\tau_2 \,
        A^{\tilde{T}}(\tau, \tau')
        B^>(\tau_1, \tau_2)
        C^<(\tau_2, \tau_1)
        g_{\eta}^{\tilde{T}}(\tau, \tau_1)
        g_{\eta}^<(\tau_2, \tau') \\
        &\quad+
        \int_{-\infty}^{\infty} d\tau_1
        \int_{-\infty}^{\infty} d\tau_2 \,
        A^{\tilde{T}}(\tau, \tau')
        B^{\tilde{T}}(\tau_1, \tau_2)
        C^{\tilde{T}}(\tau_2, \tau_1)
        g_{\eta}^{\tilde{T}}(\tau, \tau_1)
        g_{\eta}^{\tilde{T}}(\tau_2, \tau') \\
    &=  -\gamma^2
        A^{\tilde{T}}(\tau, \tau')
        \Big[
            B^T(\tau, \tau')
            C^T(\tau', \tau)
            -
            B^<(\tau, \tau')
            C^>(\tau', \tau)
            -
            B^>(\tau, \tau')
            C^<(\tau', \tau)
            +
            B^{\tilde{T}}(\tau, \tau')
            C^{\tilde{T}}(\tau', \tau)
        \Big].
\end{split}
\end{equation}
Thus, we obtain
\begin{align}
    \Sigma_{\ref{diagram_cross_B_dephase}}^K(\tau, \tau')
    &=  A^K(\tau, \tau')
        \Big(
            B^R(\tau, \tau') C^R(\tau', \tau)
            +
            B^A(\tau, \tau') C^A(\tau', \tau)
        \Big)
    =   0, \\
    \Sigma_{\ref{diagram_cross_B_dephase}}^R(\tau, \tau')
    &=  A^R(\tau, \tau')
        \Big(
            B^R(\tau, \tau') C^R(\tau', \tau)
            +
            B^A(\tau, \tau') C^A(\tau', \tau)
        \Big)
    =   0, \\
    \Sigma_{\ref{diagram_cross_B_dephase}}^A(\tau, \tau')
    &=  A^A(\tau, \tau')
        \Big(
            B^R(\tau, \tau') C^R(\tau', \tau)
            +
            B^A(\tau, \tau') C^A(\tau', \tau)
        \Big)
    =   0.
\end{align}
Thus,  it turns out that 
the bath renormalization does not appear.
We note that the absence of the bath renormalization
is a particular feature in the dephasing model,
and is not expected in the case of two-body loss.

\begin{figure}
    \includegraphics[width=\linewidth]{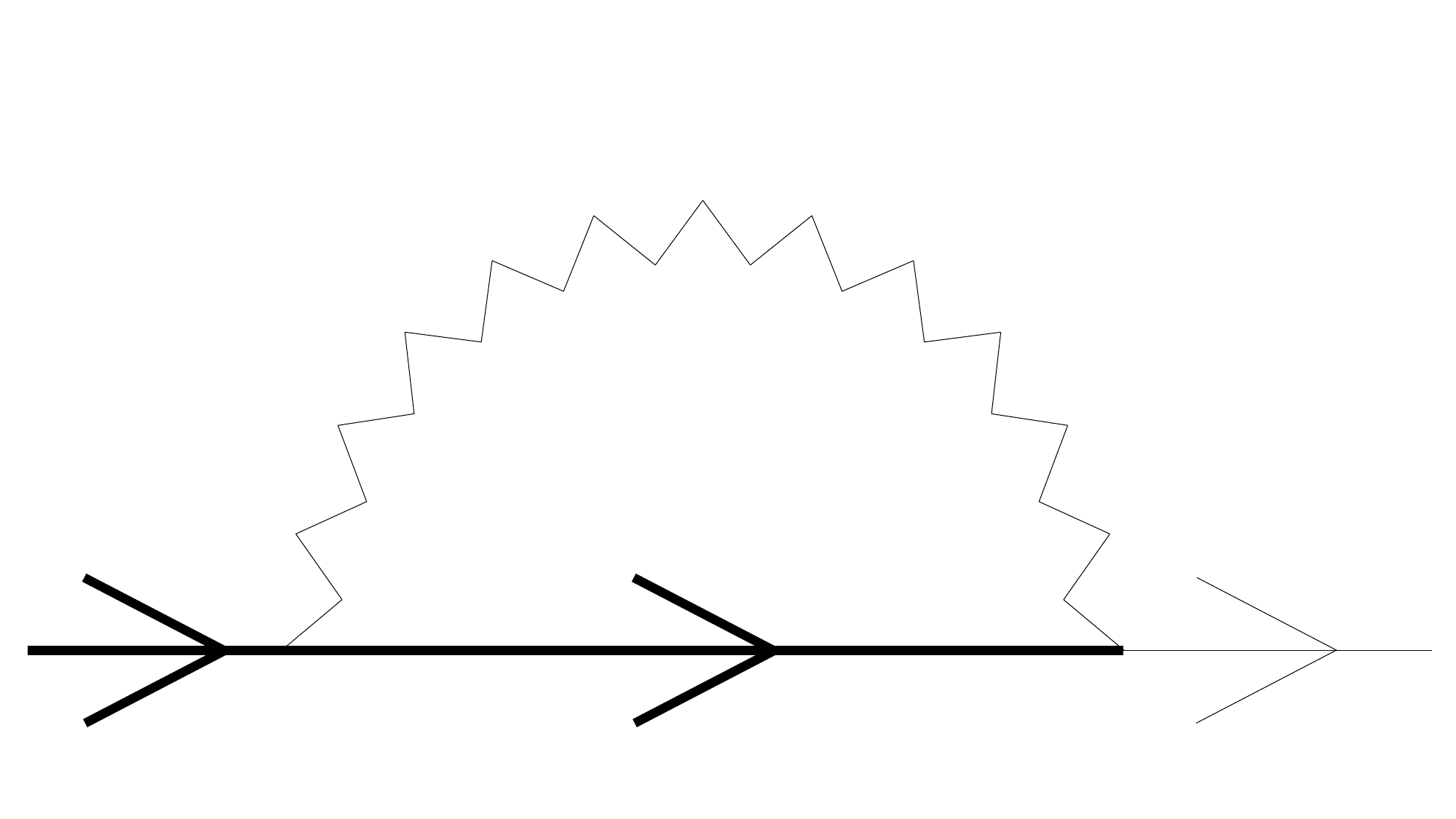}
    \caption{
       Feynman diagram corresponding to the SCBA in the dephasing model.
    }
    \label{SCBA_dephase}
\end{figure}

The above lessons imply that 
a number of diagrams in the dephasing model 
vanish due to causality and
symmetry of noise Green's functions, namely,
$g_{\eta}^T(\tau, \tau')
=   g_{\eta}^<(\tau, \tau')
=   g_{\eta}^>(\tau, \tau')
=   g_{\eta}^{\tilde{T}}(\tau, \tau')$.
At the end of the day, the remaining diagrams in the dephasing
model is ones corresponding to the 
self-cinsistent Born approximation (SCBA) whose 
the Feynman diagram is shown in~Fig.~\ref{SCBA_dephase}.
By using the symmetry of noise Green's functions,
the self-energy of the SCBA is obtained as
\begin{equation}
    \Sigma^C(\tau, \tau')
    =   ig_{\eta}^C(\tau, \tau')
        G^C(\tau, \tau').
\end{equation}

From the lessons of the dephasing model, we can claim why genuine interactions cannot be solved exactly.
As in the case of the noise-field formalism,
we can introduce auxiliary fields for genuine interactions by assuming
$g^<(\tau, \tau')=g^>(\tau, \tau')=0$ for auxiliary fields. Under this condition, however,
auxiliary-field Green's functions do not show symmetric properties
unlike noise-field Green's functions in the dephasing model.
For this reason, cancellation of a number of Feynman diagrams 
is not expected in generic many-body systems.

\section{Two-particle loss: single-site case}
\label{TwoBodyLossSingleSite}
In this section, we derive the technical details of the main text
by focusing on the single-site case.
The Hamiltonian of the system is
\begin{equation}
    H
    =   \sum_{s=L, R} \sum_{\sigma=\uparrow, \downarrow}
        H_{s\sigma}
        +
        H_T
        +
        H_{1D}^0,
\end{equation}
where
\begin{align}
    H_{s\sigma}
    &=  \sum_{\mathbf{k}}
        (\epsilon_{\mathbf{k}} - \mu_{s\sigma})
        \psi_{s\mathbf{k}\sigma}^{\dagger}
        \psi_{s\mathbf{k}\sigma}, \\
    H_T
    &=  -\sum_{s\mathbf{k}\sigma}
        \Big(
            t_{\mathbf{k}}
            \psi_{s\mathbf{k}\sigma}^{\dagger} d_{0\sigma}
            +
            t_{\mathbf{k}}^*
            d_{0\sigma}^{\dagger} \psi_{\mathbf{k}s\sigma}
        \Big), \\
    H_{1D}^0
    &=  \sum_{\sigma} \epsilon_0
        d_{0\sigma}^{\dagger} d_{0\sigma}.
\end{align}
%When focusing on fermions, an on-site interaction,
%such as those in the Anderson model, can be considered,
%and enables to be treated as energy shift
%within the mean-field approximation
%\begin{equation}
%    H_{1D}^0
%    =   \sum_{\sigma} \epsilon_0
%        d_{0\sigma}^{\dagger} d_{0\sigma}
%        +
%        U d_{0\uparrow}^{\dagger} d_{0\uparrow}
%       d_{0\downarrow}^{\dagger} d_{0\downarrow}
%    \approx
%        \sum_{\sigma}
%        (\epsilon_0 + Un_{0\bar{\sigma}})
%        d_{0\sigma}^{\dagger} d_{0\sigma}.
%\end{equation}
We now turn to the general form and
introduce the noise field term
\begin{equation}
    V_{\eta}
    =   d_{0\uparrow}^{\dagger}
        d_{0\downarrow}^{\dagger}
        \eta
        +
        \eta^{\dagger}
        d_{0\downarrow}
        d_{0\uparrow}.
\end{equation}
To obtain the same results as those
obtained by the Lindblad master equation,
we impose the following conditions:
\begin{equation}
    \Braket{\eta(\tau)\eta^{\dagger}(\tau')}_{\eta}
    =   \gamma\delta(\tau-\tau'),
    \quad
    \Braket{\eta^{\dagger}(\tau)\eta(\tau')}_{\eta}
    =   \Braket{\eta(\tau)\eta(\tau')}_{\eta}
    =   \Braket{
            \eta^{\dagger}(\tau)\eta^{\dagger}(\tau')
        }_{\eta}
    =   \Braket{\eta(\tau)}_{\eta}
    =   \Braket{\eta^{\dagger}(\tau)}_{\eta}
    =   0.
\end{equation}

Current operators are then given by
\begin{align}
    I_{s\sigma}
    &=  -\frac{\partial}{\partial \tau} N_{s\sigma}
    =   -i\Big[
            H_{\text{tot}}, N_{s\sigma}
        \Big]
    =   -i\Big[
            H_T, N_{s\sigma}
        \Big]
    =   -i\sum_{\mathbf{k}}
        \Big(
            t_{\mathbf{k}} \psi_{s\mathbf{k}\sigma}^{\dagger} d_{0\sigma}
            -
            t_{\mathbf{k}}^* d_{0\sigma}^{\dagger} \psi_{s\mathbf{k}\sigma}
        \Big), \\
    I_{E, s\sigma}
    &=  -\frac{\partial}{\partial \tau}
        \Big(
            H_{s\sigma} + \mu_{s\sigma} N_{s\sigma}
        \Big)
    =   -i\sum_{\mathbf{k}} \epsilon_{\mathbf{k}}
        \Big(
            t_{\mathbf{k}} \psi_{s\mathbf{k}\sigma}^{\dagger} d_{0\sigma}
            -
            t_{\mathbf{k}}^* d_{0\sigma}^{\dagger} \psi_{s\mathbf{k}\sigma}
        \Big).
\end{align}
The expectation values of current operators are expressed as follows:
\begin{align}
    \Braket{I_{s\sigma}}
    &=  \zeta \sum_{\mathbf{k}}
        \operatorname{Re}
        \Big[
            t_{\mathbf{k}} G_{ds\sigma}^K(\mathbf{k}, \tau, \tau)
        \Big]
    =   \zeta \int \frac{d\omega}{2\pi} \sum_{\mathbf{k}}
        \operatorname{Re}
        \Big[
            t_{\mathbf{k}} G_{ds\sigma}^K(\mathbf{k}, \omega)
        \Big], \\
    \Braket{I_{E, s\sigma}}
    &=  \zeta \sum_{\mathbf{k}} \epsilon_{\mathbf{k}}
        \operatorname{Re}
        \Big[
            t_{\mathbf{k}} G_{ds\sigma}^K(\mathbf{k}, \tau, \tau)
        \Big]
    =   \zeta \int \frac{d\omega}{2\pi} \sum_{\mathbf{k}}
        \epsilon_{\mathbf{k}}
        \operatorname{Re}
        \Big[
            t_{\mathbf{k}} G_{ds\sigma}^K(\mathbf{k}, \omega)
        \Big], \\
\end{align}
where
\begin{equation}
    G_{ds\sigma}^K(\mathbf{k}, \tau, \tau')
    =   -i \Braket{
        \Big[
            d_{0\sigma}(\tau), \psi_{s\mathbf{k}\sigma}^{\dagger}(\tau')
        \Big]_ {\zeta}
        }.
\end{equation}

To calculate the Green's function $G_{s\sigma}^K$,
we construct the interaction picture
by treating $H_T+V_{\eta}$ as a perturbation,
and we calculate $G_{ds\sigma}^K$ by Wick expansion
\begin{equation}
\begin{split}
    &G_{ds\sigma}^C(\mathbf{k}, \tau, \tau') \\
    &=  -i\Braket{T_C\Big[
            d_{0\sigma}(\tau)
            \psi_{s\mathbf{k}\sigma}^{\dagger}(\tau')
        \Big]} \\
    &=  \sum_{n=0}^{\infty}
        \frac{(-i)^{n+1}}{n!}
        \int_C d\tau_1 \cdots \int_C d\tau_n
        \Braket{T_C\Big[
            \{ H_T(\tau_1) + V_{\eta}(\tau_1) \}
            \cdots
            \{ H_T(\tau_n) + V_{\eta}(\tau_n) \}
            d_{0\sigma}(\tau)
            \psi_{s\mathbf{k}\sigma}^{\dagger}(\tau')
        \Big]}_0 \\
    &=  g_{ds\sigma}^C(\mathbf{k}, \tau, \tau') \\
        &\quad+
        \sum_{n=1}^{\infty}
        \frac{(-i)^{n+1}}{n!}
        \int_C d\tau_1 \cdots \int_C d\tau_n
        \sum_{j=1}^n \sum_{s_j \mathbf{k}_j \sigma_j}
        (-t_{\mathbf{k}_j}^*) \delta_{ss_j}
        \delta_{\mathbf{k}\mathbf{k}_j} \delta_{\sigma\sigma_j}
        g_{s\sigma}^C(\mathbf{k}, \tau_j, \tau') \\
        &\quad\times
        \Braket{T_C\Big[
            \{ H_T(\tau_1) + V_{\eta}(\tau_1) \}
            \cdots
            \{ H_T(\tau_n) + V_{\eta}(\tau_n) \}
            d_{0\sigma}(\tau)
            d_{0\sigma_j}^{\dagger}(\tau_j)
        \Big]}_0, \\
\end{split}
\end{equation}
where
\begin{equation}
    g_{ds\sigma}^C(\mathbf{k}, \tau, \tau')
    =   -i\Braket{T_C\Big[
            d_{0\sigma}(\tau)
            \psi_{s\mathbf{k}\sigma}^{\dagger}(\tau')
        \Big]}_0
    =   0,
\end{equation}
\begin{align}
    -i\Braket{T_C\Big[
        \psi_{s_j\mathbf{k}_j\sigma_j}^{\dagger}(\tau_j)
        \psi_{s\mathbf{k}\sigma}^{\dagger}(\tau')
    \Big]}_0
    &=
    -i\Braket{T_C\Big[
        \psi_{s\mathbf{k}\sigma}^{\dagger}(\tau_j)
        \psi_{s\mathbf{k}\sigma}^{\dagger}(\tau')
    \Big]}_0
    \delta_{ss_j} \delta_{\mathbf{k}\mathbf{k}_j}
    \delta_{\sigma\sigma_j}
    =
    g_{s\sigma}^C(\mathbf{k}, \tau_j, \tau')
    \delta_{ss_j} \delta_{\mathbf{k}\mathbf{k}_j}
    \delta_{\sigma\sigma_j}.
\end{align}
The contribution of each $j$ is equal and
the Green's function is given by
\begin{align}
    G_{ds\sigma}^C(\mathbf{k}, \tau, \tau')
    &=  \sum_{n=1}^{\infty}
        \frac{(-i)^{n+1}}{n!}
        \int_C d\tau_1 \cdots \int_C d\tau_n
        (-t_{\mathbf{k}}^*) \sum_{j=1}^n
        g_{s\sigma}^C(\mathbf{k}, \tau_j, \tau') \\
        &\quad\times
        \Braket{T_C\Big[
            \{ H_T(\tau_1) + V_{\eta}(\tau_1) \}
            \cdots
            \{ H_T(\tau_n) + V_{\eta}(\tau_n) \}
            d_{0\sigma}(\tau)
            d_{0\sigma_j}^{\dagger}(\tau_j)
        \Big]}_0 \\
    &=  \sum_{n=1}^{\infty}
        \frac{(-i)^{n+1}}{n!}
        \int_C d\tau_1 \cdots \int_C d\tau_n
        (-t_{\mathbf{k}}^*) \sum_{j=1}^n
        g_{s\sigma}^C(\mathbf{k}, \tau'', \tau') \\
        &\quad\times
        \Braket{T_C\Big[
            \{ H_T(\tau_1) + V_{\eta}(\tau_1) \}
            \cdots
            \{ H_T(\tau_n) + V_{\eta}(\tau_n) \}
            d_{0\sigma}(\tau)
            d_{0\sigma_j}^{\dagger}(\tau'')
        \Big]}_0 \\
    &=  -t_{\mathbf{k}}^* \int_C d\tau'' \,
        G_{11\sigma}^C(\tau, \tau'')
        g_{s\sigma}^C(\mathbf{k}, \tau'', \tau'),
\end{align}
where
\begin{align}
    G_{11\sigma}^C(\tau, \tau')
    &=  -i\Braket{T_C\Big[
            d_{0\sigma}(\tau)
            d_{0\sigma}^{\dagger}(\tau')
        \Big]} \\
    &=  \sum_{n=0}^{\infty} \frac{(-i)^{n+1}}{n!}
        \int_C d\tau_1 \cdots \int_C d\tau_n
        \Braket{T_C\Big[
            \{ H_T(\tau_1) + V_{\eta}(\tau_1) \}
            \cdots
            \{ H_T(\tau_n) + V_{\eta}(\tau_n) \}
            d_{0\sigma}(\tau) d_{0\sigma}^{\dagger}(\tau')
        \Big]}_0.
\end{align}
The Green's function $G_{11\sigma}^C$ corresponds
$G_{\frac{M+1}{2}\frac{M+1}{2}\sigma}^C$ in the multi-site case.
By using the Langreth rule
\begin{equation}
    G_{ds\sigma}^K(\mathbf{k}, \tau, \tau')
    =   -t_{\mathbf{k}}^* \int_C d\tau'' \,
        \Big(
            G_{11\sigma}^R(\tau, \tau'')
            g_{s\sigma}^K(\mathbf{k}, \tau'', \tau')
            +
            G_{11\sigma}^K(\tau, \tau'')
            g_{s\sigma}^A(\mathbf{k}, \tau'', \tau')
        \Big),
\end{equation}
we obtain
\begin{align}
    \Braket{I_{s\sigma}}
    &=  -\zeta \int d\tau' \sum_{\mathbf{k}} |t_{\mathbf{k}}|^2
        \operatorname{Re}
        \Big[
            G_{11\sigma}^R(\tau, \tau') g_{s\sigma}^K(\mathbf{k}, \tau', \tau)
            +
            G_{11\sigma}^K(\tau, \tau') g_{s\sigma}^A(\mathbf{k}, \tau', \tau)
        \Big] \notag \\
    &=  -\zeta \int \frac{d\omega}{2\pi} \sum_{\mathbf{k}} |t_{\mathbf{k}}|^2
        \operatorname{Re}
        \Big[
            G_{11\sigma}^R(\omega) g_{s\sigma}^K(\mathbf{k}, \omega)
            +
            G_{11\sigma}^K(\omega) g_{s\sigma}^A(\mathbf{k}, \omega)
        \Big], \\
    \Braket{I_{E, s\sigma}}
    &=  -\zeta \int d\tau' \sum_{\mathbf{k}}
        \epsilon_{\mathbf{k}} |t_{\mathbf{k}}|^2
        \operatorname{Re}
        \Big[
            G_{11\sigma}^R(\tau, \tau') g_{s\sigma}^K(\mathbf{k}, \tau', \tau)
            +
            G_{11\sigma}^K(\tau, \tau') g_{s\sigma}^A(\mathbf{k}, \tau', \tau)
        \Big] \notag \\
    &=  -\zeta \int d\frac{d\omega}{2\pi} \sum_{\mathbf{k}}
        \epsilon_{\mathbf{k}} |t_{\mathbf{k}}|^2
        \operatorname{Re}
        \Big[
            G_{11\sigma}^R(\omega) g_{s\sigma}^K(\mathbf{k}, \omega)
            +
            G_{11\sigma}^K(\omega) g_{s\sigma}^A(\mathbf{k}, \omega)
        \Big].
\end{align}
To calculate $G_{11\sigma}^C$,
we first eliminate the degree of freedom $\psi_{s\mathbf{k}\sigma}$.
One of the simplest way to achieve it is to adopt the path integral representation
of the partition function~\cite{kamenev2023field}.
Since the reservoir action is bilinear in  $\psi_{s\mathbf{k}\sigma}$,
we can exactly integrate out the reservoir degrees of freedom as follows:
\begin{equation}
\begin{split}
    &\int \mathcal{D}[\bar{\psi}_ {s\mathbf{k}\sigma}, \psi_{s\mathbf{k}\sigma}] \,
    \exp\left[
        -\int \frac{d\omega}{2\pi} \sum_{s\mathbf{k}}
        \left( \begin{matrix}
            \bar{\psi}_ {s\mathbf{k}\sigma}^{cl} & \bar{\psi}_ {s\mathbf{k}\sigma}^{q}
        \end{matrix} \right)
        \left( \begin{matrix}
            0 & -i\big[ g_{s\sigma}^{-1}(\mathbf{k}, \omega) \big]^A \\
            -i\big[ g_{s\sigma}^{-1}(\mathbf{k}, \omega) \big]^R & -i\big[ g_{s\sigma}^{-1}(\mathbf{k}, \omega) \big]^K
        \end{matrix} \right)
        \left( \begin{matrix}
            \psi_{s\mathbf{k}\sigma}^{cl} \\
            \psi_{s\mathbf{k}\sigma}^{q}
        \end{matrix} \right)
    \right] \\
    &\quad\times
    \exp
    \left[
        \int \frac{d\omega}{2\pi} \sum_{s\mathbf{k}}
        \left( \begin{matrix}
            \bar{\psi}_ {s\mathbf{k}\sigma}^{cl} & \bar{\psi}_ {s\mathbf{k}\sigma}^{q}
        \end{matrix} \right)
        it_{\mathbf{k}} \hat{\sigma}_ 1
        \left( \begin{matrix}
            d_{\sigma}^{cl} \\
            d_{\sigma}^{q}
        \end{matrix} \right)
        +
        \int \frac{d\omega}{2\pi} \sum_{s\mathbf{k}}
        \left( \begin{matrix}
            \bar{d}_ {\sigma}^{cl} & \bar{d}_ {\sigma}^{q}
        \end{matrix} \right)
        it_{\mathbf{k}}^* \hat{\sigma}_ 1
        \left( \begin{matrix}
            \psi_{s\mathbf{k}\sigma}^{cl} \\
            \psi_{s\mathbf{k}\sigma}^{q}
        \end{matrix} \right)
    \right] \\
    &=  \exp\left[
            \int \frac{d\omega}{2\pi} \sum_{s\mathbf{k}}
            \left( \begin{matrix}
                \bar{d}_ {\sigma}^{cl} & \bar{d}_ {\sigma}^{q}
            \end{matrix} \right)
            it_{\mathbf{k}}^* \hat{\sigma}_ 1
            \left( \begin{matrix}
                ig_{s\sigma}^K(\mathbf{k}, \omega) & ig_{s\sigma}^R(\mathbf{k}, \omega) \\
                ig_{s\sigma}^A(\mathbf{k}, \omega) & 0
            \end{matrix} \right)
            it_{\mathbf{k}} \hat{\sigma}_ 1
            \left( \begin{matrix}
                d_{\sigma}^{cl} \\
                d_{\sigma}^{q}
            \end{matrix} \right)
        \right] \\
    &=  \exp\left[
            -i\int \frac{d\omega}{2\pi}
            \left( \begin{matrix}
                \bar{d}_ {\sigma}^{cl} & \bar{d}_ {\sigma}^{q}
            \end{matrix} \right)
            \sum_{\mathbf{k}s} |t_{\mathbf{k}}|^2
            \left( \begin{matrix}
                0                                 & g_{s\sigma}^A(\mathbf{k}, \omega) \\
                g_{s\sigma}^R(\mathbf{k}, \omega) & g_{s\sigma}^K(\mathbf{k}, \omega)
            \end{matrix} \right)
            \left( \begin{matrix}
                d_{\sigma}^{cl} \\
                d_{\sigma}^{q}
            \end{matrix} \right)
        \right].
\end{split}
\end{equation}
By defining the energy shift induced by the reservoirs
\begin{align}
    R_{\sigma}(\omega)
    &=  2 \sum_{\mathbf{k}} |t_{\mathbf{k}}|^2
        \operatorname{Re} \Big[ g_{L\sigma}^R(\mathbf{k}, \omega) \Big]
    =   2 \sum_{\mathbf{k}} |t_{\mathbf{k}}|^2
        \operatorname{Re} \Big[ g_{R\sigma}^R(\mathbf{k}, \omega) \Big], \\
    \Gamma_{\sigma}(\omega)
    &=  -2 \sum_{\mathbf{k}} |t_{\mathbf{k}}|^2
        \operatorname{Im} \Big[ g_{L\sigma}^R(\mathbf{k}, \omega) \Big]
    =   -2 \sum_{\mathbf{k}} |t_{\mathbf{k}}|^2
        \operatorname{Im} \Big[ g_{R\sigma}^R(\mathbf{k}, \omega) \Big],
\end{align}
it follows that unperturbed Green's functions are given by
\begin{alignat}{2}
    g_{11\sigma}^C(\tau, \tau')
    &=  -i\Braket{T_C\Big[
            d_{0\sigma}(\tau) d_{0\sigma}^{\dagger}(\tau')
        \Big]}_0 \\
    g_{d\sigma}^{R/A}(\omega)
    &=  \frac{1}
        {
            \big[
                \omega
                -
                \epsilon_0
                -
                R_{\sigma}(\omega)
            \big]
            \pm
            i \Gamma_{\sigma}(\omega)},
    \quad\quad
    &&\Big[ g_{11\sigma}^{-1}(\omega) \Big]^{R/A}
    =   \omega
        -
        \epsilon_0
        -
        R_{\sigma}(\omega)
        \pm
        i \Gamma_{\sigma}(\omega), \\
    g_{11\sigma}^K(\omega)
    &=  \frac{
            -2i\Gamma_{\sigma}(\omega)
            (
                1 + \zeta n_{L\sigma}(\omega) + \zeta n_{R\sigma}(\omega)
            )
        }
        {
            \big[
                \omega
                -
                \epsilon_0
                -
                R_{\sigma}(\omega)
            \big]^2
            +
            \big[ \Gamma_{\sigma}(\omega) \big]^2
        },
    \quad\quad
    &&\Big[ g_{11\sigma}^{-1}(\omega) \Big]^K
    =   2i\Gamma_{\sigma}(\omega)
        (
            1 + \zeta n_{L\sigma}(\omega) + \zeta n_{R\sigma}(\omega)
        ),
\end{alignat}
where the distribution function $n_{s\sigma}(\omega)$ is
\begin{equation}
    n_{s\sigma}(\omega)
    =   \frac{1}{e^{(\omega-\Delta\mu_{s\sigma})/T_s}-\zeta},
\end{equation}
with $\mu = \sum_{s\sigma} \mu_{s\sigma}/4$ and
$\Delta\mu_{s\sigma}=\mu_{s\sigma}-\mu$.

We next examine effects of $V_{\eta}$.
In contrast to the dephasing model, the present setup does not allow the cancellation of complex diagrams, including crossing diagrams and bath renormalization diagrams. Therefore, we focus on the weak-coupling regime, where dissipation can be treated perturbatively. Furthermore, to maintain consistency with the Lindblad dynamics, diagrams associated with bath renormalization must be excluded. To this end, we adopt the SCBA, which fulfills these conditions. Under this approximation, the contour-ordered Green's function satisfies
\begin{equation}
    G_{11\sigma}^C(\tau, \tau')
    =   g_{11\sigma}^C(\tau, \tau')
        +
        \int_C d\tau_1 \int_C d\tau_2 \,
        g_{11\sigma}^C(\tau, \tau_1)
        \Sigma_{\sigma}^C(\tau_1, \tau_2)
        G_{11\sigma}^C(\tau_2, \tau'),
\end{equation}
where
\begin{equation}
    \Sigma_{\sigma}^C(\tau, \tau')
    =   i\zeta g_{\eta}^C(\tau, \tau')
        G_{11\bar{\sigma}}^C(\tau', \tau),
\end{equation}
with the noise Green's functions
\begin{align}
    g_{\eta}^C(\tau, \tau')
    &=  -i\Braket{T_C\Big[
            \eta(\tau) \eta^{\dagger}(\tau')
        \Big]}, \\
    g_{\eta}^{R/A}(\tau, \tau')
    &=  \mp\frac{i}{2} \gamma\delta(\tau-\tau'), \\
    g_{\eta}^K(\tau, \tau')
    &=  -i\gamma\delta(\tau-\tau')
    =   g_{\eta}^>(\tau, \tau')
    =   2g_{\eta}^T(\tau, \tau')
    =   2g_{\eta}^{\tilde{T}}(\tau, \tau'),\\
    g_{\eta}^<(\tau, \tau')
    &=  0.
\end{align}
The Langreth rules of the function
$C^C(\tau, \tau') = A^C(\tau, \tau') B^C(\tau', \tau)$
\begin{align}
    C^R(\tau, \tau')
    &=  A^R(\tau, \tau') B^<(\tau', \tau)
        +
        A^<(\tau, \tau') B^A(\tau', \tau), \\
    C^A(\tau, \tau')
    &=  A^<(\tau, \tau') B^R(\tau', \tau)
        +
        A^A(\tau, \tau') B^<(\tau', \tau), \\
    C^K(\tau, \tau')
    &=  A^K(\tau, \tau') B^<(\tau', \tau)
        +
        A^<(\tau, \tau')
        \Big(
            B^R(\tau', \tau)
            -
            B^A(\tau', \tau)
        \Big),
\end{align}
and $g_{\eta}^<(\tau, \tau')=0$ lead to
\begin{equation}
    \Sigma_{\sigma}^{R/A/K}(\tau, \tau')
    =   g_{\eta}^{R/A/K}(\tau, \tau')
        n_{0\bar{\sigma}},
\end{equation}
where
\begin{equation}
    n_{0\sigma}
    =   \frac{1}{-i\zeta} G_{11\sigma}^<(\tau, \tau) 
    =   \frac{1}{-i\zeta} \int_{-\infty}^{\infty}
        \frac{d\omega}{2\pi} G_{11\sigma}^<(\omega).
\end{equation}
Each component of the Green's function is given by
\begin{align}
    G_{11\sigma}^{R/A}(\omega)
    &=  \frac{1}
        {
            \Big[ g_{d\sigma}^{-1}(\omega) \Big]^{R/A}
            -
            \Sigma_{\sigma}^{R/A}(\omega)
        }
    =   \frac{1}
        {
            \omega - \epsilon - R_{\sigma}(\omega)
            \pm
            i\big[
                \Gamma_{\sigma}(\omega)
                +
                \frac{\gamma}{2} n_{0\bar{\sigma}}
            \big]
        }, \\
    G_{11\sigma}^K(\omega)
    &=  -G_{11\sigma}^R(\omega)
        \Big(
            \Big[ g_{d\sigma}^{-1}(\omega) \Big]^K
            -
            \Sigma_{\sigma}^K(\omega)
        \Big)
        G_{11\sigma}^A(\omega)
    =   \frac{
            -2i\big( \Gamma_{\sigma}(\omega) + \frac{\gamma}{2} n_{d\bar{\sigma}} \big)
            -
            2i\zeta\Gamma_{\sigma}(\omega)(n_{L\sigma}(\omega) + n_{R\sigma}(\omega))
        }{
            \big[
                \omega - \epsilon - R_{\sigma}(\omega)
            \big]^2
            +
            \big[
                \Gamma_{\sigma}(\omega)
                +
                \frac{\gamma}{2} n_{0\bar{\sigma}}
            \big]^2
        }, \\
    G_{11\sigma}^<(\omega)
    &=  \frac{
            -i\zeta\Gamma_{\sigma}(\omega)(n_{L\sigma}(\omega) + n_{R\sigma}(\omega))
        }{
            \big[
                \omega - \epsilon - R_{\sigma}(\omega)
            \big]^2
            +
            \big[
                \Gamma_{\sigma}(\omega)
                +
                \frac{\gamma}{2} n_{0\bar{\sigma}}
            \big]^2
        },
\end{align}
with the self-consistent integral equation
\begin{equation}
    n_{0\sigma}
    =   \int_{-\infty}^{\infty} \frac{d\omega}{2\pi}
        \frac{
            \Gamma_{\sigma}(\omega)
            (n_{L\sigma}(\omega) + n_{R\sigma}(\omega))
        }{
            \big[
                \omega - \epsilon - R_{\sigma}(\omega)
            \big]^2
            +
            \big[
                \Gamma_{\sigma}(\omega)
                +
                \frac{\gamma}{2} n_{0\bar{\sigma}}
            \big]^2
        }.
\end{equation}
The imaginary part of the reservoir's Green's function
is proportional to a delta function
\begin{align}
    g_{s\sigma}^{R/A}(\mathbf{k}, \omega)
    &=  \frac{1}
        {
            (\omega-\Delta\mu_{s\sigma})
            -
            (\epsilon_{\mathbf{k}}-\mu_{s\sigma})
            \pm
            i0^+
        }
    =   \mathcal{P}
        \left( \frac{1}{\omega + \mu - \epsilon_{\mathbf{k}}} \right)
        \mp i\pi\delta(\omega + \mu - \epsilon_{\mathbf{k}}), \\
    g_{s\sigma}^K(\mathbf{k}, \omega)
    &=   -2\pi i\delta(\omega + \mu - \epsilon_{\mathbf{k}})
        \big( 1 + 2\zeta n_{s\sigma}(\omega) \big),
\end{align}
leading to
\begin{equation}
    \epsilon_{\mathbf{k}}
    \operatorname{Im}
    \Big[ g_{s\sigma}^{R/A/K}(\mathbf{k}, \omega) \Big]
    =   (\omega + \mu)
        \operatorname{Im}
        \Big[ g_{s\sigma}^{R/A/K}(\mathbf{k}, \omega) \Big].
\end{equation}
Then, we obtain
\begin{align}
    \Braket{I_{s\sigma}}
    &=  \int \frac{d\omega}{2\pi}
        \Big[
            \mathcal{T}_{\sigma}(\omega)
            \big(
                2n_{s\sigma}(\omega)
                -
                n_{L\sigma}(\omega)
                -
                n_{R\sigma}(\omega)
            \big)
            +
            \mathcal{L}_{\sigma}(\omega) n_{s\sigma}(\omega)
        \Big], \\
    \Braket{I_{E, s\sigma}}
    &=  \int \frac{d\omega}{2\pi} (\omega + \mu)
        \Big[
            \mathcal{T}_{\sigma}(\omega)
            \big(
                2n_{s\sigma}(\omega)
                -
                n_{L\sigma}(\omega)
                -
                n_{R\sigma}(\omega)
            \big)
            +
            \mathcal{L}_{\sigma}(\omega) n_{s\sigma}(\omega)
        \Big],
\end{align}
where $\mathcal{T}_{\sigma}(\omega)$ and $\mathcal{L}_{\sigma}(\omega)$
respectively represent the transmittance and loss probability defined as
\begin{align}
    \mathcal{T}_{\sigma}(\omega)
    &=  \frac{
            \big[ \Gamma_{\sigma}(\omega) \big]^2
        }{
            \big[
                \omega - \epsilon - R_{\sigma}(\omega)
            \big]^2
            +
            \big[
                \Gamma_{\sigma}(\omega)
                +
                \frac{\gamma}{2} n_{0\bar{\sigma}}
            \big]^2
            }, \\
    \mathcal{L}_{\sigma}(\omega)
    &=  \frac{
            \Gamma_{\sigma}(\omega) \gamma n_{0\bar{\sigma}}
        }{
            \big[
                \omega - \epsilon - R_{\sigma}(\omega)
            \big]^2
            +
            \big[
                \Gamma_{\sigma}(\omega)
                +
                \frac{\gamma}{2} n_{0\bar{\sigma}}
            \big]^2
        }.
\end{align}
The particle and energy currents are given by
\begin{align}
    I
    &=  \frac{1}{2} \sum_{\sigma}
        \Big(
            \Braket{I_{L\sigma}}
            -
            \Braket{I_{R\sigma}}
        \Big)
    =   \sum_{\sigma}
        \int \frac{d\omega}{2\pi}
        \left[
            \mathcal{T}_ {\sigma}(\omega)
            +
            \frac{\mathcal{L}_ {\sigma}(\omega)}{2}
        \right]
        \big[ n_{L\sigma}(\omega) - n_{R\sigma}(\omega) \big], \\
    I_E
    &=  \frac{1}{2} \sum_{\sigma}
        \Big(
            \Braket{I_{E, L\sigma}}
            -
            \Braket{I_{E, R\sigma}}
        \Big)
    =   \sum_{\sigma}
        \int \frac{d\omega}{2\pi} (\omega + \mu)
        \left[
            \mathcal{T}_ {\sigma}(\omega)
            +
            \frac{\mathcal{L}_ {\sigma}(\omega)}{2}
        \right]
        \big[ n_{L\sigma}(\omega) - n_{R\sigma}(\omega) \big].
\end{align}
In addition, the particle loss is obtained by
\begin{equation}
    -\dot{N}
    =   \sum_{\sigma}
        \Big(
            \Braket{I_{L\sigma}}
            +
            \Braket{I_{R\sigma}}
        \Big)
    =   \sum_{\sigma} \int \frac{d\omega}{2\pi}
        \mathcal{L}_ {\sigma}(\omega)
        \big[ n_{L\sigma}(\omega) + n_{R\sigma}(\omega) \big]
    =   \gamma \sum_{\sigma} n_{0\sigma} n_{0\bar{\sigma}}
    =   2\gamma n_{0\uparrow} n_{0\downarrow}.
\end{equation}

\section{Two-particle loss: multi-site case}
\label{TwoBodyLossMultiSite}
In this section, we investigate the multi-site Hamiltonian
\begin{equation}
    H
    =   \sum_{s=L, R} \sum_{\sigma=\uparrow, \downarrow}
        H_{s\sigma}
        +
        H_T
        +
        \sum_{\alpha = 0, \pm}
        H_{1D}^{\alpha},
\end{equation}
where
\begin{align}
    H_{s\sigma}
    &=  \sum_{\mathbf{k}}
        (\epsilon_{\mathbf{k}} - \mu_{s\sigma})
        \psi_{s\mathbf{k}\sigma}^{\dagger}
        \psi_{s\mathbf{k}\sigma}, \\
    H_T
    &=  -\sum_{s\mathbf{k}\sigma}
        \Big(
            t_{\mathbf{k}}
            \psi_{s\mathbf{k}\sigma}^{\dagger} d_{0\sigma}
            +
            t_{\mathbf{k}}^*
            d_{0\sigma}^{\dagger} \psi_{\mathbf{k}s\sigma}
        \Big), \\
    H_{1D}^0
    &=  \sum_{\sigma} \sum_{i=-N}^N
        \epsilon_i
        d_{i\sigma}^{\dagger} d_{i\sigma}, \\
    H_{1D}^{\pm}
    &=  -\sum_{\sigma} \sum_{i=1}^N
        t_{\pm i} \Big(
            d_{\pm i\sigma}^{\dagger} d_{\pm(i-1)\sigma}
            +
            d_{\pm(i-1)\sigma}^{\dagger} d_{\pm i\sigma}
        \Big).
\end{align}
We introduce a two-particle loss in a quantum dot
by the noise field
\begin{equation}
    V_{\eta}
    =   d_{0\uparrow}^{\dagger}
        d_{0\downarrow}^{\dagger}
        \eta
        +
        \eta^{\dagger}
        d_{0\downarrow}
        d_{0\uparrow}.
\end{equation}
By employing the sane method as single-site case,
we obtain
\begin{align}
    G_{dL(R)\sigma}^K(\mathbf{k}, \tau, \tau')
    &=  -i \Braket{
        \Big[
            d_{-N(N)\sigma}(\tau),
            \psi_{\mathbf{k}L(R)\sigma}^{\dagger}(\tau)
        \Big]_ {\zeta}} \\
    &=  -t_{\mathbf{k}}^* \int_{-\infty}^{\infty} d\tau''
        \Big[
            G_{11(MM)\sigma}^R(\tau, \tau'')
            g_{L(R)\sigma}^K(\mathbf{k}, \tau'', \tau')
            +
            G_{11(MM)\sigma}^K(\tau, \tau'')
            g_{L(R)\sigma}^A(\mathbf{k}, \tau'', \tau')
        \Big], \\
    \Braket{ I_{L(R)\sigma} }
    &=  \zeta \sum_{\mathbf{k}}
        \operatorname{Re}
        \Big[
            t_{\mathbf{k}} G_{dL(R)\sigma}^K(\mathbf{k}, \tau, \tau)
        \Big] \notag \\
    &=  -\zeta \int d\tau' \sum_{\mathbf{k}} |t_{\mathbf{k}}|^2
        \operatorname{Re}
        \Big[
            G_{11(MM)\sigma}^R(\tau, \tau')
            g_{L(R)\sigma}^K(\mathbf{k}, \tau', \tau)
            +
            G_{11(MM)\sigma}^K(\tau, \tau')
            g_{L(R)\sigma}^A(\mathbf{k}, \tau', \tau)
        \Big], \\
    \Braket{ I_{E, L(R)\sigma} }
    &=  \zeta \sum_{\mathbf{k}} \epsilon_{\mathbf{k}}
        \operatorname{Re}
        \Big[
            t_{\mathbf{k}} G_{dL(R)\sigma}^K(\mathbf{k}, \tau, \tau)
        \Big] \notag \\
    &=  -\zeta \int d\tau' \sum_{\mathbf{k}}
        \epsilon_{\mathbf{k}} |t_{\mathbf{k}}|^2
        \operatorname{Re}
        \Big[
            G_{11(MM)\sigma}^R(\tau, \tau')
            g_{L(R)\sigma}^K(\mathbf{k}, \tau', \tau)
            +
            G_{11(MM)\sigma}^K(\tau, \tau')
            g_{L(R)\sigma}^A(\mathbf{k}, \tau', \tau)
        \Big],
\end{align}
where
\begin{align}
    G_{ij\sigma}^C(\tau, \tau')
    &=  -i
        \Braket{T_C\Big[
            d_{(i-N-1)\sigma}(\tau) d_{(j-N-1)\sigma}^{\dagger}(\tau')
        \Big]}.
\end{align}
The retarded and advanced components of
the matrix of the Green's function
\begin{align}
    \mathbf{G}_ {d\sigma}^C
    &=  \left( \begin{matrix}
            G_{11\sigma}^C & G_{12\sigma}^C & \cdots & G_{1M\sigma}^C \\
            G_{21\sigma}^C & G_{22\sigma}^C & \cdots & G_{2M\sigma}^C \\
            \vdots        & \vdots          & \ddots & \vdots         \\
            G_{M1\sigma}^C & G_{M2\sigma}^C & \cdots & G_{MM\sigma}^C \\
        \end{matrix} \right),
\end{align}
is given by
\begin{align}
    \Big[ \mathbf{G}_ {d\sigma}^{-1}(\omega) \Big]^{R/A}
    =   \Big[ \mathbf{g}_ {d\sigma}^{-1}(\omega) \Big]^{R/A}
        -
        \mathbf{\Sigma}_ {d\sigma}^{R/A},
\end{align}
where
\begin{align}
    \Big[ \mathbf{g}_ {d\sigma}^{-1}(\omega) \Big]^{R/A}
    &=  \left( \begin{matrix}
            \omega - \epsilon_{-N} \pm i0^+ & 0 & \cdots & 0 & \cdots & 0 \\
            0 & \omega - \epsilon_{-N-1} \pm i0^+ & \cdots & 0 & \cdots & 0 \\
            \vdots & \vdots & \ddots & \vdots & \ddots & \vdots \\
            0 & 0 & \cdots & \omega - \epsilon_0 \pm i0^+ & \cdots & 0 \\
            \vdots & \vdots & \ddots & \vdots & \ddots & \vdots \\
            0 & 0 & \cdots & 0 & \cdots & \omega - \epsilon_N \pm i0^+
        \end{matrix} \right),
\end{align}
\begin{align}
    \mathbf{\Sigma}_ {d\sigma}^{R/A}(\omega)
    =   \left( \begin{matrix}
            \frac{R_{\sigma}(\omega) \mp i\Gamma_{\sigma}(\omega)}{2} & -t_{-N} & \cdots & 0 & \cdots & \cdots & 0 \\
            -t_{-N} & 0 & \cdots & \cdots & \cdots & \cdots & 0 \\
            \vdots & \vdots & \ddots & -t_{-1} & \ddots & \cdots & \vdots \\
            0 & \cdots & -t_{-1} & \mp i\frac{\gamma}{2} n_{0\bar{\sigma}} & -t_1 & \cdots & 0 \\
            \vdots & \vdots & \ddots & -t_1 & \ddots & \ddots & \vdots \\
            \vdots & \vdots & \ddots & \vdots & \ddots & \ddots & -t_N \\
            0 & 0 & \cdots & \cdots & \cdots & -t_N & \frac{R_{\sigma}(\omega) \mp i\Gamma_{\sigma}(\omega)}{2}
        \end{matrix} \right).
\end{align}
The inverse of a tridiagonal matrix
\begin{align}
    T
    =   \left( \begin{matrix}
            a_1 & b_1 & \cdots & 0 \\
            c_1 & a_2 & \ddots & \vdots \\
            \vdots & \ddots & \ddots & b_{L-1} \\
            0 & \cdots & c_{L-1} & a_L
        \end{matrix} \right),
\end{align}
is obtained by~\cite{Fonseca2007, Jin2022, Usmani1994, Turkeshi2021}
\begin{align}
    \big[ T^{-1} \big]_ {ij}
    =   \left\{ \begin{array}{ll}
            (-1)^{i+j} b_i \cdots b_{j-1} \theta_{i-1} \phi_{j+1} / \theta_L & (i < j) \\
            \theta_{i-1} \phi_{j+1} / \theta_L & (i=j) \\
            (-1)^{i+j} c_j \cdots c_{i-1} \theta_{j-1} \phi_{i+1} / \theta_L & (i > j)
        \end{array} \right.,
\end{align}
where
\begin{align*}
    &\theta_i = a_i \theta_{i-1} - b_{i-1} c_{i-1} \theta_{i-2} \,\, (i=2, 3, \cdots, L), \quad \theta_0 = 1, \quad \theta_1 = a_1, \\
    &\phi_i = a_i \phi_{i+1} - b_i c_i \phi_{i+2} \,\, (i=1, 2, \cdots, L-1), \quad \phi_{L} = a_L, \quad \phi_{L+1} = 1.
\end{align*}
The Keldysh component of
the matrix of the Green's function is given by
\begin{align}
    \mathbf{G}_ {d\sigma}^K(\omega)
    =   \mathbf{G}_ {d\sigma}^R(\omega) \mathbf{\Sigma}_ {d\sigma}^K(\omega) \mathbf{G}_ {d\sigma}^A(\omega),
\end{align}
where
\begin{align*}
    \mathbf{\Sigma}_ {d\sigma}^K(\omega)
    =   \left( \begin{matrix}
            -i\Gamma_{\sigma}(\omega)[1 + 2\zeta n_{L\sigma}(\omega)] & & & & \\
             & \ddots & & & \\
             & & -i\gamma n_{0\bar{\sigma}} & & \\
             & & & \ddots & \\
             & & & & -i\Gamma_{\sigma}(\omega)[1 + 2\zeta n_{R\sigma}(\omega)]
        \end{matrix} \right).
\end{align*}
The $(ij)$ elements is obtained by
\begin{align*}
    G_{ij\sigma}^K(\omega)
    &=  G_{i1\sigma}^R(\omega) \Big\{ -i\Gamma_{\sigma}(\omega)[1 + 2\zeta n_{L\sigma}(\omega)] \Big\} G_{1j\sigma}^A(\omega) \\
        &+
        G_{i\frac{M+1}{2}}^R(\omega) \Big\{ -i\gamma n_{0\bar{\sigma}} \Big\} G_{\frac{M+1}{2}j\sigma}^A(\omega) \\
        &+
        G_{iM\sigma}^R(\omega) \Big\{ -i\Gamma_{\sigma}(\omega)[1 + 2\zeta n_{R\sigma}(\omega)] \Big\} G_{Mj\sigma}^A(\omega).
\end{align*}
In particular, the $(ij)=(11)$ element is
\begin{align}
    G_{11\sigma}^K(\omega)
    &=  \Big\{ -i\Gamma_{\sigma}(\omega)[1 + 2\zeta n_{L\sigma}(\omega)] \Big\} \Big| G_{11\sigma}^R(\omega) \Big|^2 \\
        &+
        \Big\{ -i\gamma n_{0\bar{\sigma}} \Big\} \Big| G_{1\frac{M+1}{2}\sigma}^R(\omega) \Big|^2 \\
        &+
        \Big\{ -i\Gamma_{\sigma}(\omega)[1 + 2\zeta n_{R\sigma}(\omega)] \Big\} \Big| G_{1M\sigma}^R(\omega) \Big|^2,
\end{align}
and the $(ij)=(MM)$ element is
\begin{align}
    G_{MM\sigma}^K(\omega)
    &=  \Big\{ -i\Gamma_{\sigma}(\omega)[1 + 2\zeta n_{L\sigma}(\omega)] \Big\} \Big| G_{M1\sigma}^R(\omega) \Big|^2 \\
        &+
        \Big\{ -i\gamma n_{0\bar{\sigma}} \Big\} \Big| G_{M\frac{M+1}{2}\sigma}^R(\omega) \Big|^2 \\
        &+
        \Big\{ -i\Gamma_{\sigma}(\omega)[1 + 2\zeta n_{R\sigma}(\omega)] \Big\} \Big| G_{MM\sigma}^R(\omega) \Big|^2,
\end{align}
with the property of the Green's function
$G^A=[G^R]^{\dagger}.$
The inverse matrix inherits the following symmetries possessed by the original matrix
\begin{alignat}{2}
    &A^T = A
    \quad &&\Rightarrow \quad
    (A^{-1})^T = A^{-1}, \\
    &A_{ij} = A_{M-j+1, M-i+1}
    \quad &&\Rightarrow \quad
    \big( A^{-1} \big)_ {ij}
    =   \big( A^{-1} \big)_ {(M-j+1)(M-i+1)},
\end{alignat}
because
\begin{align}
    I
    =   I^T
    =   (AA^{-1})^T
    =   (A^{-1})^T A^T
    =   (A^{-1})^T A,
\end{align}
\begin{align}
    &\delta_{ij}
    =   \sum_{k=1}^M A_{ik} (A^{-1})_ {kj}
    =   \sum_{k=1}^M A_{(M-k+1)(M-i+1)} (A^{-1})_ {kj}
    =   \sum_{l=1}^M (A^{-1})_ {lj} A_{(M-l+1)(M-i+1)} \\
    &=  \delta_{(M-j+1)(M-i+1)}
    =   \sum_{k=1}^M (A^{-1})_ {(M-j+1)k} A_{k(M-i+1)}
    =   \sum_{l=1}^M (A^{-1})_ {(M-j+1)(M-l+1)} A_{(M-l+1)(M-i+1)}.
\end{align}
By using the above symmetries
\begin{align}
    G_{11\sigma}^R(\omega)
    =   G_{MM\sigma}^R(\omega),
    \quad
    G_{1\frac{M+1}{2}\sigma}^R(\omega)
    =   G_{\frac{M+1}{2}1\sigma}^R(\omega)
    =   G_{M\frac{M+1}{2}\sigma}^R(\omega)
    =   G_{\frac{M+1}{2}M\sigma}^R(\omega),
\end{align}
we obtain
\begin{align}
    \Braket{I_{L(R)\sigma}}
    &=  -\zeta \int \frac{d\omega}{2\pi}\Gamma_{\sigma}(\omega)
        \operatorname{Im}\Big[ G_{11\sigma}^R(\omega) \Big] [1 + 2\zeta n_{L(R)\sigma}(\omega)] \notag \\
        &\quad
        -\zeta \int \frac{d\omega}{2\pi}\Gamma_{\sigma}(\omega)
        \frac{
            \Gamma_{\sigma}(\omega)
            \Big(
                [1 + 2\zeta n_{L(R)\sigma}(\omega)] \Big| G_{11\sigma}^R(\omega) \Big|^2
                +
                [1 + 2\zeta n_{R(L)\sigma}(\omega)] \Big| G_{1M\sigma}^R(\omega) \Big|^2
            \Big)
        }{2} \notag \\
        &\quad
        -\zeta \int \frac{d\omega}{2\pi}\Gamma_{\sigma}(\omega)
        \frac{\gamma n_{0\bar{\sigma}} \Big| G_{1\frac{M+1}{2}\sigma}^R(\omega) \Big|^2}{2}, \\
    \Braket{I_{E, L(R)\sigma}}
    &=  -\zeta \int \frac{d\omega}{2\pi} (\omega+\mu) \Gamma_{\sigma}(\omega)
        \operatorname{Im}\Big[ G_{11\sigma}^R(\omega) \Big] [1 + 2\zeta n_{L(R)\sigma}(\omega)] \notag \\
        &\quad
        -\zeta \int \frac{d\omega}{2\pi} (\omega+\mu) \Gamma_{\sigma}(\omega)
        \frac{
            \Gamma_{\sigma}(\omega)
            \Big(
                [1 + 2\zeta n_{L(R)\sigma}(\omega)] \Big| G_{11\sigma}^R(\omega) \Big|^2
                +
                [1 + 2\zeta n_{R(L)\sigma}(\omega)] \Big| G_{1M\sigma}^R(\omega) \Big|^2
            \Big)
        }{2} \notag \\
        &\quad
        -\zeta \int \frac{d\omega}{2\pi} (\omega+\mu) \Gamma_{\sigma}(\omega)
        \frac{\gamma n_{0\bar{\sigma}} \Big| G_{1\frac{M+1}{2}\sigma}^R(\omega) \Big|^2}{2}.
\end{align}
The particle and energy currents, and particle loss are given by
\begin{align}
    I
    &=  \sum_{\sigma} \int \frac{d\omega}{2\pi} \Gamma_{\sigma}(\omega)
        \left[
            -\operatorname{Im}\Big[ G_{11\sigma}^R(\omega) \Big]
            -
            \frac{
                \Gamma_{\sigma}(\omega)
                \Big(
                    \Big| G_{11\sigma}^R(\omega) \Big|^2
                    -
                    \Big| G_{ML\sigma}^R(\omega) \Big|^2
                \Big)
            }{2}
        \right]
        [n_{L\sigma}(\omega) - n_{R\sigma}(\omega)], \\
    I_E
    &=  \sum_{\sigma} \int \frac{d\omega}{2\pi} (\omega + \mu) \Gamma_{\sigma}(\omega)
        \left[
            -\operatorname{Im}\Big[ G_{11\sigma}^R(\omega) \Big]
            -
            \frac{
                \Gamma_{\sigma}(\omega)
                \Big(
                    \Big| G_{11\sigma}^R(\omega) \Big|^2
                    -
                    \Big| G_{ML\sigma}^R(\omega) \Big|^2
                \Big)
            }{2}
        \right]
        [n_{L\sigma}(\omega) - n_{R\sigma}(\omega)], \\
    -\dot{N}
    &=  -\zeta \sum_{\sigma} \int \frac{d\omega}{2\pi}\Gamma_{\sigma}(\omega)
        \left[
            2\operatorname{Im}\Big[ G_{11\sigma}^R(\omega) \Big]
            +
            \Gamma_{\sigma}(\omega)
            \Big(
                \Big| G_{11\sigma}^R(\omega) \Big|^2
                +
                \Big| G_{1M\sigma}^R(\omega) \Big|^2
            \Big)
            +
            \gamma n_{0\bar{\sigma}} \Big| G_{1\frac{M+1}{2}\sigma}^R(\omega) \Big|^2
        \right] \notag \\
        &\quad-
        \sum_{\sigma} \int \frac{d\omega}{2\pi}\Gamma_{\sigma}(\omega)
        \left[
            2\operatorname{Im}\Big[ G_{11\sigma}^R(\omega) \Big]
            +
            \Gamma_{\sigma}(\omega)
            \Big(
                \Big| G_{11\sigma}^R(\omega) \Big|^2
                +
                \Big| G_{1M\sigma}^R(\omega) \Big|^2
            \Big)
        \right]
        [n_{L\sigma}(\omega) + n_{R\sigma}(\omega)].
\end{align}
The identity
\begin{align}
    \mathbf{G}_ {d\sigma}^R - \mathbf{G}_ {d\sigma}^A
    =   \mathbf{G}_ {d\sigma}^R \Big[ \Big( \mathbf{G}_ {d\sigma}^A \Big)^{-1} - \Big( \mathbf{G}_ {d\sigma}^R \Big)^{-1} \Big] \mathbf{G}_ {d\sigma}^A
    =   \mathbf{G}_ {d\sigma}^R \Big[ \mathbf{\Sigma}_ {d\sigma}^R - \mathbf{\Sigma}_ {d\sigma}^A \Big] \mathbf{G}_ {d\sigma}^A,
\end{align}
leads to
\begin{align}
    &2i\operatorname{Im}\Big[ G_{11\sigma}^R(\omega) \Big] \notag \\
    &=  G_{11\sigma}^R(\omega) - G_{11\sigma}^A(\omega) \notag \\
    &=  G_{11\sigma}^R(\omega)
        \Big(
            -i\Gamma(\omega)
        \Big)
        G_{11\sigma}^A(\omega)
        +
        G_{1\frac{M+1}{2}\sigma}^R(\omega)
        \Big(
            -i\gamma n_{0\bar{\sigma}}
        \Big)
        G_{\frac{M+1}{2}1\sigma}^A(\omega)
        +
        G_{1M\sigma}^R(\omega)
        \Big(
            -i\Gamma(\omega)
        \Big)
        G_{1M\sigma}^A(\omega) \notag \\
    &=  -i\Big[
            \Gamma(\omega)
            \Big\{
                \Big| G_{11\sigma}^R(\omega) \Big|^2
                +
                \Big| G_{1M\sigma}^R(\omega) \Big|^2
            \Big\}
            +
            \gamma n_{0\bar{\sigma}} \Big| G_{1\frac{M+1}{2}\sigma}^R(\omega) \Big|^2
        \Big],
\end{align}
Thus, we obtain
\begin{align}
  %  \therefore \quad
    2\operatorname{Im}\Big[ G_{11\sigma}^R(\omega) \Big]
    +
    \Gamma(\omega)
    \Big(
        \Big| G_{11\sigma}^R(\omega) \Big|^2
        +
        \Big| G_{1M\sigma}^R(\omega) \Big|^2
    \Big)
    +
    \gamma n_{0\bar{\sigma}} \Big| G_{1\frac{M+1}{2}\sigma}^R(\omega) \Big|^2
    =   0.
\end{align}
In total, we reach the following expressions:
\begin{align}
    I
    &=  \sum_{\sigma} \int \frac{d\omega}{2\pi}
        \Big[
            \mathcal{T}_ {\sigma}^M(\omega)
            +
            \frac{
                \mathcal{L}_ {\sigma}^M(\omega)
            }{2}
        \Big]
        [n_{L\sigma}(\omega) - n_{R\sigma}(\omega)], \\
    I_E
    &=  \sum_{\sigma} \int \frac{d\omega}{2\pi} (\omega + \bar{\mu})
        \Big[
            \mathcal{T}_ {\sigma}^M(\omega)
            +
            \frac{
                \mathcal{L}_ {\sigma}^M(\omega)
            }{2}
        \Big]
        [n_{L\sigma}(\omega) - n_{R\sigma}(\omega)], \\
    -\dot{N}
    &=  \sum_{\sigma} \int \frac{d\omega}{2\pi}
        \mathcal{L}_ {\sigma}^M(\omega)
        [n_{L\sigma}(\omega) + n_{R\sigma}(\omega)],
\end{align}
where
\begin{align}
    \mathcal{T}_ {\sigma}^M(\omega) = \Big[ \Gamma_{\sigma}(\omega) \Big]^2 \Big| G_{1M\sigma}^R(\omega) \Big|^2
    , \quad
    \mathcal{L}_ {\sigma}^M(\omega) = \Big[ \Gamma_{\sigma}(\omega) \gamma n_{0\bar{\sigma}} \Big] \Big| G_{1\frac{M+1}{2}\sigma}^R(\omega) \Big|^2.
\end{align}
We can also obtain the convenient expression 
of the particle loss.
Since the lesser Green's function possesses the 
following structure:
\begin{align}
    \mathbf{G}_ {d\sigma}^<(\omega)
    &=  \frac{
            \mathbf{G}_ {d\sigma}^K(\omega)
            -
            \big(
                \mathbf{G}_ {d\sigma}^R(\omega)
                -
                \mathbf{G}_ {d\sigma}^A(\omega)
            \big)
        }{2} \notag \\
    &=  \mathbf{G}_ {d\sigma}^R(\omega)
        \frac{
            \mathbf{\Sigma}_ {d\sigma}^K(\omega)
            -
            \big(
                \mathbf{\Sigma}_ {d\sigma}^R(\omega)
                -
                \mathbf{\Sigma}_ {d\sigma}^A(\omega)
            \big)
        }{2}
        \mathbf{G}_ {d\sigma}^A(\omega) \notag \\
    &=  -i\zeta\Gamma_{\sigma}(\omega)
        \mathbf{G}_ {d\sigma}^R(\omega)
        \left( \begin{matrix}
            n_{L\sigma}(\omega) & & \\
             & \ddots & \\
             & & n_{R\sigma}(\omega)
        \end{matrix} \right)
        \mathbf{G}_ {d\sigma}^A(\omega),
\end{align}
we have
\begin{align}
    G_{\frac{M+1}{2}\frac{M+1}{2}\sigma}^<(\omega)
    &=  -i\zeta\Gamma_{\sigma}(\omega)
        \Big[
            \Big| G_{\frac{M+1}{2}1}^R(\omega) \Big|^2 n_{L\sigma}(\omega)
            +
            \Big| G_{\frac{M+1}{2}M}^R(\omega) \Big|^2 n_{R\sigma}(\omega)
        \Big] \notag \\
    &=  -i\zeta\Gamma_{\sigma}(\omega)
        \Big| G_{1\frac{M+1}{2}}^R(\omega) \Big|^2
        [n_{L\sigma}(\omega)+n_{R\sigma}(\omega)].
\end{align}
By using the above equation, we obtain
\begin{align}
    n_{0\sigma}
    =   \frac{1}{-i\zeta} \int \frac{d\omega}{2\pi} G_{\frac{M+1}{2}\frac{M+1}{2}}^<(\omega)
    =   \int \frac{d\omega}{2\pi} \Gamma_{\sigma}(\omega)
        \Big| G_{1\frac{M+1}{2}}^R(\omega) \Big|^2
        [n_{L\sigma}(\omega)+n_{R\sigma}(\omega)].
\end{align}
The same expression in the single-site case is given by
\begin{align}
    -\dot{N} = 2\gamma n_{0\uparrow} n_{0\downarrow}.
\end{align}

\twocolumngrid

\nocite{*}

\bibliography{main_v3}% Produces the bibliography via BibTeX.

\end{document}